\documentclass[preprint,twocolumn,5p,times,authoryear]{elsarticle}

\graphicspath{{fig/}}
\usepackage[utf8]{inputenc}
\inputencoding{utf8}
\usepackage{siunitx}
\usepackage{amsmath}
\usepackage{amsfonts}
\usepackage{amssymb}
\usepackage{wrapfig}
\usepackage{subcaption}
\usepackage{soul}
\usepackage{multicol}
\usepackage{multirow}
\usepackage{url}
\usepackage{tikz}
\usepackage[hidelinks, hypertexnames=false]{hyperref}
\usepackage{acronym}
\usepackage{longtable}
\usepackage{hhline}
\usepackage[colorinlistoftodos,prependcaption,textsize=small]{todonotes}
\usepackage{lmodern}
\usepackage{dblfloatfix}
\usepackage{mathabx}
\usepackage{moresize}

\DeclareSIUnit\cpd{cpd}
\DeclareSIUnit\ppi{ppi}
\DeclareSIUnit\FPS{FPS}
\DeclareSIUnit[number-unit-product = ]\inch{\char`"}
\DeclareSIUnit\pix{px}
\DeclareSIUnit\pixel{px}
\DeclareSIUnit\Hz{Hz}

\NewDocumentCommand\angRange{O{} m m}{\SIrange[parse-numbers=false, #1]{\ang[parse-numbers=true]{#2}}{\ang[parse-numbers=true]{#3}}{}}

\newcolumntype{P}[1]{>{\centering\arraybackslash}p{#1}}

\newcommand{\new}[1]{\textcolor{black}{#1}}

\makeatletter
\newcommand*{\addFileDependency}[1]{
\typeout{(#1)}
\@addtofilelist{#1}
\IfFileExists{#1}{}{\typeout{No file #1.}}
}\makeatother

\usepackage{xr}
\journal{Vision Research}

\begin{document}

\begin{frontmatter}



\title{NEST: Neural Estimation by Sequential Testing}


\renewcommand{\thefootnote}{\fnsymbol{footnote}}

\author[inst1]{Sjoerd Bruin\corref{cor1}}
\cortext[cor1]{Corresponding author. Email address: s.bruin@rug.nl}

\affiliation[inst1]{organization={Bernoulli Institute, University of Groningen},
            addressline={Nijenborgh 9},
            city={Groningen},
            postcode={9747AG},
            country={The Netherlands}}

\author[inst1]{Ji\v{r}\'{i} Kosinka}
\author[inst1]{Cara Tursun}


\begin{abstract}
Adaptive psychophysical procedures aim to increase the efficiency and reliability of measurements. With increasing stimulus and experiment complexity in the last decade, estimating multi-dimensional psychometric functions has become a challenging task for adaptive procedures. If the experimenter has limited information about the underlying psychometric function, it is not possible to use parametric techniques developed for the multi-dimensional stimulus space. Although there are non-parametric approaches that use Gaussian process methods and specific hand-crafted acquisition functions, their performance is sensitive to proper selection of the kernel function, which is not always straightforward. In this work, we use a neural network as the psychometric function estimator and introduce a novel acquisition function for stimulus selection. We thoroughly benchmark our technique both using simulations and by conducting psychovisual experiments under realistic conditions. We show that our method outperforms the state of the art without the need to select a kernel function and significantly reduces the experiment duration.
\end{abstract}




\end{frontmatter}


\section{Introduction} \label{sec:introduction}

The relationship between the physical qualities of a stimulus and the perceptual performance can be numerically expressed through psychometric functions \citep{klein2001}. Adaptive psychometric procedures are one of the most frequently used methodologies to estimate psychometric functions by determining the sensory thresholds or response characteristics of the \ac{HVS}. These procedures adjust the experiment and stimulus parameters in response to the participant's performance, optimizing the data collection process by choosing the most informative stimulus and experiment parameters to be tested in each trial.

Over the last several decades, a large variety of adaptive psychometric procedures have emerged and become one of the fundamental tools in vision research to study aspects of the \ac{HVS} such as visual acuity, contrast sensitivity, and color perception \citep[for a review of earlier adaptive psychopsysical procedures, please see][]{Treutwein1995,leek2001}. Despite the success of these procedures, there are still significant challenges that remain to be addressed when the experiment and stimuli are defined by a high-dimensional parameter space. These challenges manifest themselves as slow convergence of the procedure and having an unknown form of the high-dimensional psychometric function. Slow convergence is mainly due to the well-known problem of the so-called ``curse of dimensionality'', which is also observed in regression analysis, and requires a sample selection strategy that aims to improve the information gained from each trial while avoiding redundancy. To this end, \citet{KontsevichTyler1999} developed a variation of the well-known quick estimation by sequential testing (\textit{QUEST}) procedure from \citet{WatsonPelli1983} called \textit{Psi} that uses entropy minimization to select the next stimulus to test, rather than simply using the mode of the posterior probability distribution. The Psi method can also handle the types of stimuli characterized by two-dimensional parameter vectors \citep{LesmesEtAl2006, LesmesEtAl2010, VulEtAl2010}. \citet{Watson2017} developed a framework called \textit{QUEST+} based on the Psi method which includes some other capabilities, such as handling an arbitrary number of trial outcomes instead of being limited to binary (e.g., detection/no detection) responses.

Although these methods provide efficient test sample selection in each trial, they require the user to specify the parametric form of the psychometric function. In practice, the parametric form may not be known, especially for complex stimuli and experiments with multiple parameters. Moreover, parametric methods require computation and storage of multidimensional probability distributions, which become intractable as the number of dimensions increases. To address this issue, a class of nonparametric methods based on the \ac{GP} \citep{MacKay1998} was developed, such as those of \citet{SongEtAl2015} and \citet{GardnerEtAl2015a}. \ac{GP} methods replace the explicit parametric form of the psychometric function with a generic kernel function, which makes them independent of the number of stimulus and experiment parameters. Samples are usually selected by optimizing an \textit{acquisition function}, which measures the amount of potential improvement in the psychometric function estimation from a newly selected sample (e.g., \ac{BALV} by \citet{Settles2009}, \ac{BALD} by \citet{HoulsbyEtAl2012}, and \ac{LSE} by \citet{GotovosEtAl2013}). Although \ac{GP} methods do not require the researcher to specify the parametric form of the psychometric function as input, the researcher must still choose an appropriate kernel and acquisition function. Unfortunately, this choice is not trivial and has a significant impact on the accuracy of the estimation of psychometric functions \citep{OwenEtAl2021}. As a result, a significant amount of analysis is required to determine the best kernel function and acquisition function depending on the properties of the underlying psychometric function.

In this paper, we introduce a non-parametric procedure called \ac{NEST} that uses a neural network-based psychometric function approximator. Similar to \ac{GP} methods, our approach keeps computational and storage complexity tractable with high-dimensional stimulus and experiment parameters. \new{Like other neural network based methods, our method involves the selection of parameters such as those in the learning scheme, regularization, and network size and architecture. However, different from \ac{GP} methods, which may be very sensitive to the choice of kernel and acquisition function for different problem domains, we demonstrate that a general set of parameter values can approximate a wide range of psychometric functions effectively across various contexts.} Furthermore, we show that \ac{NEST} has an estimation performance that is on par with or better than the best results produced by the set of \ac{GP} methods previously benchmarked by \citet{OwenEtAl2021} and \citet{LethamEtAl2022} on a set of synthetic psychometric functions. In addition to simulation-based benchmarks, we validate our method by showing two real use cases, which are estimating the eccentricity-dependent \ac{CSF} and the visibility of spatio-temporal \ac{DCT} bases. We compare our results with two previous studies, \citet{Barten1999} and \citet{TursunDidyk2022}. In our analysis, we observe that \ac{NEST} achieves a comparable level of accuracy in estimating psychometric functions with significantly fewer trials. We expect that \ac{NEST} will allow studying more complex types of stimuli and multi-dimensional psychometric functions in vision research, which was not feasible or convenient with existing adaptive psychometric procedures.
\begin{figure}[h!]
    \centering
    \includegraphics[width=\linewidth]{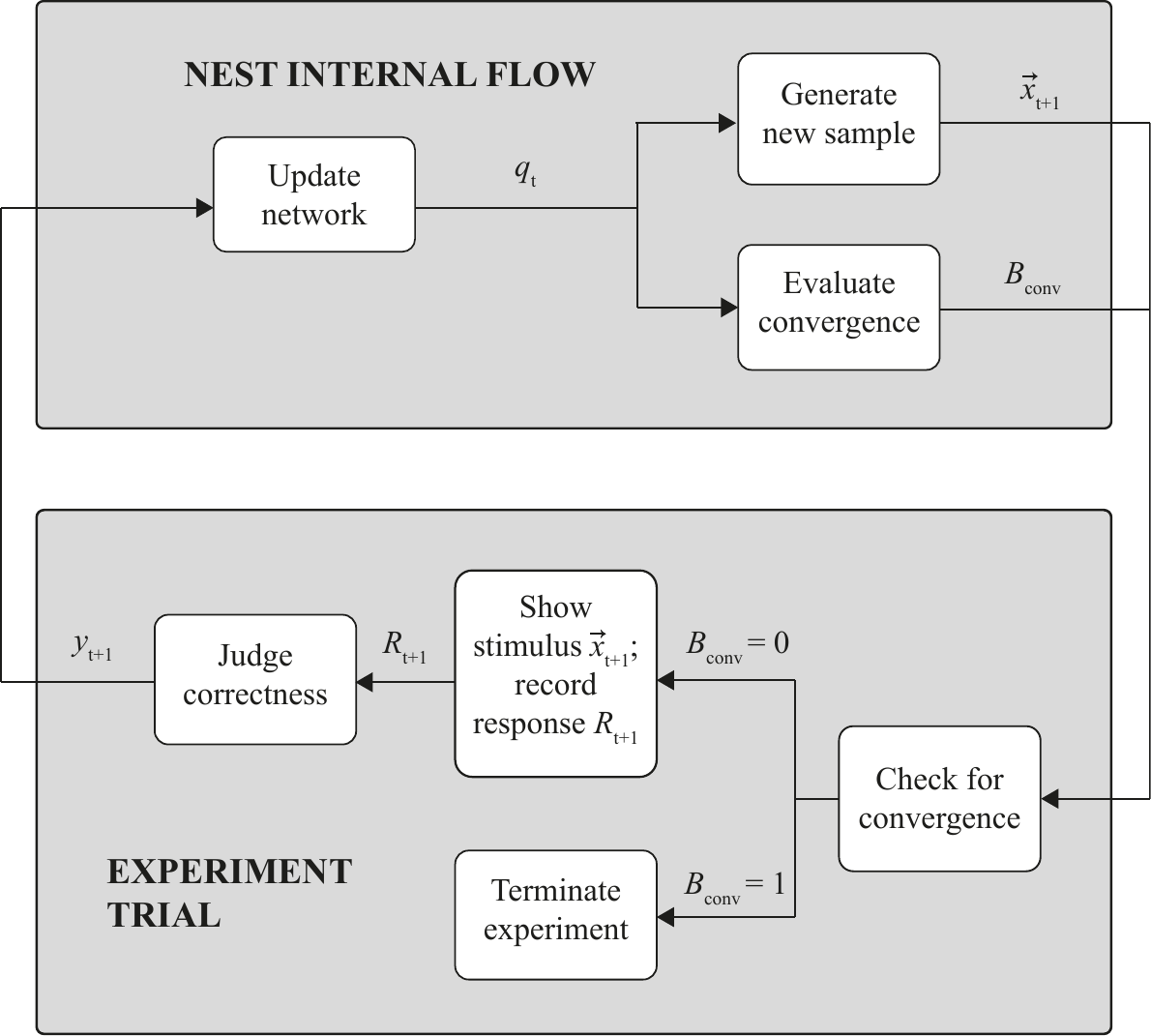}
    \caption{The \ac{NEST} experiment flow. \ac{NEST} retrains its network $q_{t}$ using the updated dataset $D_{t} = D_{t-1} \cup (\vec{x}_{t}, y_{t})$, then generates a new query $\vec{x}_{t+1}$ by optimizing the acquisition function and evaluates the Fisher convergence criterion to get a binary outcome $B_{\mathrm{conv}}$. The experimenter uses $B_{\mathrm{conv}}$ to choose whether to terminate or continue the experiment. If the experiment continues, then a new stimulus is shown based on stimulus parameters $\vec{x}_{t+1}$, the correctness $y_{t+1}$ of the response is determined, and $y_{t+1}$ is fed back into the \ac{NEST} algorithm. This process repeats until convergence.}
    \label{fig:NEST_diagram}
\end{figure}

\section{Method} \label{sec:methods}

The \ac{NEST} procedure approximates a multi-dimensional psychometric function using a neural network. The network is trained to map the \emph{stimulus space} to the probability of a correct response from a human observer for a visual task, assuming a stationary psychometric function and statistically independent responses. The general workflow of \ac{NEST} is shown in Figure \ref{fig:NEST_diagram}. The \ac{NEST} method involves training the neural network using the set of all responses collected after each trial and evaluating whether the experiment has reached convergence for termination of the experiment. Until convergence, the sample for the next trial in the experiment is selected based on an active sampling scheme driven by the \textit{acquisition function}, inspired by Bayesian optimization \citep{shahriari2015}. \new{By the end of the procedure, the trained \ac{NEST} network is the estimate of the psychometric function, which is capable of computing the probability of a correct response at each combination of input stimulus dimensions. The experimenter can use this network for further analysis of the estimated psychometric function, if desired.} 

In Section \ref{sec:architecture}, we describe the detailed architecture of the neural network used to estimate the psychometric function. In Section \ref{sec:acquisition_NEST}, we discuss the sample selection scheme for each trial. Section \ref{sec:implementation} specifies a set of design choices and fixed parameter values in the implementation of the \ac{NEST} procedure. Finally, in Section \ref{sec:termination_NEST}, we explain our novel convergence measure and the experiment termination criterion.

\subsection{Neural network architecture} \label{sec:architecture}
As the psychometric function estimator, our method uses a feedforward artificial neural network, also known as a \ac{MLP}, introduced by \citet{Rosenblatt1961}. \ac{MLP}s are an appropriate choice for the nonparametric estimation of psychometric functions because they are regarded as universal approximators \citep{hornik1989}. An \ac{MLP} consists of a sequence of consecutively applied layers of matrix operations defined as
\begin{equation}
    \begin{aligned}
    &q(y | \vec{x}, W) = \sigma_f (W_{L} \sigma( ... W_{2} (\sigma(W_{1} \vec{x} + \vec{b}_{1} ) + \vec{b}_{2}) ...) + \vec{b}_{L}),
  \end{aligned}
  \label{eq:neural_network_Rosenblatt}
\end{equation}
where $y$ represents the probability of a successful trial outcome to a presented stimulus $\vec{x}$ in a detection or discrimination task, $L$ is the total number of network layers, $W = (W_{1}, ..., W_{L})$ is the sequence of weight matrices with $W_{\ell} \in \mathbb{R}^{K_{\ell-1} \times K_{\ell}}$ the $\ell$-th weight matrix between the $(\ell -1)$-th and $\ell$-th layer, $\vec{b}_{\ell} \in \mathbb{R}^{K_{\ell}}$ is the $\ell$-th bias matrix, and $\sigma(\cdot)$ is an operator providing nonlinearity. The network defined by Equation~\ref{eq:neural_network_Rosenblatt} is a general \ac{MLP}. The final nonlinearity, $\sigma_f$, is selected from the family of logistic functions, such that the output of the network defines a proper probability measurement in the range $[0,1]$. The individual entries of the weight matrices $W_{\ell}$ are learned from the stimulus-trial outcome pairs $D = \lbrace \left( \vec{x}_{i}, y_{i} \right) \rbrace_{i=1}^{N}$ from the experiment by minimizing the training loss function $L_{\text{train}}(X,Y)$, where $X = (\vec{x}_{1}, ..., \vec{x}_{N})$ are the input data consisting of stimuli and experiment parameters selected in each trial and $Y = (y_{1}, ..., y_{N})$ are the corresponding binary training labels that consist of trial outcomes (e.g., detection/no detection in a detection task).

\subsection{Sample selection scheme} \label{sec:acquisition_NEST}
The problem of developing a quick and efficient testing scheme has been extensively studied as part of the theory of optimal experiments \citep{fedorov2013, mackay1992}. Our method shares the same objectives as set by the theory of optimal experiments, which is to select and test the stimulus and experiment parameters that are the most informative about the psychometric function we are trying to estimate. To achieve this goal, we propose a new \emph{acquisition function}, $\hat{P}_{\mathrm{acq}}$, that maps the samples in stimulus space to their intrinsic value for achieving fast convergence. $\hat{P}_{\mathrm{acq}}$ is the weighted geometric mean of four components:
\begin{equation}
    \label{eq:maximisation_target}
        \hat{P}_{\mathrm{acq}} = \left( P_{\mathrm{grad}} ^{a} \cdot
         P_{\mathrm{prox}} ^{b}
          \cdot P_{\mathrm{unc}} ^{c} \cdot P_{\mathrm{la}}^{d} \right)^{\frac{1}{a+b+c+d}},
\end{equation}
where $P_{\mathrm{grad}}$ favors the selection of samples that maximize the \textbf{\emph{gradient}} of the psychometric function, $P_{\mathrm{prox}}$ discourages the \textbf{\emph{proximity}} to the previously selected and queried samples, $P_{\mathrm{unc}}$ prioritizes selecting the samples where the model's \textbf{\emph{uncertainty}} in accurately predicting the outcome of a trial is high, and $P_{\mathrm{la}}$ aims to select a sample that has a significant influence on the prediction, based on \textbf{\emph{lookahead}} estimation of retraining the model. The weights $a$, $b$, $c$, and $d$ are optimized for the fastest overall convergence with different types of psychometric functions (Section~\ref{sec:hyperoptimisation}).

 During the experiment, the value of $\hat{P}_{\mathrm{acq}}$ depends on both the current estimate of the psychometric function and the history of responses accrued in the experiment. Therefore, we compute $\hat{P}_{\mathrm{acq}}$ and select the sample that maximizes it in each trial using a gradient descent optimization.

\paragraph{\textbf{Gradient}} The gradient component, $P_{\mathrm{grad}}$, aims to drive the sample selection towards the threshold of the psychometric function, where the gradient magnitude of the estimated function is large. \new{For psychometric functions, the gradient term takes on non-zero values in the transition region, where the probability of detection goes from 0 to 1. Since refinement of the transition region is an important goal in psychometric function estimation, the gradient term helps guide the sampling strategy towards samples in the transition region.} This strategy is also related to the commonly used heuristics in previous adaptive methods that select samples near the threshold of the psychometric function \citep{WatsonPelli1983,Emerson1986,KingSmithEtAl1994}. 
In order to make our computation agnostic to the changes in the maximum gradient magnitude of different functions, we normalize the gradient magnitude as
\begin{equation}
    \label{eq:gradient_probability}
    P_{\mathrm{grad}}(\vec{x}) = \frac{|\nabla q(\vec{x})|}{\max\limits_{\vec{z} \in \mathcal{I}}{|\nabla q(\vec{z})|}},
\end{equation}
where $\mathcal{I}$ represents the $K$-dimensional input space.

\paragraph{\textbf{Proximity}} For non-monotonic psychovisual functions, the gradient term alone tends to oversample around the decision boundary that is discovered at an early stage of the experiment. In order to reduce redundancy and to encourage exploration of the multi-dimensional stimulus and parameter space, the gradient component, $P_{\mathrm{prox}}$, aims to penalize the potential lack of variance in sample selection by assigning a lower intrinsic value to the samples that are in close proximity of prior trials. \new{\cite{Bemporad2023} previously used inverse distance weighting as a successful active learning strategy for regression problems.} $P_{\mathrm{prox}}$ is inspired from the \ac{KDE} (i.e., \textit{Parzen windows}) method, that maps each new sample $\vec{x}$ to a value that represents its proximity to prior samples $\vec{x}_{i} \in X$. We use a kernel method instead of other distance measures because it has the advantage of being differentiable in our optimization. We select the Gaussian kernel for our method, which leads to the \ac{KDE} given by
\begin{equation}
    \label{eq:parzen_windows}
    f_{\mathrm{prox}}(\vec{x}, X, h) = \frac{1}{Nh^{K} (2\pi)^{K/2}} \sum\limits_{\vec{x}_{i} \in X} e^{-\frac{(\vec{x} - \vec{x}_{i})^{\top}(\vec{x} - \vec{x}_{i})}{2h^{2}}},
\end{equation}
where $h$ is the size of the window and corresponds to the standard deviation of the Gaussian kernel. Using Equation~\ref{eq:parzen_windows}, we define the gradient component as
\begin{equation}
    \label{eq:heat_probability}
    P_{\mathrm{prox}}(\vec{x}, X, h) = 1 - \frac{f_{\mathrm{prox}}(\vec{x}, X, h)}{\max\limits_{\vec{z} \in \mathcal{I}} f_{\mathrm{prox}}(\vec{z}, X, h)}.
\end{equation}

\paragraph{\textbf{Uncertainty}} A common strategy used in active learning is \emph{uncertainty sampling}. The fundamental concept of uncertainty sampling is to allow the model to refrain from choosing samples that it is certain about and instead target the areas of the stimuli space that may be perplexing \citep{Settles2012uncertainty, lewis1994}. \new{Commonly used acquisition functions such as \ac{BALV} \citep{Settles2009} and \ac{BALD} \citep{HoulsbyEtAl2012} use uncertainty quantification for sample selection.} A neural network does not provide a built-in way to compute the uncertainty, but \citet{GalGhahramani2016} showed that the well-known dropout technique \citep{srivastava2014} against over-fitting can also approximate the network prediction uncertainty.

The \emph{Monte Carlo dropout method} introduced by \citet{GalGhahramani2016} works by replacing the weights $W$ of a network $\hat{q}(y | \vec{x}, W)$ with $\widehat{W} = (\widehat{W}_{1}, ..., \widehat{W}_{L} )$, where
\begin{equation}
\label{eq:dropout_weights}
    \widehat{W}_{\ell} = \mathrm{diag}\! \left( [z_{\ell, j}]_{j=1}^{K_{\ell}} \right) W_{\ell}.
\end{equation}
With this change, Monte Carlo dropout replaces weights of the network with a function $W_{\ell}$ of the random variable $z_{\ell, j}$ with a Bernoulli distribution. Given the output of the dropout neural network $\hat{y}(\vec{x}, \widehat{W}) = \hat{q}(\vec{x}, \widehat{W})$, it is possible to compute the expected output of the network for stimulus $\vec{x}$ as
\begin{equation}
    \label{eq:first_moment_dropout}
    \mathbb{E}_{\vec{x}}[\hat{y}(\vec{x}, \widehat{W})] \approx \frac{1}{M} \sum\limits_{m=1}^{M} \hat{y}(\vec{x}, \widehat{W}^{m}),
\end{equation}
where $M$ is the number of evaluations of the dropout neural network with random sampling of $z_{\ell, j}$. We use the computed variance of the output of the network,
\begin{equation}
    \label{eq:variance_dropout}
    \begin{aligned}
        \mathbb{V}(\hat{y}(\vec{x}, \widehat{W})) &= \mathbb{E}(\hat{y}(\vec{x}, \widehat{W})^{2}) - \mathbb{E}(\hat{y}(\vec{x}, \widehat{W}))^{2}, 
    \end{aligned}
\end{equation}
as a measure of the prediction uncertainty for the input sample. Since the output of our neural network is a scalar value, the variance directly translates into the standard deviation $\sigma(\hat{y}(\vec{x}, \widehat{W})) = \sqrt{\mathbb{V}(\hat{y}(\vec{x}, \widehat{W}))}$, and we define our acquisition function term normalized by the maximum value standard deviation takes across the stimulus space as
\begin{equation}
    \label{eq:std_probability}
    P_{\mathrm{unc}}(\vec{x}) = \frac{\sigma(\hat{y}(\vec{x}, \widehat{W}))}{\max\limits_{\vec{z} \in \mathcal{I}} \sigma(\hat{y}(\vec{z}, \widehat{W}))}.
\end{equation}

\paragraph{\textbf{Lookahead}}
During our initial experiments with the acquisition function components introduced so far, we observed an over-exploration of the stimuli space boundary during sample selection, which was also noted in several previous studies \citep{SiivolaEtAl2018, SongEtAl2018, OwenEtAl2021}. The model prediction exhibits relatively high uncertainty for the samples along the boundary, which makes them a likely candidate for confidence-based acquisition strategies. Boundary samples are also more likely to be located far from other sample points, and thus favored by proximity-based acquisition schemes, too. The gradient term, meanwhile, does not favor the boundary points, but it does not discourage their selection in general, either. Overall, this trend results in frequently selecting samples from the boundary, which can provide only limited information about the full function domain from the extremes of the stimulus space. We also observed that this issue is exacerbated as the dimensionality of the stimulus space increases.

\new{Lookahead methods have previously been tried for parametric methods, including variance minimisation of the next trial \citep{KingSmithEtAl1994} and the use of dynamic programming to look multiple steps ahead \citep{KimEtAl2017} in one-dimensional parametric estimation.} \citet{LethamEtAl2022} showed that lookahead acquisition functions can perform significantly better in high-dimensional problems by refitting the model on a candidate point and observing the change in the model.
For our neural network model, we use the empirical \ac{NTK} approximation to provide a lookahead mechanism. We follow the definition of \citet{MohamadiEtAl2022} to approximate the behavior of the neural network after retraining with a potential sample from the stimulus space. The lookahead approximation of the network output is given by
\begin{equation}
    \label{eq:lookahead_definition}
    \begin{aligned}
        q_{D^{+}}(\vec{x}) = &q_{D}(\vec{x}) + \Theta_{D}(\vec{x}, X^{+}) \cdot \Theta_{D}(X^{+}, X^{+})^{-1} \cdot \\
        &\left(Y^{+} - q_{D}(X^{+}) \right),
    \end{aligned}
\end{equation}
where $\vec{x}'$ is the potential new sample, $D$ is the training set from the trials conducted so far, $D^{+} = D \cup \lbrace (\vec{x}', y' ) \rbrace$, $X^{+} = X \cup \{\vec{x}'\}$, $Y^{+} = Y \cup \{y'\}$, and $\Theta_{D}(a, b)$ is the empirical \ac{NTK} of two inputs given by
\begin{equation}
    \label{eq:empirical_NTK}
    \Theta_{D}(a, b) = \nabla_{\theta}q_{D}(a) \nabla_{\theta}q_{D}(b)^{T},
\end{equation}
where $\theta$ denotes the parameters of the neural network. Equation \ref{eq:lookahead_definition} allows us to approximate the effect of iteratively training the neural network on the output of the network.
In order to maximize the information gained from each trial, we aim to select the sample that maximizes the change in the neural network output.

In order to evaluate the model change, we subsample the stimulus space using \emph{blue noise}, and compute the model outputs for the samples, $\vec{u} \in U$. To consider the worst-case scenario regarding the binary outcomes of the trial, $y' \in \{0, 1\}$, we compute the change in the output of the network for both cases and take the minimum:
\begin{equation}
    \label{eq:lookahead_acquisition_value}
    \begin{array}{rcl}
        f_{\mathrm{la}}(\vec{x}, D, U) &=& \frac{1}{|U|} \min_{y' \in \{0, 1\}} S_{y'} \quad \text{with} \\[1mm]
        S_{y'} &=& \sum\limits_{\vec{u} \in U} (q_{D \cup \{ (\vec{x}, y') \}}(\vec{u}) - q_{D} (\vec{u}))^{2}.
    \end{array}
\end{equation}
Based on this definition, the lookahead component $f_{\mathrm{la}}(\vec{x}, D, U)$ is normalized by the maximum value, similar to other components of the acquisition function as
\begin{equation}
    \label{eq:normalised_lookahead}
    P_{\mathrm{la}}(\vec{x}, D, U) = \frac{f_{\mathrm{la}}(\vec{x}, D, U)}{\max\limits_{\vec{z} \in \mathcal{I}} f_{\mathrm{la}}(\vec{z}, D, U)}.
\end{equation}

\subsection{Implementation} \label{sec:implementation}

\paragraph{Network architecture}
As we mentioned earlier in Section~\ref{sec:architecture}, we use a fully connected \ac{MLP} network with three hidden layers consisting of 256, 128, and 32 neurons, respectively. In our experiments, we observed that this network size is sufficient even for learning complex and high-dimensional psychophysical functions, which are introduced later in Section~\ref{sec:data_NEST}. Each hidden layer uses \ac{ReLU} activation \citep{NairHinton2010}.

\paragraph{Initialization} At the beginning of the experiment, we randomly initialize the weights of our network using He initialization \citep{HeEtAl2015}. We start by performing a random exploration of the stimulus space using Sobol sampling with the probability
\begin{equation}
    \label{eq:random_sampling_probability}
    p_{\mathrm{random}}(t) = \max \left(p_{\mathrm{base}}, p_{0} f^{t-1} \right),
\end{equation}
where $p_{0} = 0.5$ is the initial random sampling probability, $f = 0.97$ is a factor by which we multiply $p_{0}$ after each trial, and $t$ is the current trial number. This formulation gradually shifts from random exploration to the acquisition-function-driven sample selection strategy as the number of trials increases. Eventually, the random sampling probability decreases to the asymptote $p_{\mathrm{base}} = 0.05$. The non-zero asymptote encourages occasional random exploration, even at the later stages of the experiment.

\paragraph{Loss function}
We use the \ac{BCE} loss  defined as
\begin{equation}
   \label{eq:bce}
   L_{\mathrm{BCE}} (X, Y) = \frac{1}{N} \sum\limits_{i=1}^{N} y_{i} \log{q(\vec{x}_{i})} + (1 - y_{i}) \log{\left( 1 - q(\vec{x}_{i}) \right)},
\end{equation}
where $X = (\vec{x}_{1}, ..., \vec{x}_{N})$ are the input data of selected stimuli, $Y = (y_{1}, ..., y_{N})$ are the associated binary training labels, and $q(\vec{x})$ is a shorthand for the prediction of the network $q(y | \vec{x}, W, \alpha, \gamma)$. In order to maintain numerical stability, we clamp the value of the logarithmic terms in Equation \ref{eq:bce} to the finite range of $[-100, 100]$.

\paragraph{Training}
We train the network after each trial of the experiment using the Adam optimizer \citep{KingmaBa2014} for 100 epochs with a learning rate $\eta_{0} = 0.0003$. We additionally apply an exponential learning rate annealing scheme given by
\begin{equation}
    \label{eq:learning_rate_annealing}
    \eta(k) = \eta_{0} \gamma^{k},
\end{equation}
where k is the epoch number and $\gamma$ is the decay rate. We set $\gamma = 0.01^{\frac{1}{100}}$ such that the learning rate will decay to $0.01\eta_0$ by the end of the training. In order to avoid overfitting and to increase the stability between trials, we use the \textit{shrink-and-perturb} trick for warm-starting introduced by \citet{AshAdams2020}.

We apply normalization to balance the scale of different stimulus space dimensions such that we have zero mean and unit variance in each dimension. However, we repeat the normalization only until the 25th trial because we observed that the magnitude of change in mean and variance is negligible after that point.

We set the Parzen window parameter $h$ used in Equation~\ref{eq:parzen_windows} to $0.25$. The number of samples $M$ used for Monte Carlo dropout is set to 100. In order to improve the running time, we only evaluate the last dropout layer multiple times \citep{MaKaewell2020}. For finding the maximum value in Equation~\ref{eq:maximisation_target}, we use the \ac{LBFGS} optimizer provided by the Scipy library with step size $\epsilon = 10^{-6}$ \citep{ScipyOptimize}.

\paragraph{Sample selection}
In order to deal with the non-convexity of the acquisition function (Equation~\ref{eq:maximisation_target}), we use multiple random initializations. The gradient required for the optimization is computed using automatic differentiation with PyTorch \citep{PyTorch}.

\paragraph{Scaling the network output}
For the final nonlinearity $\sigma_{f}$ in Equation \ref{eq:neural_network_Rosenblatt}, we use the Weibull \ac{CDF} defined as
\begin{equation}
    \label{eq:Weibull_function}
    \Psi(x, \rho, T) = 1 - e^{-10^{\rho \frac{x-T}{20}}},
\end{equation}
where $T$ is the \textit{threshold} and $\rho$ is the $slope$ of the function \citep{Weibull1951}. We set $\rho = 1$ and $T = 0$ in our implementation.

In order to scale the output of the network to the range $[0,1]$ representing a probability, we use
\begin{equation}
    \label{eq:neural_network}
    \begin{aligned}
    q(y | \vec{x}, W, &\alpha, \gamma) = \frac{1}{2} \left( 1 + \alpha - \gamma \right) \\
    &+ \frac{1}{2} \left(1 - \alpha - \gamma \right) q(y | \vec{x}, W),
  \end{aligned}
\end{equation}
where the \textit{lower asymptote} $\alpha$ represents the probability of success for a random response and $\gamma$ is the \textit{lapse rate}. In our implementation, we selected the Weibull function because it is representative of probability summation among independent detection mechanisms in the \ac{HVS} \citep{green1975, nachmias1981}; however, the use of other S-shaped functions such as logistic function or the Gaussian integral is also possible.

\paragraph{Regularization}
In order to prevent \textit{overfitting}, we apply two types of regularization. The first one is the \textit{dropout} technique introduced by \citet{HintonEtAl2012}, whereas the second one is the \textit{input noise injection} introduced by \citet{SietsmaDow1991}. We set the dropout probability to $0.1$, and set $\sigma = 0.01$ for the Gaussian noise while applying input noise injection.

\subsection{Termination condition} \label{sec:termination_NEST}
\new{When conducting a perceptual experiment, the experimenter must choose the number of trials to perform. A straightforward method is to set a fixed number $N$ of trials based on time constraints or resource availability and conclude after exactly $N$ trials. This approach simplifies planning but may overestimate or underestimate the necessary accuracy for the psychometric function.}

\new{Alternatively, an adaptive termination condition can be employed where the experiment continues until a predetermined level of convergence is achieved. This allows for more efficient use of resources by terminating once the desired accuracy threshold is met.}

\new{In this context, we propose a specific termination criterion for \ac{NEST} that monitors the experiment progress and stops when the required accuracy is attained. We apply this condition in the live experiments described in Section~\ref{sec:experiments_NEST}.}

In order to detect the convergence after each trial $i\in[1,\ldots,N]$, we monitor the magnitude of change in the weights of the neural network from the previous trial.
More specifically, we compute $\nabla L_{\mathrm{BCE}}(X, Y)$ by following the approach of \citet{LiaoEtAl2019} and using an approximation of the \ac{FIM} defined as
\begin{equation}
    \label{eq:approximate_Fisher_information}
    \widetilde{\mathbf{F}}_{D} = \mathbf{J}_{D}^{\top} \mathbf{J}_{D} = \left( \nabla L_{\mathrm{BCE}}(X, Y) \right)^{\top} \left( \nabla L_{\mathrm{BCE}}(X, Y) \right),
\end{equation}
where $\mathbf{J}_{D}$ is the Jacobian of the neural network parameters computed based on the dataset $D$. \new{The Fisher Information matrix quantifies the amount of information about the weights of the neural network present in the dataset $D$. By monitoring changes in the value of this metric between trials, we can determine how much new information the newly selected samples contribute.} $\widetilde{\mathbf{F}}_{D}$ is an approximation of the true \ac{FIM} given by $\mathbf{F}_{D} = \mathbf{J}_{D} \mathbf{J}_{D}^{\top}$, which is prohibitively expensive to compute \citep{LiaoEtAl2019}. Equation \ref{eq:approximate_Fisher_information} computes the amount of information that the training data contribute to the neural network parameters. We use the cumulative sum of the \textit{Fisher energy} $E(D, K)$ from \cite{LiaoEtAl2019}, which is defined as
\begin{equation}
    \label{eq:Fisher_energy}
    E(D, K) = \sum\limits_{k=1}^{K} e(D, k) = \frac{\eta(k)}{NK\eta_{0}} \sum\limits_{k=1}^{K} \text{Tr}(\widetilde{\mathbf{F}}_{D}),
\end{equation}
where $K$ represents the number of training epochs in each trial, $e(D, k)$ is the Fisher energy at epoch $k$, and $\text{Tr}(\cdot)$ is the trace operator.

Over the course of the experiment, the value of $E(D, K)$ stabilizes when the introduction of new training data no longer provides significant new information to the training. In order to detect convergence, we monitor a 15-trial windowed moving average of the difference between $E(D, K)$ after trial $i-1$ and after trial $i$. If the average value in the window drops below a selected threshold value $E_{\mathrm{thres}}$, then we assume convergence. A lower $E_{\mathrm{thres}}$ represents a more strict stability requirement; however, it will be less likely to detect convergence in the presence of noisy trial outcomes.

We start checking for convergence after at least one negative and one positive sample have been recorded. We provide a detailed analysis of our convergence criterion in the supplementary material.
\section{Dataset of synthetic psychometric functions}  \label{sec:simulations_NEST}

\begin{figure*}[t]
    \centering
    \includegraphics[width=\textwidth]{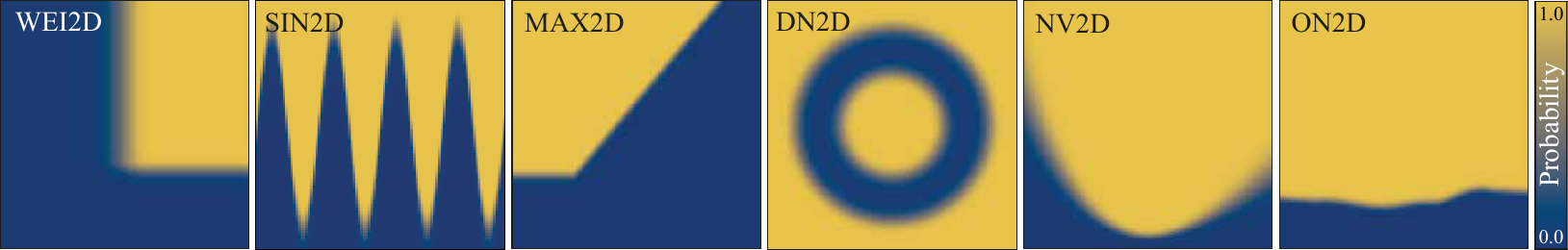}
    \caption{Representative plots of 2D synthetic psychometric functions that we used for calibration and benchmarking. Horizontal and vertical axes represent the input stimulus parameters. We also used higher dimensional psychometric functions that are not visualized here (see  Section~\ref{sec:simulations_NEST}).}
    \label{fig:2D_ground_truths}
\end{figure*}

We use different types of synthetic psychometric functions for learning the weights of the acquisition function (Equation~\ref{eq:maximisation_target}) and for a comprehensive performance benchmark of our method with Monte Carlo simulations. Figure \ref{fig:2D_ground_truths} shows representative plots of the 2D psychometric functions used in the analysis.

\paragraph{Multidimensional Weibull (WEI1D-4D)} The multi-dimensional version of the \textit{Weibull \ac{CDF}} is given by
\begin{equation}
    \label{eq:Weibull}
    \Psi(\vec{x}, \vec{\beta}, \vec{T}, \alpha, \gamma) = \alpha + (1 - \gamma - \alpha) \prod\limits_{k=1}^{K} e^{-10^{\beta_{k} \frac{x_{k} - T_{k}}{20}}},
\end{equation}
where $\vec{\beta}$ and $\vec{T}$ are the vectors representing the slope and threshold in each dimension, respectively. The thresholds of the simple multi-dimensional Weibull \ac{CDF} demarcate a $K$-dimensional hypercuboid, and transitions along each dimension may have different slopes. \new{This function is an extension of traditional Weibull \ac{CDF}, commonly used as a one-dimensional psychophysics function, to higher number of dimensions.} It is also a good test case for whether the neural network can model sharp changes in the target function since it needs to approximate the corners of the hypercuboid. We have performed simulations for the \ac{WEI1D}, \ac{WEI2D}, \ac{WEI3D}, and \ac{WEI4D} functions.

\paragraph{\ac{SIN2D}} The \ac{SIN2D} function is a variant of the Weibull \ac{CDF} for which the threshold value is described by
\begin{equation}
    \label{eq:threshold_sinusoid}
    T(x_{0}, A, f) = A \sin{\left( 2 \pi f x_{0} \right)},
\end{equation}
where $x_{0}$ is the first element of two input dimensions, $A$ represents the amplitude, and $f$ represents the frequency of the sinusoidal pattern. We subsequently apply a 1D Weibull \ac{CDF} using $T(x_{0}, A, f)$ as the threshold and $x_{1}$ as the stimulus dimension. A representative plot of this function is shown in Figure \ref{fig:2D_ground_truths}. \new{The sinusoidal shape of this test function exhibits a high level of periodic change in the threshold. Insufficient exploration capacity can cause a method to miss the existence of the peaks and valleys of this test function.}

\paragraph{\ac{MAX2D}} The \ac{MAX2D} function is another variant of the Weibull \ac{CDF} with a threshold given by
\begin{equation}
    \label{eq:max_threshold}
    T(x_{0}, c_{0}, c_{f}, t) = \max(t, c_{0} + c_{f}x_{0}),
\end{equation}
which produces a 2D maximum function with two piecewise linear segments. \citet{WatsonAhumada2016} proposed that the log-contrast sensitivity of human vision to spatial frequency may be represented with the \ac{MAX2D}.

\paragraph{\ac{DN2D}} The \ac{DN2D} function is a variant of the 2D Weibull \ac{CDF}, whose definition is
\begin{align}
    \label{eq:donut_threshold}
    \Psi(\vec{x}, \vec{\beta}, r_{1}, r_{2}, \alpha, \gamma) &= \alpha + (1 - \gamma - \alpha) e^{-10^{\beta \frac{T_{\mathrm{d}}(\vec{x}, r_{1}, r_{2})}{20}}}, \\
    T_{\mathrm{d}}(\vec{x}, r_{1}, r_{2}) &= \max(||\vec{x}||_{2} - r_{2}, r_{1} - ||\vec{x}||_{2}),
\end{align}
where $r_{1} < r_{2}$ are the inner and outer radii of the threshold. As shown in Figure \ref{fig:2D_ground_truths}, \ac{DN2D} represents a non-monotonic probability function and therefore tests the capability of modeling a rather complex region \citep{garcia2014}. \new{This challenges the methods to do sufficient exploration throughout the input space; otherwise, they risk identifying the annulus as a disk.}

\paragraph{\ac{NV2D}} This function was introduced by \citet{OwenEtAl2021} and defined as
\begin{align}
    \label{eq:Owen_2D_function}
    \Psi(\vec{x}, \alpha, \gamma) &= \alpha - (1 - \alpha - \gamma) \Phi(T_{\mathrm{nt}}(\vec{x}, \alpha, \gamma)), \\
    T_{\mathrm{nt}}(\vec{x}, \alpha, \gamma) &= \frac{4(1-\alpha) (1+x_{1})}{0.1 + 0.8 (0.2 x_{0} - 1)^{2} x_{0}^{2}} - 4(1 - 2 \alpha),
\end{align}
where $\Phi(\cdot)$ is the standard Gaussian \ac{CDF}.

\ac{NV2D} represents a general 2D psychometric function and was used by \citet{OwenEtAl2021} and \citet{LethamEtAl2022} to test the accuracy of their Gaussian process methods, so we include this function as a common testbed. \new{It also provides a good benchmark on a relatively simple, well-defined synthetic psychometric function compared to more complex functions such as DN2D and SIN2D.}

\paragraph{\ac{ON2D}} \citet{DubnoEtAl2013} defined a number of well-known empirical audiometric functions, one of which is the \ac{ON2D} function. It has a relatively constant value with only slight upward and downward deviations (Figure~\ref{fig:2D_ground_truths}).

\paragraph{\ac{HART6}} Formulated by \citet{LyuEtAl2021}, \ac{HART6} is a well-established 6D psychometric function from the family of logistic functions. \new{It has five local minima and one global minimum. This property means that the psychometric function has six regions where the probability changes from high to low. This makes it an interesting test case since it tests the methods on their ability to segment the psychometric function into its disjoint regions of changing probability.}

\ac{HART6} allows us to test the capabilities of our method in a higher-dimensional setting. The definition of \ac{HART6} is given in Section \ref{sec:hartmann6}

\paragraph{\ac{PS8D}} \ac{PS8D} is an 8D function, introduced by \citet{LethamEtAl2022}. We include it in our benchmarks to compare with \ac{GP} methods. The \ac{PS8D} function is defined as
\begin{equation}
\label{eq:Letham_8D_function}
\Psi(\vec{x}, \alpha, \gamma) = \alpha + (1 - \alpha - \gamma) \Phi\left( \frac{x_{1} - c(\vec{x})}{x_{5}(2 + c(\vec{x}))} \right),
\end{equation}
where
\begin{equation}
    \begin{aligned}
        c(\vec{x}) = &\left( \frac{x_{3}}{2} \left( 1 - \cos{\left( \frac{3}{5} \pi x_{2} x_{8} + x_{7} \right)} \right) + x_{4} \right) \cdot \\
        &\left( 2 - x_{6} \left( 1 + \sin{\left( \frac{3}{10} \pi x_{2} x_{8} + x_{7} \right)} \right) \right) - 1.
    \end{aligned}
\end{equation}
The input space is $\vec{x} \in [-1, 1]^{8}$.

\section{Learning the acquisition function weights} \label{sec:hyperoptimisation}
\begin{table*}[b]
    \centering
    \caption{Optimal parameter settings for individual functions and for all functions combined. The last column shows the percentage improvement in performance when using the optimal acquisition parameter values instead of the values from the combined optimization.}
    \label{tab:hyperoptimisation_results}
    \begin{tabular}{c|c|c|c|c|c|c|c}
        \textbf{Function} & \textbf{a} & \textbf{b} & \textbf{c} & \textbf{d} & $\mathbf{AUC}_{\mathrm{func}}$ & $\mathbf{AUC}_{\mathrm{comb}}$ & \textbf{$\Delta$Error} \\ \hline
        \ac{NV2D} & 4.6 & 2.6 & 1.4 & 3.6 & $22.60 \pm 0.48$ & $25.67 \pm 0.28$ & 13.58\% \\
        \ac{DN2D} & 8.6 & 16.0 & 8.0 & 7.6 & $40.51 \pm 1.07$ & $47.71 \pm 0.87$ & 17.78\% \\
        \ac{SIN2D} & 2.8 & 8.0 & 3.4 & 6.0 & $56.80 \pm 0.87$ & $58.84 \pm 0.42$ &  3.59\% \\
        \ac{WEI4D} & 3.6 & 15.4 & 6.6 & 0.6 & $41.56 \pm 0.82$ & $45.28 \pm 0.36$ & 8.97\% \\
        \ac{HART6} & 0.4 & 6.0 & 7.0 & 7.2 & $247.77 \pm 1.44$ & $255.17 \pm 1.74$ & 2.98\% \\ \hline
        Combined & 0.8 & 10.6 & 6.0 & 4.0 & -- & -- & --
    \end{tabular}
\end{table*}
We optimize the acquisition function weights $a$, $b$, $c$, and $d$ in Equation \ref{eq:maximisation_target} individually for \ac{NV2D}, \ac{DN2D}, \ac{SIN2D}, \ac{WEI4D}, and \ac{HART6} functions and provide the results in Table \ref{tab:hyperoptimisation_results}. We also provide the optimal weights for the set of all psychometric functions, which may be used when the underlying psychometric function type is not known. As expected, the learning and prediction performance improves when we use acquisition function weights specifically optimized for the given psychometric function, but the improvement is limited especially for high-dimensional functions such as \ac{HART6}. This shows that the acquisition function weights optimized for the combined set of all psychometric functions is still a good choice when the experimenter has no assumptions on the underlying psychometric function.

\section{Monte Carlo simulations}
\label{sec:data_NEST}
For benchmarking, we performed Monte Carlo simulations and also carried out real experiments with multidimensional psychometric functions. In this section, we provide our performance metrics (Section~\ref{sec:error_metrics_NEST}) and then the results of the simulations (Section~\ref{sec:error_metric_results}).

In our analysis, we use the synthetic psychometric functions introduced in Section~\ref{sec:simulations_NEST} as our dataset and perform 100 repetitions for each psychometric function with random selections of its parameters. In addition to our method, we test all of the \ac{GP} methods from \citet{OwenEtAl2021} and the global lookahead methods \ac{EAVC} and \ac{GlobalMI} from \citet{LethamEtAl2022}. Furthermore, we also benchmark the QUEST+ procedure using \ac{WEI1D}--\ac{WEI4D}, \ac{SIN2D}, and \ac{MAX2D} psychometric functions. The remaining functions do not have a parametric form that is required by QUEST+. The methods and psychometric functions that we used in our benchmarks are summarized in Table~\ref{tab:comparisons}.
\begin{table}[b]
    \centering
    \caption{An overview of benchmarks for tested methods.}
    \scriptsize \begin{tabular}{P{2.0cm}|P{1.2cm}|P{2.7cm}|P{1.0cm}}
        \centering\textbf{Function} & \textbf{QUEST+} & \textbf{Gaussian Processes} & \textbf{\ac{NEST}} \\ \hhline{-|-|-|-}
        \ac{WEI1D}-\ac{WEI4D} & \checkmark & \checkmark & \checkmark \\
        \ac{SIN2D} & \checkmark & \checkmark & \checkmark \\
        \ac{MAX2D} & \checkmark & \checkmark & \checkmark \\
        \ac{DN2D} & $\times$ & \checkmark & \checkmark \\
        \ac{NV2D} & $\times$ & \checkmark & \checkmark \\
        \ac{ON2D} & $\times$ & \checkmark & \checkmark \\
        \ac{HART6} & $\times$ & \checkmark & \checkmark \\
        \ac{PS8D} & $\times$ & \checkmark & \checkmark \\
    \end{tabular}
    \label{tab:comparisons}
\end{table}

\begin{table*}[hb]
    \caption{The  mean \ac{AUC} $\pm$ standard error of the \ac{RMSE} and Brier metrics for the \ac{NV2D}, \ac{WEI4D}, \ac{HART6}, and \ac{PS8D} functions. The best method(s) for each metric are shown in bold typeface. Our method performs better ($p < 0.03$, Games-Howell test) than the other methods on the \ac{NV2D}, \ac{WEI4D}, and \ac{HART6} functions. For \ac{PS8D}, our method has slightly higher \ac{AUC}, but it achieves the best score for both error metrics at the end of the experiment (see Figure \ref{fig:PS8D_results} for details). The psychometric functions marked as N/A are not tested with QUEST+ due to the lack of a parametric form (Section~\ref{sec:data_NEST}).}
    \centering
    \setlength{\tabcolsep}{0.25em}
    \footnotesize
    \begin{tabular}{l|c|c|c|c|c|c|c|c}
     \textbf{Method} & \multicolumn{2}{c|}{\textbf{\ac{NV2D}}} & \multicolumn{2}{c|}{\textbf{\ac{WEI4D}}} & \multicolumn{2}{c|}{\textbf{\ac{HART6}}} & \multicolumn{2}{c}{\textbf{\ac{PS8D}}} \\
         & \textbf{AUC$_{\mathrm{RMSE}}$} & \textbf{AUC$_{\mathrm{Brier}}$} & \textbf{AUC$_{\mathrm{RMSE}}$} & \textbf{AUC$_{\mathrm{Brier}}$} & \textbf{AUC$_{\mathrm{RMSE}}$} & \textbf{AUC$_{\mathrm{Brier}}$} & \textbf{AUC$_{\mathrm{RMSE}}$} & \textbf{AUC$_{\mathrm{Brier}}$} \\
        \hline
        NEST (ours) & $\mathbf{22.69 \pm 0.30}$  & $\mathbf{6.03 \pm 0.15}$  & $\mathbf{44.04 \pm 0.84}$ & $\mathbf{8.25 \pm 0.38}$ & $\mathbf{246.52 \pm 2.64}$ & $\mathbf{108.75 \pm 2.19}$ & $371.39 \pm 0.97$ & $204.77 \pm 1.12$ \\
        QUEST+ & N/A & N/A & $110.74 \pm 3.72$ & N/A & N/A & N/A & N/A & N/A \\
        BALD RBF & $28.30 \pm 0.18$ & $11.27 \pm 0.14$ & $78.26 \pm 0.27$ & $19.05 \pm 0.18$ & $287.09 \pm 1.13$ & $307.12 \pm 5.16$ & $375.23 \pm 2.20$ & $291.06 \pm 2.39$ \\
        BALD mon.-RBF & $27.78 \pm 0.19$ & $13.08 \pm 0.17$ & $77.01 \pm 0.35$ & $23.81 \pm 0.23$ & $288.47 \pm 3.25$ & $307.05 \pm 8.81$ & $403.35 \pm 2.19$ & $314.91 \pm 2.17$ \\
        BALD lin.-add. & $35.78 \pm 0.24$ & $17.36 \pm 0.25$ & $106.62 \pm 0.62$ & $23.80 \pm 0.06$ & $342.06 \pm 0.80$ & $360.31 \pm 4.30$ & $394.67 \pm 1.10$ & $282.80 \pm 1.42$ \\
        BALV RBF & $27.05 \pm 0.17$ & $10.45 \pm 0.14$ & $100.63 \pm 0.34$ & $24.61 \pm 0.29$ & $287.18 \pm 1.34$ & $280.11 \pm 27.98$ & $367.21 \pm 4.08$ & $282.56 \pm 2.08$ \\
        BALV mon.-RBF & $26.05 \pm 0.17$ & $12.12 \pm 0.15$ & $58.06 \pm 0.85$ & $10.60 \pm 0.30$ & $284.36 \pm 1.74$ & $281.93 \pm 8.21$ & $397.58 \pm 2.06$ & $308.84 \pm 2.32$ \\
        BALV lin.-add. & $40.70 \pm 0.24$ & $19.31 \pm 0.28$ & $130.00 \pm 0.74$ & $39.19 \pm 0.22$ & $329.77 \pm 1.21$ & $347.69 \pm 4.03$ & $396.92 \pm 1.55$ & $286.47 \pm 1.58$ \\
        LSE RBF & $37.68 \pm 0.23$ & $8.35 \pm 0.07$ & $74.24 \pm 0.37$ & $\mathbf{7.39 \pm 0.07}$ & $342.44 \pm 1.37$ & $256.03 \pm 4.48$ & $406.67 \pm 2.75$ & $244.55 \pm 3.51$ \\
        LSE mon.-RBF & $40.84 \pm 0.22$ & $8.33 \pm 0.09$ & $119.14 \pm 3.57$ & $28.07 \pm 1.19$ & $335.37 \pm 2.36$ & $266.34 \pm 19.77$ & $407.75 \pm 3.50$ & $242.43 \pm 4.99$ \\
        LSE lin.-add. & $51.06 \pm 0.19$ & $7.45 \pm 0.12$ & $146.34 \pm 2.62$ & $42.51 \pm 1.26$ & $340.71 \pm 1.46$ & $327.82 \pm 6.36$ & $380.17 \pm 1.38$ & $203.16 \pm 1.91$ \\
        EAVC RBF & $48.12 \pm 0.34$ & $11.23 \pm 0.16$ & $79.18 \pm 0.56$ & $10.65 \pm 0.17$ & $333.23 \pm 0.81$ & $216.01 \pm 21.97$ & $\mathbf{328.80 \pm 0.58}$ & $\mathbf{184.38 \pm 0.82}$ \\
        GlobalMI RBF & $39.09 \pm 0.31$ & $11.85 \pm 0.15$ & $102.67 \pm 0.43$ & $22.34 \pm 0.35$ & $311.79 \pm 0.51$ & $253.46 \pm 30.79$ & $332.62 \pm 0.60$ & $205.70 \pm 1.04$
    \end{tabular}
    \label{tab:AUC_results}
\end{table*}

\subsection{Performance metrics} \label{sec:error_metrics_NEST}
In order to assess the quality of the results, we use two metrics. The first one is the \ac{RMSE}, which measures the accuracy across the stimulus space and is defined as
\begin{equation}
    \label{eq:RMSE}
    L_{\mathrm{RMSE}}\left(\mathbf{Y}_{\mathrm{true}}, \mathbf{X}_{\mathrm{test}} \right) = \sqrt{ \frac{1}{N} \sum\limits_{i=1}^{N} \left( y_{i} - \Psi(\vec{x}_{i}) \right)^{2} },
\end{equation}
where $\mathbf{Y}_{\mathrm{true}} = (y_{1}, ..., y_{N})$ and $\mathbf{X}_{\mathrm{test}} = ( \vec{x}_{1}, ..., \vec{x}_{N} )$.

The second metric is the \textit{Brier score} developed by \citet{Brier1950} and also used by \citet{LethamEtAl2022}, defined as
\begin{equation}
    \label{eq:Brier_score}
    L_{\mathrm{Brier}} (\mathbf{Y}_{\mathrm{true}}, \mathbf{X}_{\mathrm{test}}, \mu^{*}) = \frac{1}{N} \sum\limits_{i=1}^{N} \left( o_{i} - P(\Psi(\vec{x}_{i}) \geq \mu^{*}) \right)^{2},
\end{equation}
where $o_{i} = (y_{i} \geq \mu^{*}) \in \{ 0, 1 \}$ is a binary variable that indicates whether the true label $y_{i}$ is larger than the target probability $\mu^{*}$, and $P(\Psi(\vec{x}_{i}) \geq \mu^{*})$ denotes the likelihood that the outcome of the approximated psychometric function is higher than $\mu^{*}$. We set $\mu^{*} = \frac{1+\alpha}{2}$, where $\alpha$ is the lower asymptote of the simulation experiment (Equation~\ref{eq:neural_network}). \ac{GP} methods automatically encode such a probability distribution over the outcome. For \ac{NEST}, we use the variance computed using Equation \ref{eq:variance_dropout} to model a Gaussian distribution and determine the required likelihood. Different from \ac{RMSE}, which measures only the accuracy, the Brier score takes into account both the accuracy and uncertainty in the estimation by penalizing high confidence in incorrect threshold placement and rewarding high confidence in correct threshold placement. We classify the test samples either as supra- or sub-threshold stimuli and use this binary target to compute the Brier score.

\subsection{Results} \label{sec:error_metric_results}

\begin{figure}[t]
    \centering
    \includegraphics[width=\linewidth]{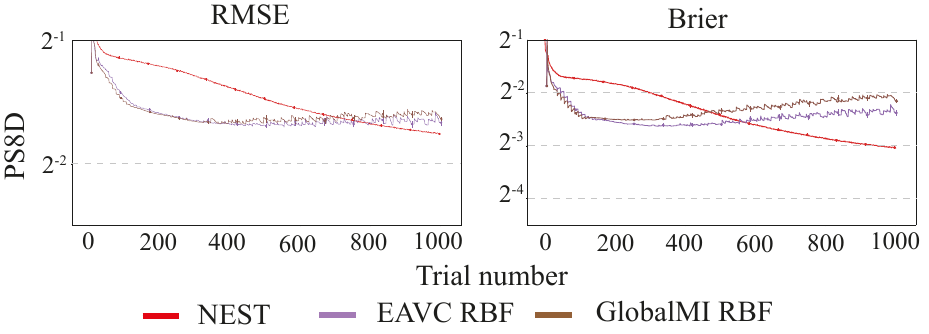}
    \caption{The results for the \ac{PS8D} function for \ac{NEST}, and the \ac{EAVC} and \ac{GlobalMI} \ac{GP} methods. Although \ac{AUC} is higher for \ac{NEST}, it can achieve better performance scores towards the end of the simulation.}
    \label{fig:PS8D_results}
\end{figure}

We benchmark our method, QUEST+, and \ac{GP} methods using the \ac{RMSE} and Brier scores for all of the synthetic psychometric functions listed in Table~\ref{tab:comparisons}. \new{We use the parameter settings for \ac{NEST} described in Section \ref{sec:implementation}, and for the \ac{GP} methods, we reproduce the settings used in \cite{OwenEtAl2021} and \cite{LethamEtAl2022}. Psychophysical methods should not require significant parameter tuning for different experiments. Therefore, it would be incorrect to extensively fine-tune parameters for each test function. Because of this, we perform the simulations on the different test functions using these pre-defined parameter values.} Due to limited space, we provide results only for \ac{NV2D}, \ac{WEI4D}, \ac{HART6}, and \ac{PS8D} in Table \ref{tab:AUC_results}. For the remaining psychometric functions, please refer to the supplementary material. We consider the performance across all trials using the \acf{AUC} for the two metrics. The table shows scores of the best method and the methods whose performance scores do not have a statistically significant difference from the best method (\citet{GamesHowell1976} post-hoc test) in bold typeface.

\new{For the \ac{SIN2D} test function, we noticed a significant gap in performance between \ac{NEST} and the \ac{GP} methods: the \ac{GP} methods appear to fail to model the underlying function. Further analysis of this issue is done in Section \ref{sec:interchange_results}, where it becomes clear that the \ac{GP} is indeed able to fit to the \ac{SIN2D} function, but only if it is provided with the better selection of samples that the \ac{NEST} acquisition function produces.}

The \ac{LSE} \ac{GP} methods perform poorly in the \ac{RMSE} error metric but very well in the Brier score since the latter score focuses on threshold accuracy above overall slope accuracy. Despite this, \ac{NEST} matches the Brier score of the \ac{LSE} methods in \ac{WEI4D} and outperforms it in \ac{NV2D} and \ac{HART6}. \ac{NEST} also outperforms the lookahead functions \ac{EAVC} and \ac{GlobalMI} in \ac{NV2D}, \ac{WEI4D}, and \ac{HART6}. On the \ac{PS8D} function, \ac{NEST} performs slightly below the \ac{GlobalMI} and \ac{EAVC} methods. Upon further investigation, we observed that although \ac{GlobalMI} and \ac{EAVC} achieve a fast decrease in \ac{RMSE} and the Brier score at the beginning of the experiment, \ac{NEST} achieves lower prediction error at the end of the experiment (Figure~\ref{fig:PS8D_results}). Overall, \ac{NEST} either outperforms the \ac{GP} methods for most tested functions or it is on par with them. QUEST+ performs the best on parametric 1D and 2D functions, but \ac{NEST} outperforms QUEST+ on higher-dimensional functions like \ac{WEI4D}.

\begin{figure*}[t!]
    \centering
    \includegraphics[width=1.0\textwidth]{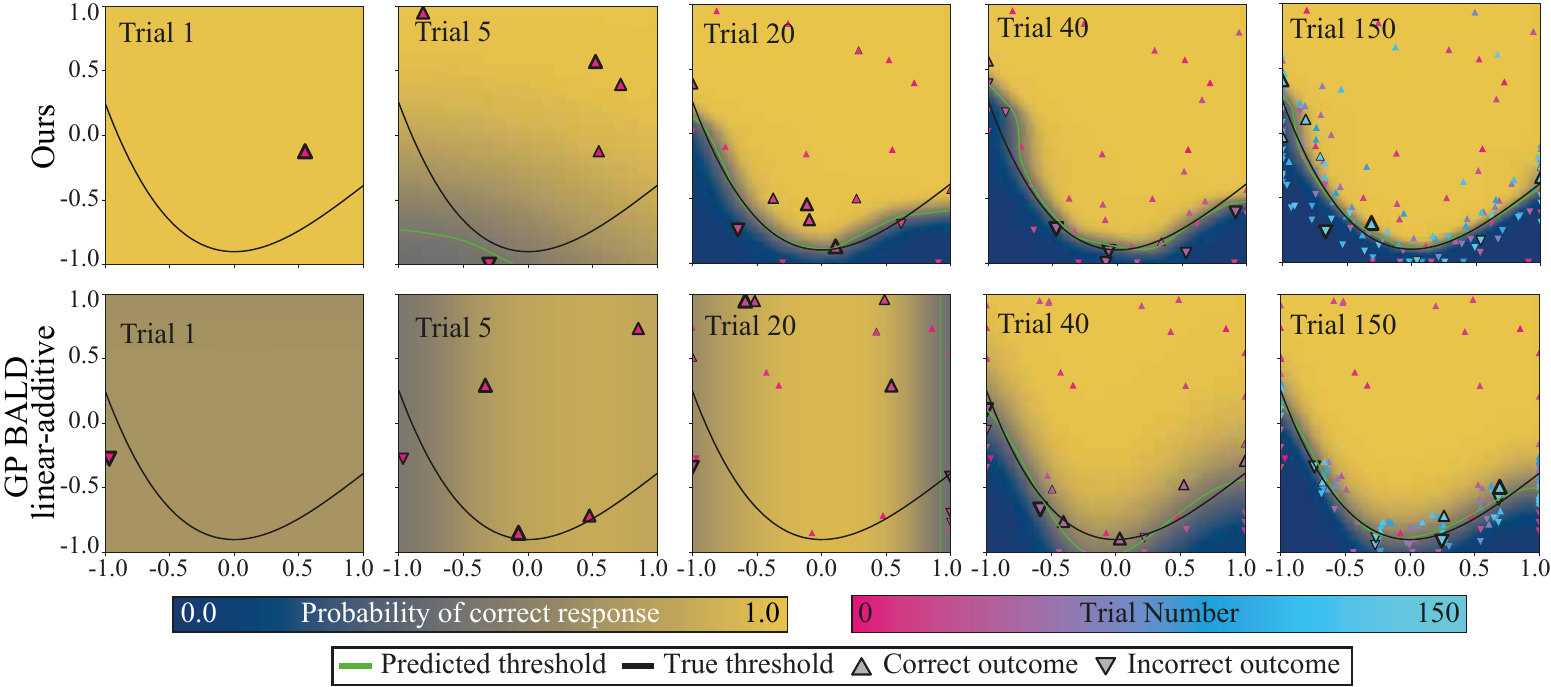}
    \caption{\ac{NEST}  and \ac{GP} BALD linear-additive psychometric function estimation for the \ac{NV2D} function after 1, 5, 20, 40, and 150 trials. Horizontal and vertical axes represent the input stimulus parameters. The ten most recent samples have a thicker border around the marker. Our method finds a relatively good approximation of the psychometric function after 20 trials.}
    \label{fig:NEST_novel_test_function_steps}
\end{figure*}

The kernel function in \ac{GP} methods should be selected according to the underlying psychometric function type for optimal performance. In Figure \ref{fig:NEST_novel_test_function_steps}, stimulus selection and estimation of the \ac{NV2D} psychometric function is shown for different numbers of trials for one experiment run. In the same figure, we also compare our method with the \ac{GP} method, using the \ac{BALD} acquisition function and the linear-additive kernel function defined in \citet{OwenEtAl2021}. Our approach surpasses GP methods by being more flexible in modeling various psychometric functions, eliminating the need for kernel selection.

\begin{figure*}[t!]
    \centering
    \includegraphics[height=0.29\linewidth]{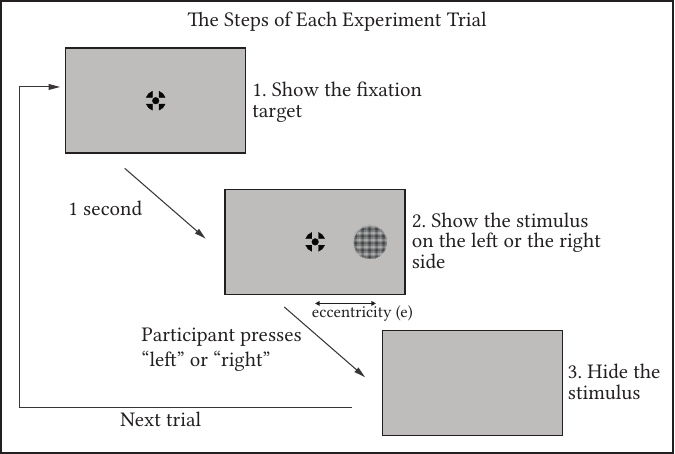} \includegraphics[height=0.29\linewidth]{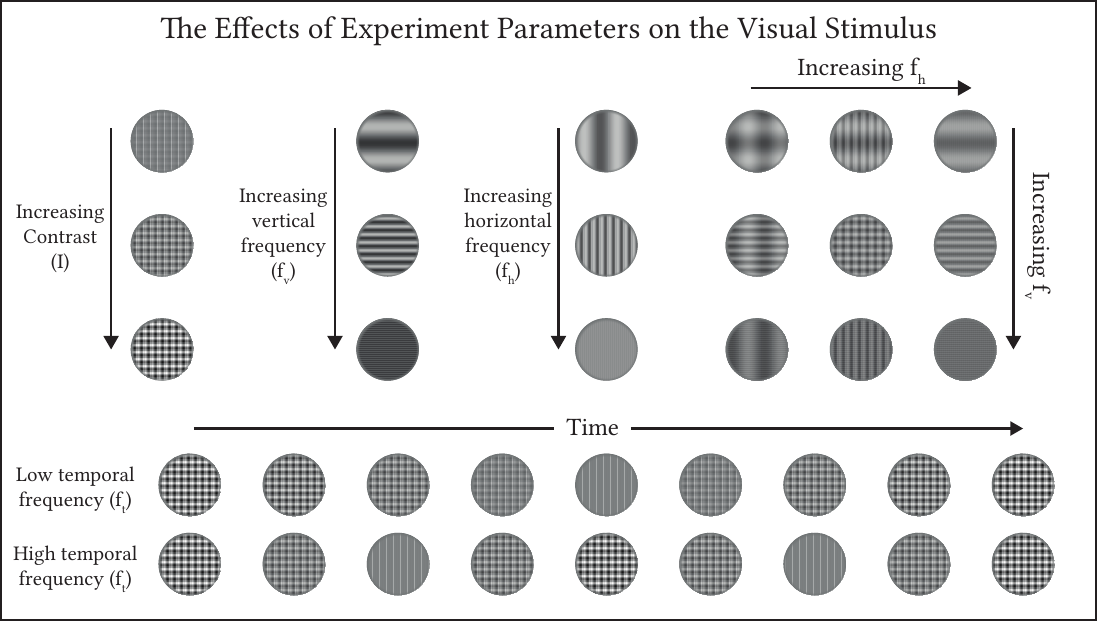}
    \caption{\new{The experimental procedure (left) and stimulus parameters (right) for the live experiments in Section \ref{sec:experiments_NEST}. The participant first fixates on the fixation target. Then, a stimulus is shown on the left or right side of the fixation target. The participant indicates (or guesses if the stimulus is not visible for them) whether the stimulus is on the left or right side using the arrow keys. Then the stimulus disappears and a new trial begins. The stimulus is shown at a distance corresponding to a given number of degrees of eccentricity. The other four studied parameters are contrast $I$, vertical frequency $f_{v}$, horizontal frequency $f_{h}$, and temporal frequency $f_{t}$.}}
    \label{fig:live_experiment_diagram}
\end{figure*}

\section{Validation with psychovisual experiments} \label{sec:experiments_NEST}
We further benchmarked our method by conducting two psychovisual experiments. The first experiment (Section~\ref{sec:Barten_CSF_results}) is a reproduction of the eccentricity-dependent spatial \ac{CSF}, which serves as a test on a thoroughly studied 3D psychometric function. The second experiment (Section~\ref{sec:Tursun_Didyk_results}) is a reproduction of the spatio-temporal visibility model as a function of eccentricity by \citet{TursunDidyk2022}, which uses \ac{vPEST} \citep{Findlay1978}. We observe that our method can estimate the psychometric functions from original studies with a high accuracy albeit with a smaller number of trials (Section~\ref{sec:Tursun_Didyk_results}). \new{A visual illustration of the experiment is shown in Figure \ref{fig:live_experiment_diagram}. The live spatial \ac{CSF} experiment has three parameters: eccentricity, contrast, and vertical frequency. The spatio-temporal experiment of \cite{TursunDidyk2022} additionally includes horizontal frequency and temporal frequency parameters.}

In both experiments, the participants were positioned at a distance of 62 cm from the display. We used a 55-inch 4K LG OLED55CX display with a 120 Hz refresh rate and linear calibration. The peak spatial frequency of the display was 18.15 \ac{cpd} and the peak luminance was set to 170 cd/m$^{2}$.

Each participant's neural network is trained individually. A combined model representing the average psychometric function is obtained using \textit{ensemble averaging} with equal weights, without additional training \citep{NaftalyEtAl1997}. The experiment procedure was approved by the research ethics committee of the hosting institution (CETO \textnumero. 95598952), and informed consent was obtained from each participant.

\subsection{Spatial CSF experiment} \label{sec:Barten_CSF_results}
In this experiment, we use sinusoidal Gabor gratings as the stimuli on a uniform gray background at 85 cd/m$^{2}$ luminance. The stimulus was presented at different retinal eccentricities $e \in [5^\degree, 30^\degree]$ with spatial frequencies $u \in [0.5, 9.07]$ \ac{cpd}. The experiment task was to indicate whether the stimulus was on the left or the right side of the visual field relative to a fixation target at the center. We compare our method with the model developed by \citet{Barten1999} using retinal ganglion cell density formula from \citet{Watson2014}.

\begin{figure*}[t!]
    \centering
    \includegraphics[width=1.0\textwidth]{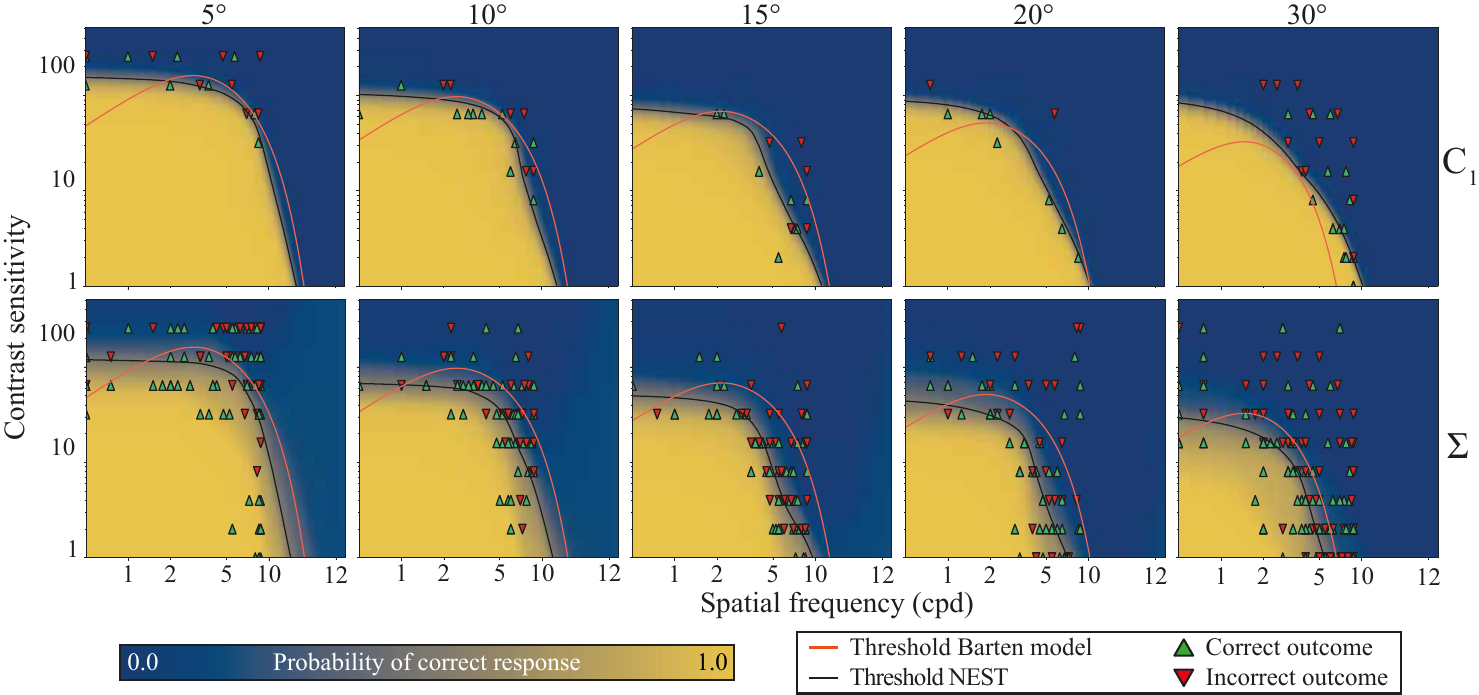}
    \caption{Psychometric function estimates from the CSF experiment (Section~\ref{sec:Barten_CSF_results}) for multiple eccentricities. The top row shows the model of participant C$_1$ and the bottom row ($\Sigma$) shows the average model of all participants. \ac{NEST} learns a function that closely aligns with the fitted theoretical model.}
    \label{fig:CSF_function_results}
\end{figure*}

Six participants (3F, 3M), including one author, with mean age $27.2 \pm 4.5$ performed this experiment. All participants reported normal or corrected-to-normal vision. Figure~\ref{fig:CSF_function_results} shows the results for participant C$_1$ as well as averaging all participants' models in the spatial \ac{CSF} experiment for eccentricities $5^{\circ}$, $10^{\circ}$, $15^{\circ}$, $20^{\circ}$, and $30^{\circ}$. The results for the other partcipants are provided in the supplementary material. The parameters of the Barten \ac{CSF} model are shown in Table \ref{tab:Barten_parameters}. The mean number of trials of our method was $353.8 \pm 51.1$ using our convergence criterion (Section \ref{sec:convergence_criterion_results}). The experiment took 20--35 minutes for each participant.

Figure~\ref{fig:CSF_function_results} shows that many of the samples selected by \ac{NEST} during the experiment are close to the final value of the detection threshold and that the samples are not clustered around a particular region (e.g., the stimulus space boundary), highlighting efficient sample selection. The results from a selected participant C$_{1}$ and the model fit to all participants' responses are generally in agreement with the Barten's \ac{CSF} model.

\begin{table}[b!]
    \centering
    \caption{Parameters and values used for the CSF model by \cite{Barten1999}.}
    \begin{tabular}{c|c||c|c}
        \textbf{Parameter} & \textbf{Value} & \textbf{Parameter} & \textbf{Value}  \\ \hline
        $\sigma_{0}$ & 1.5 & $r_{e}$ & 7.633 \\
        $e_{g}$ & 3.3 & $\eta$ & 0.04 \\
        $k$ & 2.3 & &
    \end{tabular}
    \label{tab:Barten_parameters}
\end{table}

\begin{figure*}[ht!]
    \centering
    \includegraphics[width=\linewidth]{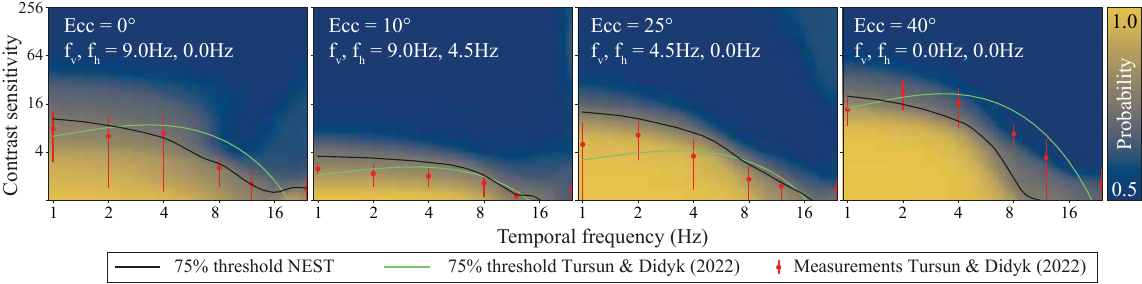}
    \caption{Selected 2D slices of the 5D psychometric function learned from the experiment discussed in Section \ref{sec:Tursun_Didyk_results}. $75\%$ detection threshold from the parameterized psychometric function introduced by \citet{TursunDidyk2022} and their mean sensitivity measurements are shown in green and red, respectively. The predicted threshold by \ac{NEST} is shown in black. We observe that \ac{NEST} is able to capture trends of the data better than the model of \cite{TursunDidyk2022}. Vertical bars represent standard error of the mean.}
    \label{fig:spatio_temporal_experiment}
\end{figure*}

\subsection{Spatio-temporal visibility model experiment} \label{sec:Tursun_Didyk_results}
The second experiment is a reproduction of the perceptual visibility model as a function of spatio-temporal frequency and eccentricity \citep{TursunDidyk2022}, which uses \ac{vPEST} \citep{Findlay1978}.
Due to the high dimensionality, their procedure limited the number of unique stimulus values in each dimension to just three. Nevertheless, the authors report an average duration of 240 minutes for each participant in multiple sessions. For each of the 162 stimulus combinations, a separate \ac{vPEST} procedure was run, and the total number of trials for each participant was approximately 5,000.

Five participants (2F, 3M), including one author, with mean age $23.4 \pm 2.5$ performed this experiment. Figure~\ref{fig:spatio_temporal_experiment} shows the results found by combining the individual results of all the participants P1--P5 in this experiment. The individual results are shown in the supplementary material. The experiment took an average of $730.8 \pm 65.6$ trials with our method to reach convergence, which took 60--80 minutes for each participant. Both the number of trials and experiment duration is significantly reduced with our method.

Figure \ref{fig:spatio_temporal_experiment} shows the psychometric function learned by \ac{NEST} for all combined participants, as well as the original data collected by \cite{TursunDidyk2022} and their model fit. Due to its non-parametric nature, we observe that \ac{NEST} has more flexibility in coming up with a better fit to the measurements with significantly fewer trials.

\begin{table}[b!]
    \ssmall
    \caption{\new{Mean RMSE and Brier AUC scores obtained from 100 runs with different neural network sizes on \ac{NV2D}, \ac{DN2D}, \ac{SIN2D}, \ac{WEI4D}, and \ac{HART6} test functions (lower is better). The best scores are shown in bold typeface. Here, $d$ indicates the total number of hidden layers and $W_0$ the width of the first hidden layer. The width of every subsequent hidden layer after the first one is halved. $\text{p}^{*}_{\mathrm{RMSE}}$ and $\text{p}^{*}_{\mathrm{Brier}}$ are the dropout parameter values used for regularization. All architectures are fully connected feedforward neural networks. $W_\text{exp}$ represents the network architecture that we used in Sections~\ref{sec:data_NEST} and \ref{sec:experiments_NEST} (please refer to the text for details).}
    }
    \centering
    \new{
    \begin{tabular}{c|c|c|c|c}
        \textbf{Network} & \textbf{$\text{AUC}_{\mathrm{RMSE}}$} & \textbf{$\text{p}^{*}_{\mathrm{RMSE}}$} & \textbf{$\text{AUC}_{\mathrm{Brier}}$} & \textbf{$\text{p}^{*}_{\mathrm{Brier}}$} \\ \hline
        $W_{0} = 128, d = 2$ & 0.1968 & 0.05 & 0.0875 & 0.05 \\
        $W_{0} = 256, d = 2$ & 0.1603 & 0.05 & 0.0625 & 0.05 \\
        $W_{0} = 512, d = 2$ & 0.1484 & 0.05 & 0.0582 & 0.05 \\
        $W_{0} = 128, d = 3$ & 0.1632 & 0.05 & 0.0635 & 0.05 \\
        $W_{0} = 256, d = 3$ & \textbf{0.1430} & 0.1 & \textbf{0.0553} & 0.1 \\
        $W_{\text{exp}}$ & 0.1489 & 0.05 & 0.0587 & 0.05 \\
        $W_{0} = 512, d = 3$ & 0.1444 & 0.25 & 0.0554 & 0.25 \\
        $W_{0} = 128, d = 4$ & 0.1676 & 0.05 & 0.0681 & 0.05 \\
        $W_{0} = 256, d = 4$ & 0.1464 & 0.1 & 0.0559 & 0.1 \\
        $W_{0} = 512, d = 4$ & 0.1506 & 0.2 & 0.0554 & 0.1 \\
        $W_{0} = 128, d = 5$ & 0.1949 & 0.05 & 0.0891 & 0.05 \\
        $W_{0} = 256, d = 5$ & 0.1513 & 0.05 & 0.0578 & 0.05 \\
        $W_{0} = 512, d = 5$ & 0.1538 & 0.25 & \textbf{0.0553} & 0.1 \\
        Best GP result & 0.2090 & N/A & 0.1324 & N/A \\
    \end{tabular}}
    \label{tab:architecture_study_results}
\end{table}

\begin{table*}[hb]
    \caption{\new{Results of the hybrid model comparison. The table shows the mean \ac{AUC} for the \ac{RMSE} and Brier error scores from 100 runs (lower is better). Values are expressed as mean $\pm$ standard error of the mean (SEM). The lowest scores and those which do not have statistically significant difference ($p<0.05$, Tukey HSD test) are shown in bold. Model/Data column represents the model and sample selection (i.e., acquisition function) pairings corresponding to each row. \ac{NEST} model using the \ac{NEST} acquisition function is the best performer. Furthermore, using \ac{NEST} data with a \ac{GP} model slightly improves the performance on \ac{SIN2D} and \ac{HART6} test functions.}}
    \centering
    \setlength{\tabcolsep}{0.25em}
    \ssmall
    \new{
    \begin{tabular}{l|c|c|c|c|c}
         & \multicolumn{5}{c}{\textbf{$\text{AUC}_{\mathrm{RMSE}}$}} \\
        \textbf{Model/Data} & \textbf{NV2D} & \textbf{SIN2D} & \textbf{DN2D} & \textbf{WEI4D} & \textbf{HART6} \\ \hline
        NEST/NEST & $\mathbf{0.0785 \pm 0.0007}$ & $\mathbf{0.2237 \pm 0.0011}$ & $\mathbf{0.1626 \pm 0.0016}$ & $\mathbf{0.0607 \pm 0.0006}$ & $\mathbf{0.2444 \pm 0.0016}$ \\
        NEST/GP & $\mathbf{0.0791 \pm 0.0007}$ & $0.4046 \pm 0.0026$ & $0.1774 \pm 0.0042$ & $0.0653 \pm 0.0017$ & $0.2805 \pm 0.0036$ \\
        GP/GP & $0.0929 \pm 0.0006$ & $0.3882 \pm 0.0017$ & $0.2036 \pm 0.0016$ & $0.0849 \pm 0.0023$ & $0.2883 \pm 0.0039$ \\
        GP/NEST & $0.1003 \pm 0.0008$ & $0.2703 \pm 0.0079$ & $0.2287 \pm 0.0058$ & $0.0765 \pm 0.0008$ & $0.2810 \pm 0.0015$ \\ \hline
        \multicolumn{6}{c}{} \\
         & \multicolumn{5}{c}{\textbf{$\text{AUC}_{\mathrm{Brier}}$}} \\
        \textbf{Model/Data} & \textbf{NV2D} & \textbf{SIN2D} & \textbf{DN2D} & \textbf{WEI4D} & \textbf{HART6} \\ \hline
        NEST/NEST & $\mathbf{0.0212 \pm 0.0003}$ & $\mathbf{0.0842 \pm 0.0009}$ & $\mathbf{0.0868 \pm 0.0011}$ & $\mathbf{0.0111 \pm 0.0003}$ & $\mathbf{0.1139 \pm 0.0012}$ \\
        NEST/GP & $\mathbf{0.0217 \pm 0.0003}$ & $0.2395 \pm 0.0027$ & $0.0973 \pm 0.0020$ & $\mathbf{0.0105 \pm 0.0005}$ & $0.1523 \pm 0.0032$ \\
        GP/GP & $0.0249 \pm 0.0004$ & $0.2231 \pm 0.0011$ & $0.0966 \pm 0.0016$ & $0.0143 \pm 0.0005$ & $0.2198 \pm 0.0037$ \\
        GP/NEST & $0.0426 \pm 0.0005$ & $0.1373 \pm 0.0055$ & $0.1467 \pm 0.0054$ & $0.0153 \pm 0.0003$ & $0.2105 \pm 0.0032$
    \end{tabular}}

    \label{tab:interchange_results}
\end{table*}

\begin{table*}[b!]
    \caption{\new{Time spent in network training and the computation of the acquisition function on \ac{NV2D} and \ac{HART6} test functions for different network sizes. The time spent in training is relatively stable across different network sizes, whereas computing the acquisition function shows more variability. Total time spent in each trial is the summation of time spent in both network training and the computation of the acquisition function.}}
    \centering
    \setlength{\tabcolsep}{0.25em}
    \ssmall
    \new{
    \begin{tabular}{c|c|c|c|c|c|c}
         & \multicolumn{6}{c}{\textbf{Time spent in network training (s)}} \\
         & \multicolumn{3}{c|}{\textbf{\ac{NV2D}}} & \multicolumn{3}{c}{\textbf{\ac{HART6}}} \\
        \textbf{Network} & \textbf{10th perc.} & \textbf{median} & \textbf{90th perc.} & \textbf{10th perc.} & \textbf{median} & \textbf{90th perc.} \\ \hline
        $W_{\mathrm{exp}}$ & 0.100 & 0.104 & 0.109 & 0.132 & 0.140 & 0.144 \\
        $W_{0} = 256, d = 3$ & 0.107 & 0.110 & 0.115 & 0.135 & 0.144 & 0.147 \\
        $W_{0} = 512, d = 5$ & 0.193 & 0.204 & 0.215 & 0.273 & 0.281 & 0.288 \\ \hline
        \multicolumn{7}{c}{} \\
         & \multicolumn{6}{c}{\textbf{Time spent in computing the acquisition function (s)}} \\
         & \multicolumn{3}{c|}{\textbf{\ac{NV2D}}} & \multicolumn{3}{c}{\textbf{\ac{HART6}}} \\
        \textbf{Network} & \textbf{10th perc.} & \textbf{median} & \textbf{90th perc.} & \textbf{10th perc.} & \textbf{median} & \textbf{90th perc.} \\ \hline
        $W_{\mathrm{exp}}$ & 0.332 & 0.517 & 1.073 & 1.029 & 1.501 & 2.485 \\
        $W_{0} = 256, d = 3$ & 0.370 & 0.677 & 1.176 & 1.095 & 1.662 & 2.902 \\
        $W_{0} = 512, d = 5$ & 0.838 & 1.436 & 2.266 & 2.210 & 3.692 & 6.456
    \end{tabular}}

    \label{tab:timing_study}
\end{table*}

\begin{table*}[b!]
    \caption{The ablation mean \ac{AUC} $\pm$ standard error of the \ac{RMSE} and Brier metrics for the \ac{NV2D}, \ac{WEI4D}, and \ac{HART6} functions. Each row indicates the included components, Uncertainty (U), Proximity (P), Gradient (G), and Lookahead (L). For the ``Random'' category, Poisson-disk random sampling \citep{Wang2020} is used as a pure random exploration behavior as a reference. The group of best scores that lack a significant difference ($p \leq 0.05$, Dunnett test) are shown in boldface.}
    \centering
    \setlength{\tabcolsep}{0.25em}
    \ssmall
    \begin{tabular}{l|c|c|c|c|c|c|c|c|c|c}
     \textbf{Method} & \multicolumn{2}{c|}{\textbf{\ac{NV2D}}} & \multicolumn{2}{c|}{\textbf{\ac{SIN2D}}} & \multicolumn{2}{c|}{\textbf{\ac{DN2D}}} & \multicolumn{2}{c|}{\textbf{\ac{WEI4D}}} & \multicolumn{2}{c}{\textbf{\ac{HART6}}} \\
         & \textbf{AUC$_{\mathrm{RMSE}}$} & \textbf{AUC$_{\mathrm{Brier}}$} & \textbf{AUC$_{\mathrm{RMSE}}$} & \textbf{AUC$_{\mathrm{Brier}}$} & \textbf{AUC$_{\mathrm{RMSE}}$} & \textbf{AUC$_{\mathrm{Brier}}$} & \textbf{AUC$_{\mathrm{RMSE}}$} & \textbf{AUC$_{\mathrm{Brier}}$} & \textbf{AUC$_{\mathrm{RMSE}}$} & \textbf{AUC$_{\mathrm{Brier}}$} \\
        \hline
        Random & $34.75 \pm 0.31$ & $9.95 \pm 0.14$ & $62.77 \pm 0.30$ & $\mathbf{22.52 \pm 0.23}$ & $\mathbf{51.45 \pm 0.29}$ & $\mathbf{22.38 \pm 0.21}$ & $75.39 \pm 0.52$ & $12.31 \pm 0.17$ & $263.11 \pm 1.38$ & $128.83 \pm 1.39$ \\
        U & $37.10 \pm 0.46$ & $11.17 \pm 0.24$ & $72.42 \pm 0.50$ & $26.15 \pm 0.39$ & $64.08 \pm 0.82$ & $29.70 \pm 0.65$ & $\mathbf{43.92 \pm 0.30}$ & $\mathbf{8.19 \pm 0.12}$ & $270.08 \pm 2.96$ & $149.43 \pm 3.40$ \\
        P & $32.63 \pm 0.34$ & $9.60 \pm 0.16$ & $67.92 \pm 0.36$ & $26.59 \pm 0.27$ & $57.76 \pm 0.41$ & $26.09 \pm 0.23$ &  $72.00 \pm 0.52$ & $11.78 \pm 0.16$ & $255.58 \pm 1.40$ & $125.17 \pm 1.39$ \\
        P + U & $\mathbf{24.63 \pm 0.24}$ & $\mathbf{6.86 \pm 0.11}$ & $62.26 \pm 0.37$ & $23.75 \pm 0.35$ & $\mathbf{50.95 \pm 0.53}$ & $\mathbf{23.20} \pm 0.28$ &  $\mathbf{43.31 \pm 0.32}$ & $\mathbf{7.92 \pm 0.12}$ & $252.70 \pm 2.67$ & $123.47 \pm 2.61$ \\
        G & $43.81 \pm 0.78$ & $12.70 \pm 0.32$ & $77.93 \pm 0.51$ & $28.74 \pm 0.35$ & $73.79 \pm 0.52$ & $34.39 \pm 0.62$ & $44.59 \pm 0.33$ & $8.76 \pm 0.15$ & $316.00 \pm 2.14$ & $161.20 \pm 2.11$ \\
        G + U & $39.52 \pm 0.47$ & $12.14 \pm 0.26$ & $71.48 \pm 0.50$ & $26.07 \pm 0.38$ & $63.93 \pm 0.44$ & $29.03 \pm 0.60$ &  $\mathbf{44.23 \pm 0.30}$ & $\mathbf{8.06 \pm 0.11}$ & $268.69 \pm 2.75$ & $144.30 \pm 2.86$ \\
        G + P & $29.03 \pm 0.35$ & $8.28 \pm 0.15$ & $65.29 \pm 0.38$ & $25.04 \pm 0.28$ & $53.93 \pm 0.44$ & $23.88 \pm 0.25$ &  $\mathbf{42.56 \pm 0.36}$ & $\mathbf{8.02 \pm 0.13}$ & $309.82 \pm 2.10$ & $157.56 \pm 2.09$ \\
        G + P + U & $\mathbf{23.41 \pm 0.20}$ & $\mathbf{6.44 \pm 0.10}$ & $\mathbf{59.98 \pm 0.30}$ & $\mathbf{21.98 \pm 0.20}$ & $\mathbf{49.83 \pm 0.40}$ & $\mathbf{22.61 \pm 0.20}$ &  $\mathbf{44.06 \pm 0.27}$ & $\mathbf{8.11 \pm 0.12}$ & $262.87 \pm 2.49$ & $129.89 \pm 2.29$ \\
        L & $47.11 \pm 1.04$ & $13.53 \pm 0.42$ & $63.86 \pm 0.41$ & $23.62 \pm 0.26$ & $55.46 \pm 0.38$ & $24.80 \pm 0.25$ &  $58.61 \pm 0.46$ & $9.42 \pm 0.14$ & $295.62 \pm 2.70$ & $132.87 \pm 2.59$ \\
        L + U & $28.08 \pm 0.32$ & $\mathbf{6.94 \pm 0.12}$ & $65.34 \pm 0.39$ & $23.68 \pm 0.28$ & $58.91 \pm 0.67$ & $26.45 \pm 0.54$ &  $45.40 \pm 0.31$ & $\mathbf{8.01 \pm 0.12}$ & $\mathbf{238.34 \pm 2.26}$ & $\mathbf{104.88 \pm 2.17}$ \\
        L + P & $29.85 \pm 0.29$ & $8.34 \pm 0.14$ & $68.34 \pm 0.39$ & $26.97 \pm 0.28$ & $58.42 \pm 0.44$ & $26.36 \pm 0.24$ &  $58.15 \pm 0.39$ & $8.88 \pm 0.12$ & $277.55 \pm 2.26$ & $119.91 \pm 2.10$ \\
        L + P + U & $\mathbf{23.85 \pm 0.22}$ & $\mathbf{6.38 \pm 0.10}$ & $62.66 \pm 0.34$ & $23.79 \pm 0.24$ & $\mathbf{50.72 \pm 0.58}$ & $\mathbf{22.88 \pm 0.30}$ &  $45.91 \pm 0.26$ & $\mathbf{8.33 \pm 0.11}$ & $\mathbf{238.75 \pm 2.30}$ & $\mathbf{106.87 \pm 2.26}$ \\
        L + G & $46.80 \pm 0.98$ & $13.43 \pm 0.42$ & $65.20 \pm 0.47$ & $23.82 \pm 0.28$ & $56.90 \pm 0.57$ & $24.28 \pm 0.32$ &  $44.83 \pm 0.29$ & $\mathbf{8.32 \pm 0.11}$ & $307.36 \pm 2.79$ & $144.13 \pm 2.61$ \\
        L + G + U & $28.85 \pm 0.31$ & $7.37 \pm 0.12$ & $65.04 \pm 0.44$ & $23.73 \pm 0.30$ & $60.17 \pm 0.76$ & $26.99 \pm 0.62$ &  $45.26 \pm 0.30$ & $\mathbf{8.29 \pm 0.11}$ & $249.04 \pm 2.35$ & $113.29 \pm 2.22$ \\
        L + G + P & $28.10 \pm 0.31$ & $7.75 \pm 0.14$ & $65.47 \pm 0.36$ & $25.00 \pm 0.26$ & $53.35 \pm 0.46$ & $\mathbf{23.46 \pm 0.25}$ &  $\mathbf{43.68 \pm 0.28}$ & $\mathbf{8.05 \pm 0.11}$ & $294.13 \pm 2.88$ & $134.17 \pm 2.77$ \\
        Full & $\mathbf{23.29 \pm 0.20}$ & $\mathbf{6.25 \pm 0.09}$ & $\mathbf{60.23 \pm 0.29}$ & $\mathbf{22.08 \pm 0.22}$ & $\mathbf{50.00 \pm 0.40}$ & $\mathbf{22.68 \pm 0.24}$ &  $\mathbf{44.05 \pm 0.25}$ & $\mathbf{8.10 \pm 0.12}$ & $\mathbf{246.52 \pm 2.64}$ & $\mathbf{108.75 \pm 2.19}$
    \end{tabular}
    \label{tab:ablation_AUC_results}
\end{table*}

\new{\section{Hybrid model comparison}}
\label{sec:interchange_results}
\new{\ac{NEST} and the \ac{GP} methods consist of two parts: an acquisition function that determines the sampling strategy and model fitting. In order to see the performance effects of those individual parts, we benchmark two hybrid models:}
\new{\begin{enumerate}
     \item We refit a \ac{GP} model on the data generated by the acquisition function of \ac{NEST};
     \item We fit a neural network on the data generated by the samples selected by the \ac{GP} model.
\end{enumerate}}

\new{In our implementation, we follow the same fitting procedure for both NEST and \ac{GP} models as described earlier in Section \ref{sec:error_metric_results}. However, the sample selection in each trial is driven by the sampling strategy of the other method. For consistency, we use the samples generated during the simulations we reported in Section \ref{sec:error_metric_results} as a stand-in for the acquisition function. We benchmark the performance using \ac{NV2D}, \ac{SIN2D}, \ac{DN2D}, \ac{WEI4D}, and \ac{HART6} test functions. For each function, we perform 100 runs with the same number of trials as in Section \ref{sec:error_metric_results}. First, we fit the \ac{NEST} model to the samples selected by the best \ac{GP} result (denoted by NEST/GP). Then, we fit a \ac{GP} model to the data generated by the acquisition function of \ac{NEST} (GP/NEST). We compute the prediction errors of those two hybrid methods with the prediction errors of the native model-acquisition function pairings (i.e., NEST/NEST and GP/GP). The resulting normalized \ac{AUC} scores for this analysis are shown in Table \ref{tab:interchange_results}, with the best performance for each column marked in boldface, along with those results that are not statistically significantly different according to the Tukey \ac{HSD} test \citep{Tukey1949}.}

\new{We observe that the combination of the \ac{NEST} model paired with the \ac{NEST} acquisition function achieves the lowest \ac{AUC} scores in each test function. The results are mixed for the NEST/GP and GP/NEST hybrid models. The NEST/GP pairing achieves better scores for NV2D, WEI4D, and HART6 test functions, whereas the GP/NEST pairing performs better for SIN2D and HART6 than the GP/GP baseline. Overall, we observe that the difference between the performance of the NEST/GP and GP/GP pairings is minor, which indicates that the influence of the sample selection strategy driven by the acquisition function is the dominant performance factor.}

\new{
\section{Network size selection}} \label{sec:architecture_study}
\new{We benchmarked different network sizes for our method by changing the width and depth of the network. Similar to our previous benchmarks, we computed the mean of RMSE and Brier scores from a Monte Carlo simulation of 100 runs on test functions (\ac{NV2D}, \ac{SIN2D}, \ac{DN2D}, \ac{WEI4D}, and \ac{HART6}) and reported the results in Table \ref{tab:architecture_study_results}. In each row, $d$ indicates the total number of hidden layers and $W_0$ indicates the width of the first hidden layer. The width of every subsequent hidden layer after the first one is halved. We performed the benchmark for $W_{0} \in [128, 256, 512]$ and $d \in [2, 3, 4, 5]$. Additionally, we show the error scores of the network size used in the simulations of Section~\ref{sec:data_NEST} and the live experiments of Section~\ref{sec:experiments_NEST}, labelled as $W_{\text{exp}}$. $W_\text{exp}$ is very similar to the network represented by $W_0 = 256$ and $d=3$ except for the width of the last layer, which is selected as $32$ in $W_\text{exp}$ instead of $64$. Since larger networks require stronger regularization to prevent overfitting, we optimized the dropout parameter of the network for each network size and indicated its value in columns labeled as $\text{p}^{*}_{\mathrm{RMSE}}$ and $\text{p}^{*}_{\mathrm{Brier}}$. Finally, the best result obtained from the \ac{GP} methods evaluated in Section \ref{sec:error_metric_results} is shown in the last row.}

\new{Table \ref{tab:architecture_study_results} shows that the combination $(W_{0} = 256, d = 3)$ is the top performer with $\text{AUC}_\text{RMSE} = 0.1430$ and $\text{AUC}_\text{Brier} = 0.0553$. $W_\text{exp}$ has a very close performance to this network size with $\text{AUC}_\text{RMSE} = 0.1489$ and $\text{AUC}_\text{Brier} = 0.0587$, indicating that our network size selection is close to ideal. We also see that the performance is not too sensitive to the choice of network size except for $W_0=128, d=2$ and $W_0=128, d=5$, where the network depth seems too low or too high. Furthermore, the performance for a wide range of network sizes is better than the best performance obtained from the \ac{GP} method.}

\new{
\section{Running time analysis}}
\new{We measured the wall-clock time during neural network training and acquisition function evaluation in each trial to assess the computational cost of our method. In order to test the complexity introduced by the number of dimensions, we collect timings from \ac{NV2D} and \ac{HART6} test functions, representing both a low- and high-dimensional test case. Furthermore, in order to evaluate the influence of the network size, we measured running times for three different network sizes: the network from Section \ref{sec:error_metric_results} with hidden layer widths $[256, 128, 32]$, the best network size in Section \ref{sec:architecture_study} with $W = 256$ and $d = 3$, and the largest network size analyzed in Section \ref{sec:architecture_study} with $W = 512$ and $d = 5$. We took our measurements on the hardware we used for the live experiments in Section \ref{sec:experiments_NEST} with an Nvidia RTX 4090 GPU (24GB memory), an Intel Core i9-13700K CPU, 2x32 GB DDR5 4800 MT/s RAM, and a 1 TB, 7000 MB/s PCI Express 4.0 NVMe SSD. We performed 50 runs of Monte Carlo simulation for each configuration to obtain a stable estimation.}

\new{Table \ref{tab:timing_study} shows the median together with 10th and 90th-percentile points of the time spent in network training and acquisition function computation for the \ac{NV2D} and \ac{HART6} test functions. As the number of trials increases during the course of the experiment, the training time increases due to the growing number of samples. In order to represent the worst-case scenario for each test function, we reported the measured timings for the aggregate data from the last 10 trials of each experiment. If the reader is interested in the plots of the training and acquisition times per trial for each test function and each architecture throughout the experiment, those are provided in the Supplementary Materials.} 

\new{From Table \ref{tab:timing_study}, we observe that the training times always remain below 0.5 seconds for different architectures and test functions. The acquisition times are mostly longer, and depend on the choice of network size and the specific test function used. For all configurations, except for the large network with the \ac{HART6} test function, the sum of the training and acquisition times is mostly below 3 seconds, which is the 90th percentile. Such durations are usually acceptable for a live experiment. Longer running times are associated with larger networks such as $W_0=512, d=5$ in our analysis. Therefore, the trend in the running times is yet another reason for selecting a moderate network size, in addition to the findings from network size selection in Section~\ref{sec:architecture_study}.}
\\
\section{Ablation study} \label{sec:ablation}

We perform Monte Carlo simulations with subsets of components from the acquisition function and report the \ac{AUC} from \ac{RMSE} and Brier score in Table~\ref{tab:ablation_AUC_results}. We group the best \ac{AUC} scores that are statistically different from the rest ($p \leq 0.05$, Dunnett test). Different subsets of the acquisition function components seem to achieve good results for estimating particular psychometric functions. For example, proximity and uncertainty components for \ac{NV2D}, the combination of gradient, proximity, and uncertainty for \ac{SIN2D}, and lookahead component for \ac{HART6} were critical for an efficient estimation of the psychometric function. In order to achieve the best scores overall for all tested psychometric functions and metrics, we observe that using all four components of the acquisition function is crucial.

\section{Discussion and limitations} \label{sec:discussion}

\new{\ac{NEST} is a flexible method for experiments involving stimuli of varying complexity when the psychometric function's form is unknown or when using parametric methods like QUEST+ becomes impractical due to excessive parameter requirements. Researchers can utilize \ac{NEST} in several additional ways. First, they can employ the samples generated during the experiment as training data for fitting other models, essentially treating \ac{NEST} as an efficient sampling tool. Alternatively, the psychometric function provided by \ac{NEST} can serve directly as a model. Using this approach, we showed that we could efficiently reproduce an existing theoretical model in Section \ref{sec:Barten_CSF_results}, and we could highlight deviations from the existing model in Section \ref{sec:Tursun_Didyk_results} that could inspire further research on the refinement of the theoretical model. Lastly, the input-output pairs generated by \ac{NEST} can be used for parameter regression in complex models with many parameters and input dimensions, which would be computationally prohibitive using parametric methods like QUEST+ or time-wise infeasible due to the excessive number of trials required.}

\new{One limitation of using neural networks instead of Gaussian processes relates to their interpretability. Both models give a mapping from the input to the probability of detection. A Gaussian process simultaneously models its uncertainty in the predicted probability of detection by means of a variance output, whereas a neural network requires a method like Monte Carlo dropout as an approximation of the uncertainty. However, \citet{VerdojaKyrki2021} showed that it requires careful consideration of dropout rate $p_d$ for calibrating with respect to the epistemic uncertainty and it requires scaling of the computed uncertainty based on the value of the network output. Our method produced good results for the Brier score metric, which penalizes overconfidence in mistakes and underconfidence in correct predictions. This suggests that our model can model uncertainty somewhat accurately. However, future work could improve the interpretability of our method by incorporating more accurate neural network uncertainty measures.}

Our benchmarks in Section~\ref{sec:data_NEST} assume that the input space is continuous, but in practice we often use discrete sampling to improve running time. We observed that coarse discrete sampling of the stimulus space for sample selection has minimal impact on the performance (see the supplementary material for details).

Being a non-parametric model, our psychometric function estimate does not in general allow for extrapolation outside the stimulus space boundaries used in experiments. In the experiment shown in Figure~\ref{fig:CSF_function_results}, our display was capable of rendering spatial frequencies up to the Nyquist limit $9.07$ \ac{cpd}. Due to our regularization scheme, we observed that the learned psychometric function outputs changed smoothly for spatial frequencies higher than the limit imposed by the display hardware. As a future work, we believe that the model generalization may be improved by optimizing the regularization parameters.

\new{We have shown that our method is able to efficiently sample a diverse set of synthetic and real-life psychometric functions. The acquisition function, which consists of a combination of four heuristic terms based on existing active learning and psychophysical literature, worked well for selecting samples. However, future research into novel acquisition approaches has the potential to further improve the accuracy of our method. Since our acquisition function consists of heuristic approaches, future research could additionally look into data-driven methods for selecting samples. For example, one approach could be to train a sample selection scheme on simulated but representative data such that the algorithm learns to sample effectively for an unknown psychometric function based on the state of the trained network and the previously selected samples. Although challenges such as the selection of an appropriate training dataset will need to be solved, such a data-driven approach has the potential to improve the acquisition function used in this study.}

\section{Conclusion} \label{sec:conclusion}
In this study, we introduced a neural network-based psychophysical procedure called \acf{NEST}. The aim of using a neural network was to: (1) expand existing psychophysical procedures for high-dimensional problems; (2) free the experimenter from the requirement of choosing a parametric form in the fitting process required by the QUEST+ procedure, or a kernel function as in the \ac{GP} methods; and (3) improve the state-of-the-art performance for high-dimensional psychometric function estimation with a Bernoulli response variable. We showed that \ac{NEST} is able to outperform QUEST+ in high-dimensional problems, and that it performs better overall on psychometric function estimation than the state-of-the-art \ac{GP} methods. Subsequently, we demonstrated that our network is practically applicable for high-dimensional psychovisual experiments: firstly, we established that we could recover the \ac{CSF} function as a function of spatial frequency and of eccentricity within a reasonable experimentation time. Secondly, we showed that for a selected high-dimensional psychovisual experiment, using \ac{NEST} significantly reduced the number of trials and the amount of time needed to perform the experiment, while achieving better fits. These results show that \ac{NEST} offers significant benefits for conducting multi-dimensional psychovisual experiments.
\\ \\
\textbf{CRediT authorship contribution statement} \\ \\
\textbf{Sjoerd Bruin}: methodology, investigation, formal analysis, data curation, visualization, validation, software, writing - original draft. \textbf{Ji\v{r}\'{i} Kosinka}: conceptualization, supervision, writing - review \& editing. \textbf{Cara Tursun}: conceptualization, supervision, writing - review \& editing
\\ \\
\textbf{Declaration of competing interest} \\
The authors declare that there are no competing financial interests or personal relationships that could have influenced the work in this paper.
\\ \\
\textbf{Data availability} \\
\new{The live experiment data (in anonymized form) and simulation results will be made available on request. We have published a Github implementation of the \ac{NEST} algorithm for public use at: \href{https://github.com/Bruin96/nest-1/tree/v0.9.1}{https://github.com/Bruin96/nest-1/tree/v0.9.1}.}


\appendix

    \acrodef{2AFC}{\textit{2-alternative forced choice}}
    \acrodef{AR}{\textit{augmented reality}}
    \acrodef{AUC}{\textit{area under curve}}
    \acrodef{BALD}{\textit{Bayesian active learning by disagreement}}
    \acrodef{BALV}{\textit{Bayesian active learning by variance}}
    \acrodef{BCE}{\textit{binary cross-entropy}}
    \acrodef{CDF}{\textit{cumulative distribution function}}
    \acrodef{CMF}{\textit{cortical magnification factor}}
    \acrodef{cpd}{cycles per degree}
    \acrodef{CSF}{\textit{contrast sensitivity function}}
    \acrodef{CSG}{\textit{cognitive search guidance}}
    \acrodef{DCT}{\textit{discrete cosine transform}}
    \acrodef{DN2D}{\textit{2D donut}}
    \acrodef{DOS}{\textit{dichotomous optimistic search}}
    \acrodef{EAVC}{\textit{expected absolute volume change}}
    \acrodef{FIM}{\textit{Fisher information matrix}}
    \acrodef{GlobalMI}{\textit{global mutual information}}
    \acrodef{GP}{\textit{Gaussian process}}
    \acrodef{HART6}{\textit{6D Hartmann}}
    \acrodef{HOO}{\textit{hierarchical optimistic optimization}}
    \acrodef{HSD}{\textit{honestly significant difference}}
    \acrodef{HVS}{\textit{human visual system}}
    \acrodef{JND}{\textit{just noticeable difference}}
    \acrodef{KDE}{\textit{kernel density estimation}}
    \acrodef{LBFGS}{\textit{limited-memory Broyden-Fletcher-Goldfarb-Shanno}}
    \acrodef{LEF}{\textit{luminous efficiency function}}
    \acrodef{LGN}{\textit{lateral geniculate nucleus}}
    \acrodef{LSE}{\textit{level set estimation}}
    \acrodef{MAP}{\textit{maximum a posteriori}}
    \acrodef{MAX2D}{\textit{2D max}}
    \acrodef{MLE}{\textit{maximum likelihood estimation}}
    \acrodef{MLESAC}{\textit{maximum likelihood estimation sample consensus}}
    \acrodef{MLP}{\textit{multilayer perceptron}}
    \acrodef{MR}{\textit{mixed reality}}
    \acrodef{MSE}{\textit{mean-squared error}}
    \acrodef{$n$AFC}{$n$-\textit{alternative forced choice}}
    \acrodef{NEST}{\textit{Neural Estimation by Sequential Testing}}
    \acrodef{NTK}{\textit{neural tangent kernel}}
    \acrodef{NV2D}{\textit{2D novel test}}
    \acrodef{OLED}{\textit{organic LED}}
    \acrodef{ON2D}{\textit{2D older normal}}
    \acrodef{OVP}{\textit{optimal viewing position}}
    \acrodef{PEST}{\textit{Parameter Estimation by Sequential Testing}}
    \acrodef{PLP}{\textit{preferred landing position}}
    \acrodef{POO}{\textit{parallel optimistic optimization}}
    \acrodef{PS8D}{\textit{8D psychometric function}}
    \acrodef{PSE}{\textit{point of subjective equality}}
    \acrodef{PSG}{\textit{peripheral search guidance}}
    \acrodef{RANSAC}{\textit{random sample consensus}}
    \acrodef{RBF}{\textit{radial basis function}}
    \acrodef{ReLU}{\textit{rectified linear unit}}

    \acrodef{RMSE}{\textit{root mean-squared error}}
    \acrodef{RSVP}{\textit{rapid serial visual presentation}}
    \acrodef{sCSF}{\textit{spatial contrast sensitivity function}}
    \acrodef{SIN2D}{\textit{2D sinusoid}}
    \acrodef{SNR}{\textit{signal-to-noise ratio}}
    \acrodef{stCSF}{\textit{spatio-temporal contrast sensitivity function}}
    \acrodef{svCSF}{\textit{spatio-velocity contrast sensitivity function}}
    \acrodef{tCSF}{\textit{temporal contrast sensitivity function}}
    \acrodef{VDP}{\textit{visual difference predictor}}
    \acrodef{vPEST}{\textit{virulent PEST}}
    \acrodef{VR}{\textit{virtual reality}}
    \acrodef{V1}{\textit{primary visual cortex}}
    \acrodef{V2}{\textit{secondary visual cortex}}
    \acrodef{V3}{\textit{tertiary visual cortex}}
    \acrodef{V4}{\textit{quaternary visual cortex}}
    \acrodef{WEI1D}{\textit{1D Weibull}}
    \acrodef{WEI2D}{\textit{2D Weibull}}
    \acrodef{WEI3D}{\textit{3D Weibull}}
    \acrodef{WEI4D}{\textit{4D Weibull}}
    \acrodef{WPM}{words per minute}

\bibliographystyle{elsarticle-harv}
\bibliography{references}






\end{document}


\begin{frontmatter}
\title{Supplementary Material for\\
NEST: Neural Estimation by Sequential Testing}

\author[inst1]{Sjoerd Bruin\corref{cor1}}
\cortext[cor1]{Corresponding author. Email address: s.bruin@rug.nl}

\affiliation[inst1]{organization={Bernoulli Institute, University of Groningen},
            addressline={Nijenborgh 9},
            city={Groningen},
            postcode={9747AG},
            country={The Netherlands}}

\author[inst1]{Ji\v{r}\'{i} Kosinka}
\author[inst1]{Cara Tursun}
\end{frontmatter}

\renewcommand*{\thesection}{\arabic{section} in Suppl.\ Mat.}
\renewcommand*{\thefigure}{\arabic{figure} in Suppl.\ Mat.}
\renewcommand*{\thetable}{\arabic{table} in Suppl.\ Mat.}

\titleformat{\section}{\normalsize\bfseries\boldmath}{~\arabic{section}.}{1em}{}
\titleformat{\subsection}{\normalfont\normalsize\itshape}{~\arabic{section}.~\arabic{subsection}.}{1em}{}
\titleformat{\figure}{\normalfont\normalsize}{~\arabic{figure}.}{1em}{}
\titleformat{\table}{\normalfont\normalsize}{~\arabic{table}.}{1em}{}
\titleformat{\equation}{\normalfont\normalsize}{~\arabic{equation}.}{1em}{}

\makeatletter
\renewcommand{\fnum@figure}{Figure \arabic{figure}}
\makeatother

\makeatletter
\renewcommand{\fnum@table}{Table \arabic{table}}
\makeatother

\acused{WEI4D}
\acused{CDF}
\acused{RMSE}
\acused{NEST}
\acused{AUC}
\acused{NV2D}
\acused{SIN2D}
\acused{DN2D}
\acused{ON2D}
\acused{MAX2D}
\acused{PS8D}
\acused{GlobalMI}
\acused{EAVC}

\section{Definition of Hartmann6 psychometric function} \label{sec:hartmann6}
We use a similar notation as \cite{LethamEtAl2022}. The \ac{HART6} function is given by
\begin{equation}
    \label{eq:Hartmann6}
    h(\vec{x}) = 1 - \sum\limits_{i=1}^{4} \alpha_{i} e^{-\sum\limits_{j=1}^{6}A_{ij} (x_{j} - P_{ij})^{2}},
\end{equation}
where $\vec{\alpha} = (2.0, 2.2, 2.8, 3.0)$,
\begin{equation*}
    A = \begin{pmatrix}
            8 & 3 & 10 & 3.5 & 1.7 & 6 \\
            0.5 & 8 & 10 & 1.0 & 6 & 9 \\
            3 & 3.5 & 1.7 & 8 & 10 & 6 \\
            10 & 6 & 0.5 & 8 & 1.0 & 9
        \end{pmatrix}, \text{ and }
\end{equation*}
\begin{equation*}
        P = 10^{-4}
        \begin{pmatrix}
            1312 & 1696 & 5569 & 124 & 8283 & 5886 \\
            2329 & 4135 & 8307 & 3736 & 1004 & 9991 \\
            2348 & 1451 & 3522 & 2883 & 3047 & 6650 \\
            4047 & 8828 & 8732 & 5743 & 1091 & 381
        \end{pmatrix}.
\end{equation*}
\ac{HART6} is defined on the range $\vec{x} \in [0, 1]^{6}$. We follow \cite{LethamEtAl2022} in rescaling \ac{HART6} as
\begin{equation}
    \label{eq:rescaled_hartmann6}
    f(\vec{x}) = 3 h(\vec{x}) - 2
\end{equation}
in order to make the comparison with \cite{LethamEtAl2022} as close as possible. We convert the function $f(\vec{x})$ into a probability of detection by passing it into the standard Gaussian \ac{CDF} $\Phi(\cdot)$ and scaling by the lower asymptote and lapse as
\begin{equation}
    \label{eq:hartmann6_probability}
    \Psi(\vec{x}, \alpha, \gamma) = \alpha + (1-\alpha-\gamma) \Phi(f(\vec{x})).
\end{equation}

\section{Sample Selection} \label{sec:sample_selection}
\begin{figure*}[ht!]
    \centering
    \includegraphics[width=\textwidth]{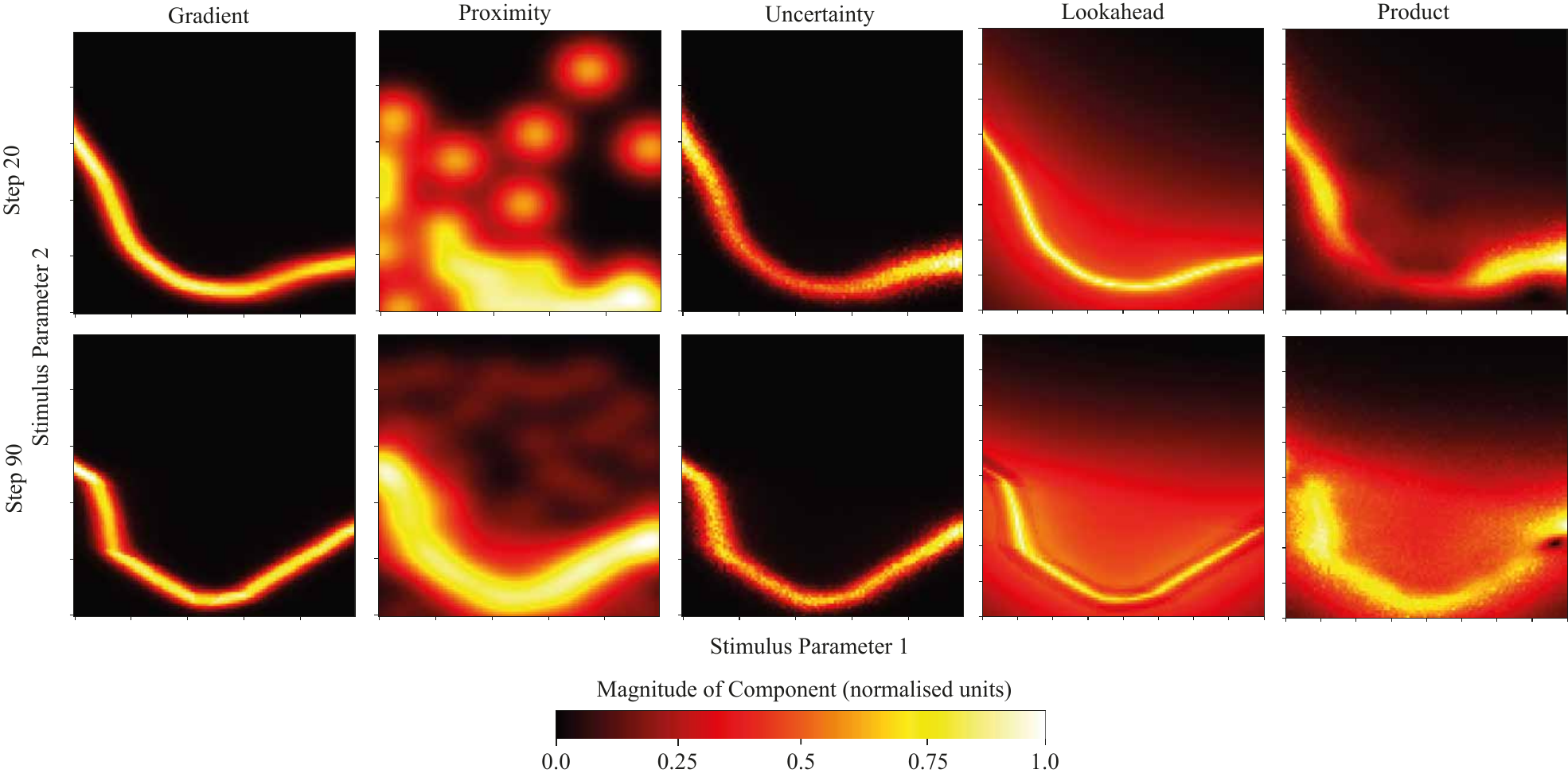}
    \caption{Visualisation of the components of the acquisition function for the \ac{NV2D} function. At 20 samples, the algorithm encourages exploration around the estimated boundary, but at 90 samples, more emphasis is given to the space at the edges of the transition region.}
    \label{fig:sample_selection_components}
\end{figure*}

Figure \ref{fig:sample_selection_components} shows the gradient, proximity, uncertainty, and lookahead components of the query generation function discussed in Section \ref{sec:acquisition_NEST}, as well as the combined product term. The gradient term visualised in Figure \ref{fig:sample_selection_components} reaches a maximum value at the steepest point along the transition region, which is the expected behaviour. The uncertainty term has a larger value near the edges of the input space, which is expected since there is more uncertainty at those places due to the unknown distribution outside the input space. The lookahead term finds its highest points along the transition region, which makes sense since the predicted minimum impact of a label of either sign is largest there. Additionally, the lookahead component picks out locations along the boundary that have not been explored as much as other regions.

At step 20, the proximity term has a somewhat random distribution. At step 90, on the other hand, it is clear that the samples have been selected along the predicted transition boundary, and as a consequence, the proximity term is large along this boundary. For the final product, this means that there is a strong discouraging effect from sampling many more samples along the boundary, which is counterbalanced by the high gradient and uncertainty terms. As a result, the final product term at step 20 is mostly focused on exploring the transition boundary, whereas at step 90 it is more likely to select points away from the transition boundary while retaining a propensity to select samples towards the middle of the transition region.

\section{Convergence Criterion} \label{sec:convergence_criterion_results}
For analyzing convergence, we computed the Fisher energy progression for 100 runs for 2D-6D Weibull functions and also for the sphere function, which defines a radius $r$ against which the size of the input vector $\vec{x}$ is tested as a measure of the threshold. The definition is given by
\begin{equation}
    \label{eq:sphere_function}
    \Psi(\vec{x}, \beta, r, \alpha, \gamma) = \alpha + (1 - \gamma - \alpha) e^{-10^{\beta \frac{||\vec{x}||_{2} - r}{20}}},
\end{equation}
where $||\vec{x}||_{2}$ indicates the Euclidean distance of the stimulus from the center of the input space. A single 1-dimensional slope parameter $\beta$ is used in this function. We used a radius equal to $\frac{1}{4}$ of the input space dimensions. Additionally, we created a baseline for the Fisher energy using a multi-dimensional psychometric function that gave random responses all the time. Rather than looking at the absolute size of the Fisher energy, we study how much the Fisher energy changes from one trial to the next. We monitor a window of 15 difference values, and study how this window converges. Our aim is to suggest a useful heuristic for terminating an experiment.

Table \ref{tab:Spearman_Fisher_energy} shows the Spearman rank correlation between the \ac{RMSE} errors in the validation set and the Fisher energy computed on the training set. We use the Spearman rank correlation because the relationship has a logarithmic character, and therefore the Pearson correlation would miss the relationship between the variables. The correlation shows that there is a strong association between the Fisher energy and the generalization error, which makes it a useful heuristic for suggesting termination.

\begin{figure*}[t!]
    \centering
    \includegraphics[width=\textwidth]{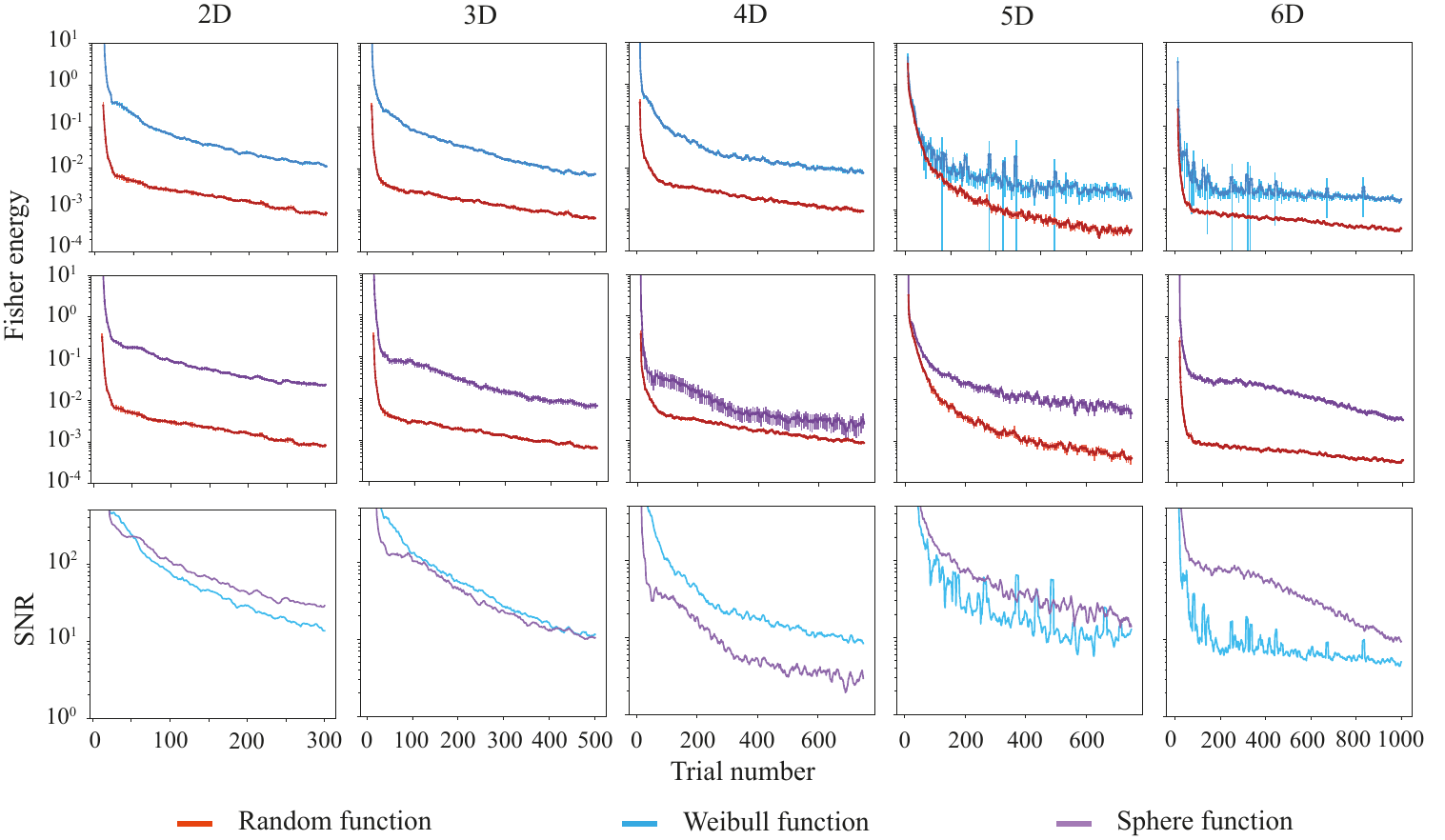}
    \caption{15-trial windowed Fisher energy and 95\% confidence interval of a psychometric function with random responses, the Weibull psychometric function, and the sphere psychometric function. The mean SNR value versus the random function trends down towards a similar value for both psychometric functions, which supports the use of the SNR value as a useful heuristic for assessing convergence.}
    \label{fig:Fisher_convergence}
\end{figure*}

In order to suggest quantitative termination criteria, we looked at the correspondence between the Fisher energy of a simulation with completely random responses and a simulation of a well-defined psychometric function using the Sphere and Weibull psychometric functions. The results are shown in Figure \ref{fig:Fisher_convergence}

\begin{table}[b!]
    \centering
    \caption{The Spearman rank correlation between the \ac{RMSE} on the validation set and the Fisher energy for the training set. The correlation is very strong, indicating that the Fisher energy is a good predictor of the achieved generalization error.}
    \label{tab:Spearman_Fisher_energy}
    \begin{tabular}{c|c|c}
        \textbf{Dimension} & \textbf{Weibull} & \textbf{Sphere}  \\ \hline
        2D & 0.9843 & 0.9790 \\
        3D & 0.9902 & 0.9848 \\
        4D & 0.9705 & 0.8431 \\
        5D & 0.7931 & 0.9107 \\
        6D & 0.8874 & 0.9717
    \end{tabular}
\end{table}

Several observations arise from Figure \ref{fig:Fisher_convergence} Firstly, the Fisher energy decreases as the number of dimensions increases. This suggests that we need to recommend a convergence criterion based on the dimensionality of the input space. Figure \ref{fig:Fisher_convergence} further shows that there is a converging ratio between the Fisher energy of the random function and those of non-random functions. The ratio, which we call the \ac{SNR}, appears to trend towards a value of 10 in most cases. This value makes for a suitable cut-off for our windowed Fisher energy value. We have used this value to assess convergence during the live experiments discussed in Sections \ref{sec:Barten_CSF_results} and \ref{sec:Tursun_Didyk_results}. We have found that the results of the experiment did not change noticeably when one of the authors continued the experiment after Fisher energy convergence had been reached. This shows that the criterion reflects a suitable level of convergence for \ac{NEST}. If a higher confidence in the final result is needed, then a lower \ac{SNR} value can be chosen. The approximate baseline Fisher energy difference levels from the random simulation that we have found are given in Table \ref{tab:baseline_Fisher_energy_levels} We found these values to be useful guides for the termination condition in the psychovisual experiments that we have performed.

\begin{table}[b!]
    \centering
    \caption{Baseline windowed Fisher energy difference levels determined for the random psychometric function.}
    \begin{tabular}{c|c}
        \textbf{Dimension} & \textbf{Windowed Fisher energy difference} \\
        \hline
        2D & $9\cdot 10^{-4}$\\
        3D & $7 \cdot 10^{-4}$\\
        4D & $6 \cdot 10^{-4}$ \\
        5D & $5 \cdot 10^{-4}$ \\
        6D & $4 \cdot 10^{-4}$
    \end{tabular}
    \label{tab:baseline_Fisher_energy_levels}
\end{table}

\begin{table*}[hb]
    \centering
    \caption{The mean \ac{AUC} $\pm$ standard error for the discrete sampling experiments using the \ac{RMSE} and Brier metrics for the \ac{NV2D}, \acs{DN2D}, \acs{SIN2D}, \acs{WEI4D}, and \ac{HART6} functions. The best method and methods that are statistically similar ($p \geq 0.05$, Games-Howell test) to it are shown in bold typeface for each metric. Using 32 samples is typically enough to get close to the performance of the continuous form of the acquisition function.}
    \label{tab:coarseness_AUC_results}
    \setlength{\tabcolsep}{0.25em}
    \ssmall
    \begin{tabular}{l|c|c|c|c|c|c|c|c|c|c}
     \textbf{Method} & \multicolumn{2}{c|}{\textbf{NV2D}} & \multicolumn{2}{c|}{\textbf{DN2D}} & \multicolumn{2}{c|}{\textbf{SIN2D}} & \multicolumn{2}{c|}{\textbf{WEI4D}} & \multicolumn{2}{c}{\textbf{HART6}}\\
         & \textbf{AUC$_{\mathrm{RMSE}}$} & \textbf{AUC$_{\mathrm{Brier}}$} & \textbf{AUC$_{\mathrm{RMSE}}$} & \textbf{AUC$_{\mathrm{Brier}}$} & \textbf{AUC$_{\mathrm{RMSE}}$} & \textbf{AUC$_{\mathrm{Brier}}$} & \textbf{AUC$_{\mathrm{RMSE}}$} & \textbf{AUC$_{\mathrm{Brier}}$} & \textbf{AUC$_{\mathrm{RMSE}}$} & \textbf{AUC$_{\mathrm{Brier}}$}\\
        \hline
        Continuous & $\mathbf{21.83 \pm 0.23}$ & $\mathbf{7.56 \pm 0.11}$ & $\mathbf{24.05 \pm 1.01}$ & $\mathbf{6.25 \pm 0.21}$ & $\mathbf{67.21 \pm 0.32}$ & $\mathbf{25.25 \pm 0.20}$ & $\mathbf{64.89 \pm 1.13}$ & $\mathbf{13.62 \pm 0.35}$ & $\mathbf{277.51 \pm 1.87}$ & $\mathbf{148.15 \pm 2.81}$\\
        4 samples & $84.56 \pm 1.69$ & $42.88 \pm 0.99$ & $71.94 \pm 0.64$ & $26.87 \pm 0.45$ & $141.14 \pm 0.51$ & $75.79 \pm 0.54$ & $196.31 \pm 3.81$ & $53.86 \pm 1.50$ & $320.20 \pm 2.54$ & $232.36 \pm 4.08$ \\
        8 samples & $36.43 \pm 0.68$ & $12.50 \pm 0.33$ & $52.87 \pm 1.10$ & $16.07 \pm 0.42$ & $103.92 \pm 0.82$ & $51.39 \pm 0.68$ & $107.62 \pm 1.88$ & $25.20 \pm 0.68$ & $311.34 \pm 2.89$ & $215.79 \pm 3.78$ \\
        16 samples & $27.50 \pm 0.34$ & $8.61 \pm 0.19$ & $32.29 \pm 0.97$ & $8.57 \pm 0.21$ & $79.24 \pm 1.16$ & $32.70 \pm 0.75$ & $77.03 \pm 1.24$ & $16.42 \pm 0.40$ & $300.69 \pm 2.00$ & $192.55 \pm 3.04$ \\
        32 samples & $22.80 \pm 0.24$ & $\mathbf{7.60 \pm 0.12}$ & $\mathbf{23.61 \pm 0.65}$ & $\mathbf{6.42 \pm 0.15}$ & $68.74 \pm 0.49$ & $\mathbf{25.09 \pm 0.29}$ & $\mathbf{69.31 \pm 1.10}$ & $\mathbf{14.80 \pm 0.36}$ & $288.59 \pm 2.09$ & $172.27 \pm 2.44$ \\
        64 samples  & $\mathbf{21.80 \pm 0.24}$ & $\mathbf{7.46 \pm 0.11}$ & $\mathbf{24.32 \pm 0.95}$ & $\mathbf{6.36 \pm 0.18}$ & $\mathbf{67.23 \pm 0.37}$ & $\mathbf{24.77 \pm 0.22}$ & $\mathbf{67.38 \pm 1.06}$ & $\mathbf{14.41 \pm 0.34}$ & $\mathbf{281.63 \pm 1.90}$ & $\mathbf{160.00 \pm 2.21}$ \\
        128 samples & $\mathbf{21.88 \pm 0.24}$ & $\mathbf{7.59 \pm 0.12}$ & $\mathbf{23.85 \pm 0.84}$ & $\mathbf{6.46 \pm 0.18}$ & $\mathbf{67.68 \pm 0.30}$ & $\mathbf{25.19 \pm 0.20}$ & $\mathbf{64.82 \pm 1.17}$ & $\mathbf{13.61 \pm 0.37}$ & $\mathbf{280.94 \pm 1.77}$ & $\mathbf{156.48 \pm 2.15}$ \\
        256 samples & $\mathbf{21.83 \pm 0.26}$ & $\mathbf{7.55 \pm 0.14}$ & $\mathbf{22.27 \pm 0.74}$ & $\mathbf{6.09 \pm 0.18}$ & $\mathbf{67.53 \pm 0.32}$ & $\mathbf{25.19 \pm 0.21}$ & $\mathbf{65.53 \pm 1.07}$ & $\mathbf{13.84 \pm 0.35}$ & $\mathbf{276.74 \pm 1.90}$ & $\mathbf{151.73 \pm 2.51}$ \\
        512 samples & $\mathbf{22.26 \pm 0.25}$ & $\mathbf{7.81 \pm 0.13}$ & $\mathbf{22.63 \pm 0.74}$ & $\mathbf{6.05 \pm 0.15}$ & $\mathbf{66.34 \pm 0.31}$ & $\mathbf{24.48 \pm 0.20}$ & $\mathbf{65.12 \pm 1.10}$ & $\mathbf{13.58 \pm 0.35}$ & $\mathbf{281.25 \pm 1.80}$ & $\mathbf{158.05 \pm 2.73}$

    \end{tabular}
\end{table*}

\section{The shrink-and-perturb operation}
As mentioned in the main paper, we use the \textit{shrink-and-perturb} trick for warm-starting, introduced by \citet{AshAdams2020}, in order to avoid overfitting and to increase the stability between trials. This method shrinks the size of each weight parameter and adds a small amount of random noise at the beginning of the training by
\begin{equation}
    \label{eq:shrink_perturb}
    W_{t}(i, j) \leftarrow \lambda W_{t}(i, j) + p,
\end{equation}
where $\lambda$ is the shrink parameter, $p \sim \mathcal{N}(0, \epsilon_{\mathrm{per}}^{2})$, and $\epsilon_{\mathrm{per}}$ is the standard deviation of the noise used to perturb the weights. In our experiments, we set $\lambda = 0.9$ and $\epsilon_{per} = 0.01$.

\section{Discrete Sample Robustness} \label{sec:coarseness}

The NEST algorithm assumes that we can select samples for evaluation from a continuous space, but this assumption does not always hold, for example if the stimuli in an experiment can only be generated at fixed step sizes. We study the impact of discrete sample selection on the accuracy of the method in this section. After finding an optimal sample location in the continuous sample space, we snap the selected location to the nearest discrete sample and evaluate that location. We set $n_{\mathrm{sample}} \in \{4, 8, 16, 32, 64, 128, 256, 512\}$, where $n_{\mathrm{sample}}$ represents the number of sample locations in each dimension.

\begin{figure*}[t!]
    \centering
    \includegraphics[width=0.9\textwidth]{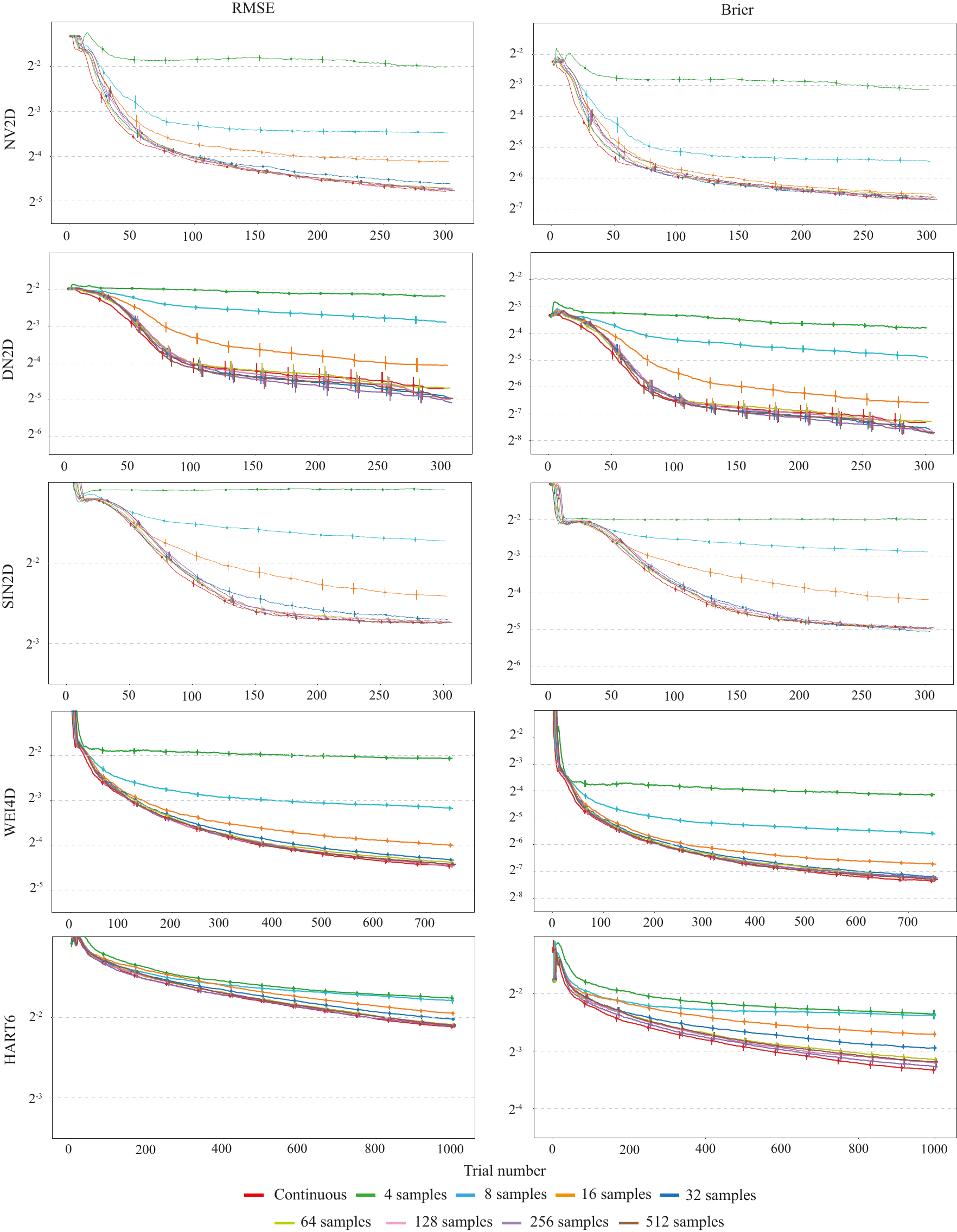}
    \caption{NEST \ac{RMSE} errors for \ac{NV2D}, \acs{DN2D}, \acs{SIN2D}, \acs{WEI4D}, and \ac{HART6} with different numbers of discrete samples. Using between 4 and 16 samples per dimension is not enough to get a performance that is comparable to the continuous version, but using 32 or more samples typically is sufficient.}
    \label{fig:coarse_sample_accuracy}
\end{figure*}

Table \ref{tab:coarseness_AUC_results}\ and Figure \ref{fig:coarse_sample_accuracy} show the errors for \ac{NV2D}, \ac{DN2D}, \ac{SIN2D}, \ac{WEI4D}, and \ac{HART6}. Using only 4, 8, or 16 samples per dimension leads to a significantly decreased performance. Starting at 32 samples, however, the results become very close to the continuous sample selection case. This shows that our method is robust against discretization of trial samples.

\section{Other Results} \label{sec:full_error_metric_results}
\begin{figure*}[ht!]
    \centering
    \includegraphics[width=\textwidth]{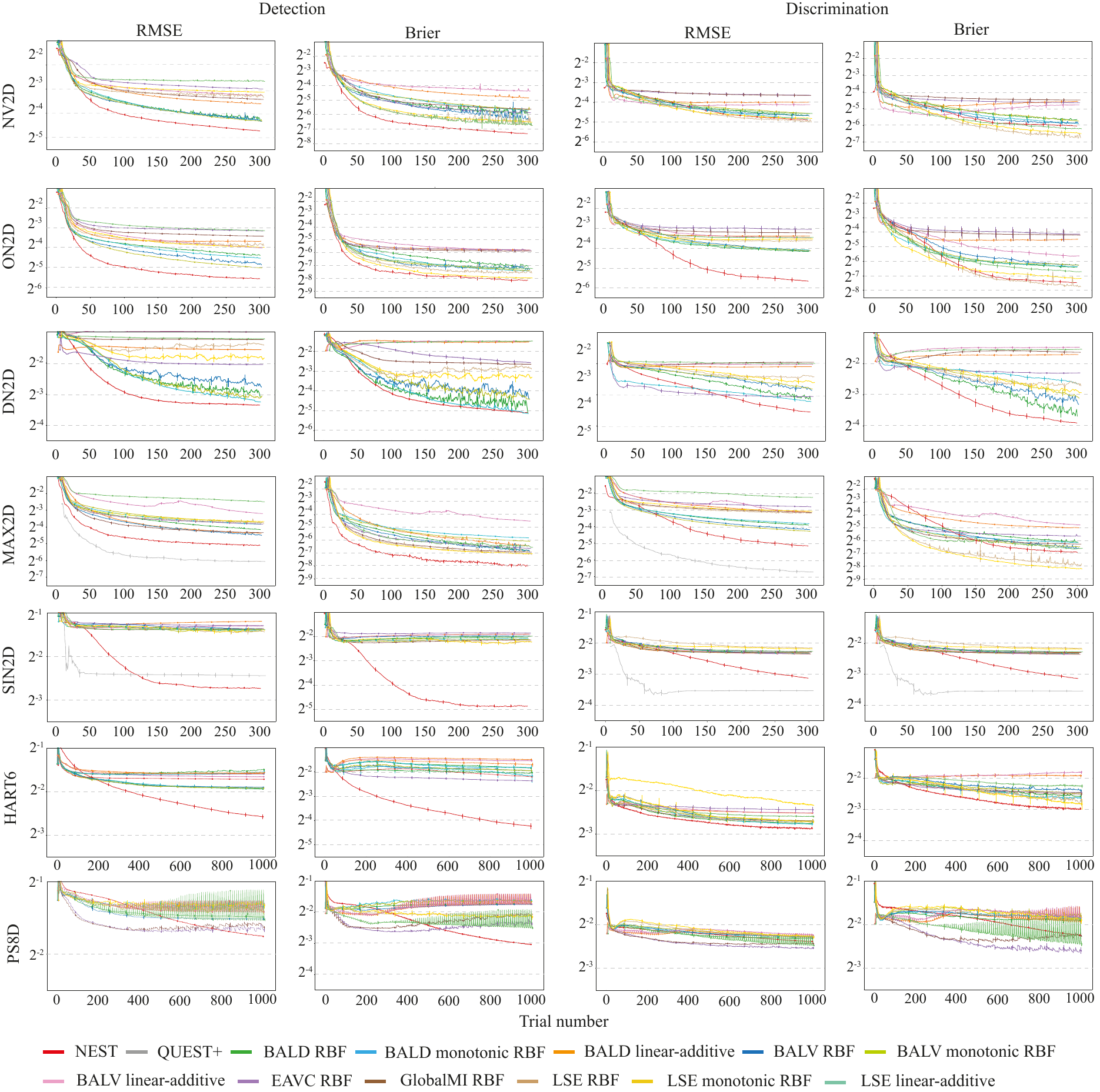}
    \caption{Simulation results for all test functions except the Weibull functions. \ac{NEST} performs equally well or better in the most test cases. As mentioned in the main paper, \ac{GlobalMI} and \ac{EAVC} methods have a fast decrease in the error metrics at the beginning of the experiment for the \ac{PS8D} function but eventually \ac{NEST} achieves lower error scores. This is observable in the last row of the plots.}
    \label{fig:full_error_metric_results}
\end{figure*}

\begin{figure*}[ht!]
    \centering
    \includegraphics[width=\textwidth]{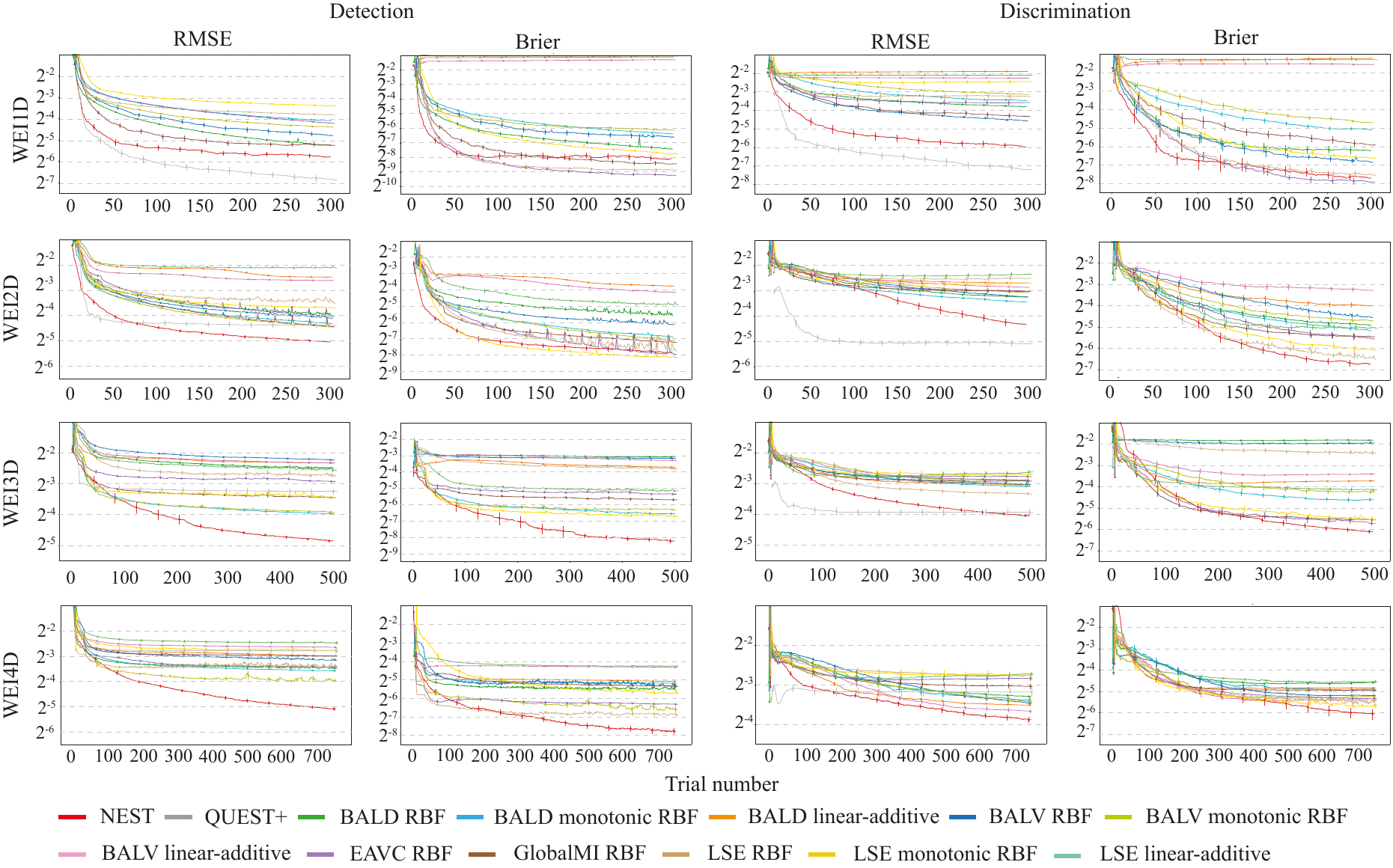}
    \caption{Simulation results for the Weibull test functions. \ac{NEST} performs best in most test cases.}
    \label{fig:weibull_error_metric_results}
\end{figure*}

\begin{figure*}[t!]
    \centering
    \includegraphics[width=0.9\textwidth]{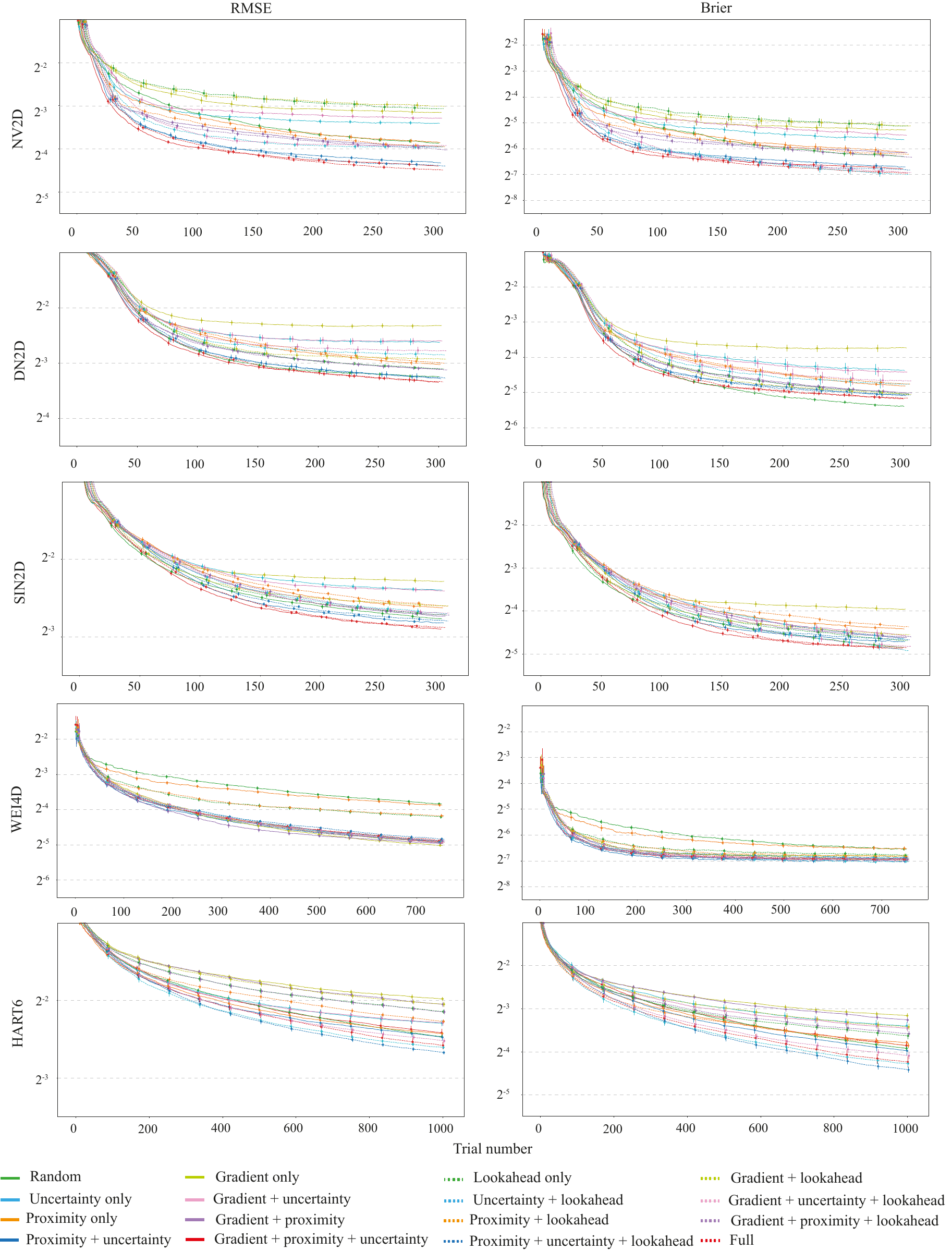}
    \caption{NEST ablation results for \ac{NV2D}, \ac{DN2D}, \ac{SIN2D}, \ac{WEI4D}, and \ac{HART6}. We see that the different components are relevant for different functions and that the combined acquisition function always performs best or similarly to the other best options. The lookahead component has the most significant impact in the estimation of higher-dimensional functions such as the \ac{HART6} function.}
    \label{fig:ablation_results}
\end{figure*}

\begin{figure*}[t!]
    \centering
    \includegraphics[width=\textwidth]{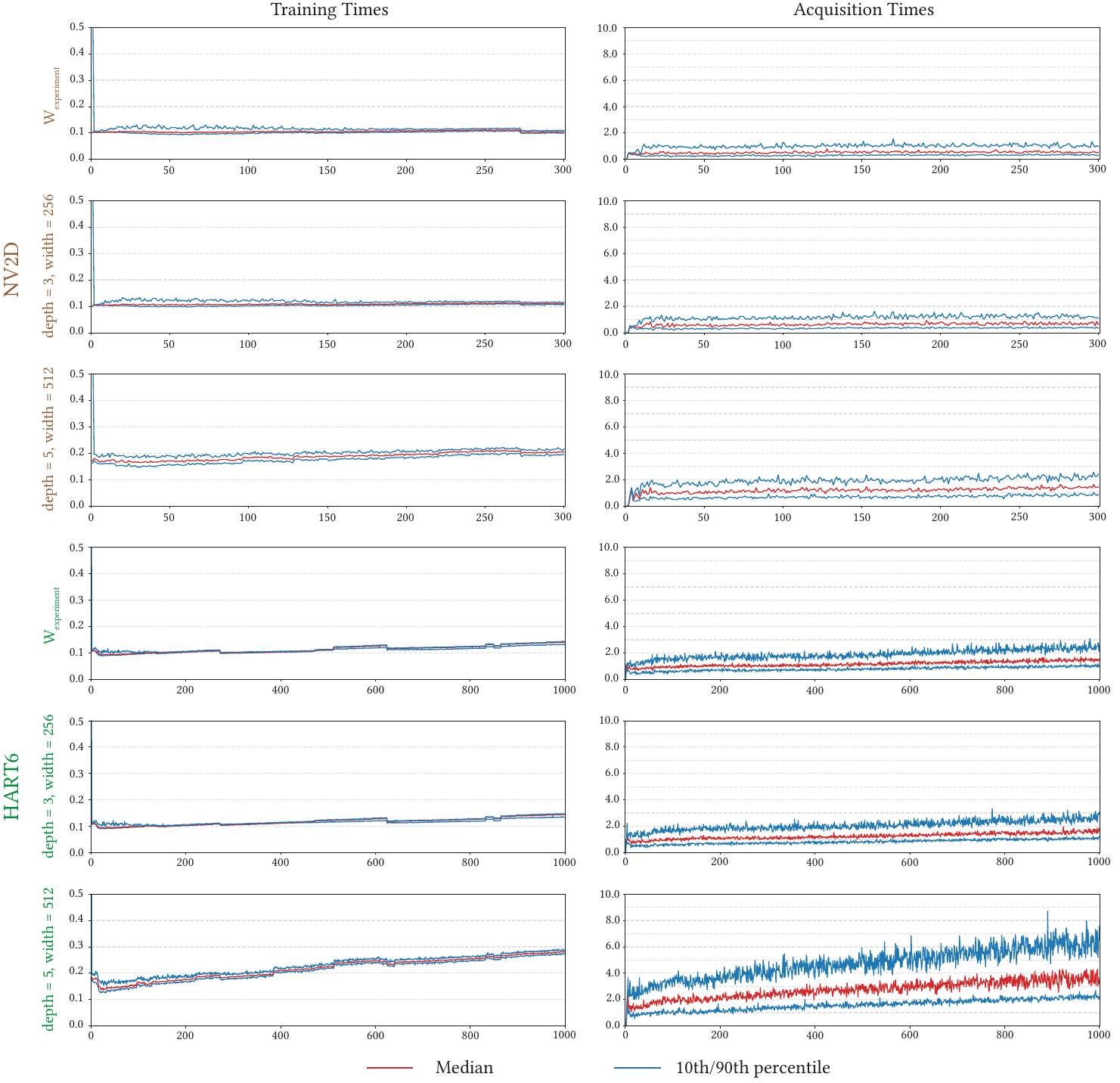}
    \caption{Training and acquisition times in seconds of the \ac{NV2D} and \ac{HART6} test functions for different architectures. The training and acquisition times are bounded for most test cases except for the large network size on the \ac{HART6} test function.}
    \label{fig:timing_results}
\end{figure*}

In the rest of the Supplementary Materials, we provide a number of additional results to support the conclusions in the main paper. Table \ref{tab:full_2D_AUC_results} and Table \ref{tab:weibull_AUC_results} show the \ac{AUC} values for the \ac{RMSE} and Brier score for all tested functions. The plotted error curves accompanying these tables are shown in Figure \ref{fig:full_error_metric_results} and Figure \ref{fig:weibull_error_metric_results} We evaluate the experiments on both \textit{detection} ($\alpha = 0.0$) and \ac{2AFC} or \textit{discrimination} ($\alpha = 0.5$) versions of the synthetic psychometric functions, in order to show that \ac{NEST} can also handle problems where the lower asymptote is different from 0. Figure \ref{fig:ablation_results} shows the graphed ablation results. Figure \ref{fig:timing_results} shows the graphed timing results for the \ac{NV2D} and \ac{HART6} test functions for different architectures.

\begin{figure*}[t!]
    \centering
    \includegraphics[height=0.9\textheight]{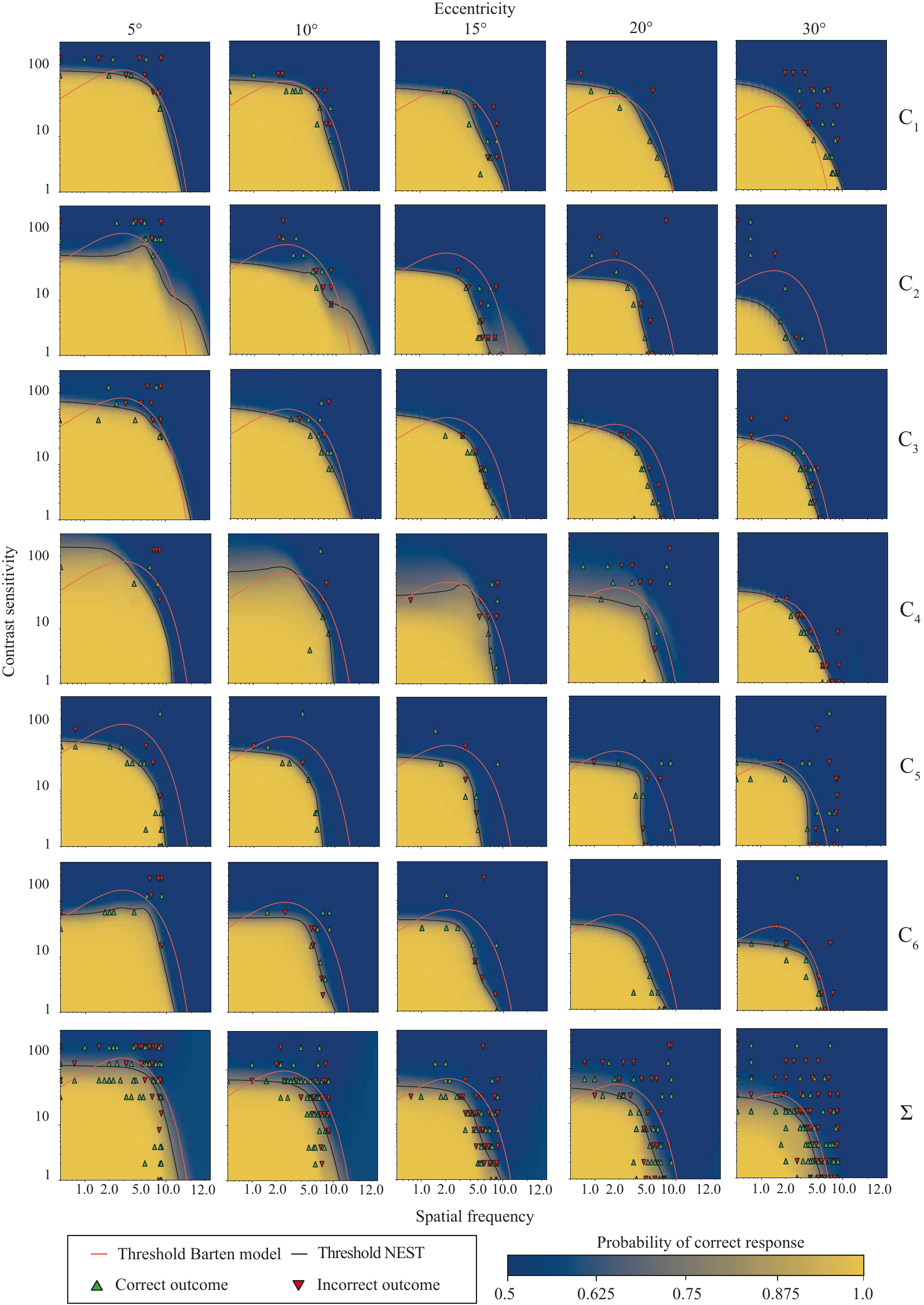}
    \caption{Results of the CSF experiment for multiple eccentricities. Each row shows the results for a different participant (labelled C$_1$ -- C$_6$), and the bottom row (labelled $\Sigma$) shows the results when training a network on the combined data of the participants. The individual trial outcomes are shown as upward-pointing (correct) or downward-pointing (incorrect) triangles, and the detection thresholds of the theoretical model and the \ac{NEST} method are shown as red and black lines, respectively. \ac{NEST} learns a function that closely aligns with the fitted theoretical model.}
    \label{fig:additional_CSF_function_results}
\end{figure*}

\begin{figure*}
    \centering
    \includegraphics[width=\textwidth]{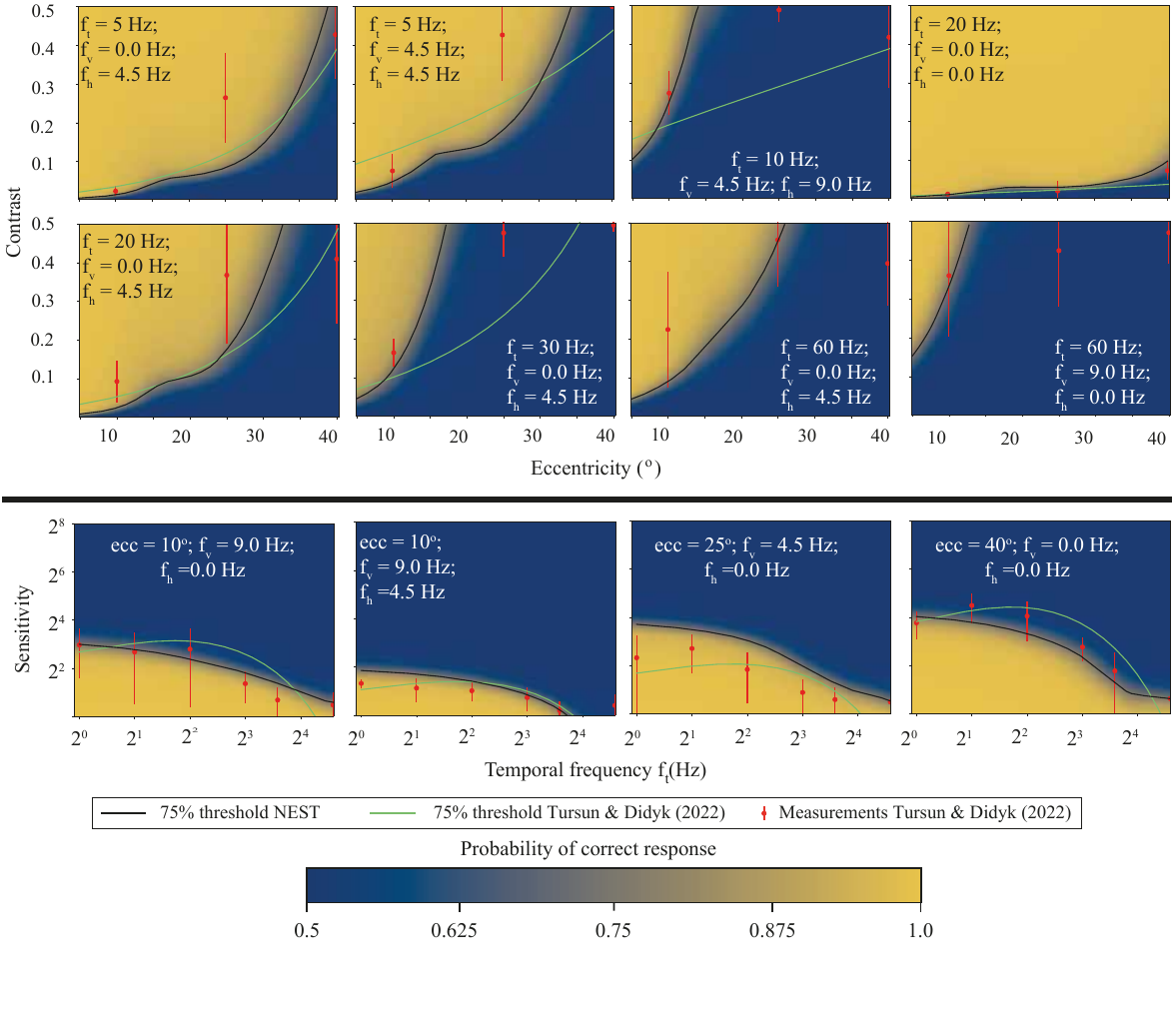}
    \caption{Results of the spatio-temporal experiment for participant P1.}
    \label{fig:spatio_temporal_P1}
\end{figure*}

Finally, Figure \ref{fig:additional_CSF_function_results} shows the estimated psychometric functions for the Barten CSF experiment described in Section \ref{sec:Barten_CSF_results}, including the individual psychometric functions of each participant as well as the combined psychometric function. \new{Tables 6--10 show the architecture study results for the individual functions.} Figures \ref{fig:spatio_temporal_P1} and onwards show the expanded results of the spatio-temporal experiment in Section \ref{sec:Tursun_Didyk_results}.

\begin{table*}[hb]
    \centering
    \setlength{\tabcolsep}{0.15em}
    \footnotesize
    \begin{tabular}{l|c|c|c|c|c|c|c|c}
     \textbf{Method} & \multicolumn{2}{c|}{\textbf{\ac{NV2D} detection}} & \multicolumn{2}{c|}{\textbf{\ac{NV2D} discrimination}} & \multicolumn{2}{c|}{\textbf{\acs{ON2D} detection}} & \multicolumn{2}{c}{\textbf{\acs{ON2D} discrimination}} \\
         & \textbf{AUC$_{\mathrm{RMSE}}$} & \textbf{AUC$_{\mathrm{Brier}}$} & \textbf{AUC$_{\mathrm{RMSE}}$} & \textbf{AUC$_{\mathrm{Brier}}$} & \textbf{AUC$_{\mathrm{RMSE}}$} & \textbf{AUC$_{\mathrm{Brier}}$} & \textbf{AUC$_{\mathrm{RMSE}}$} & \textbf{AUC$_{\mathrm{Brier}}$} \\
        \hline
        NEST (ours) & $\mathbf{22.69 \pm 0.30}$ & $\mathbf{6.03 \pm 0.15}$ & $16.78 \pm 0.17$ & $8.44 \pm 0.16$ & $\mathbf{13.94 \pm 0.18}$ & $\mathbf{3.70 \pm 0.08}$ & $\mathbf{16.85 \pm 0.59}$ & $5.37 \pm 0.36$ \\
        QUEST+ & N/A & N/A & N/A & N/A & N/A & N/A & N/A & N/A \\
        BALD RBF & $28.30 \pm 0.18$ & $11.27 \pm 0.14$ & $18.27 \pm 0.21$ & $10.63 \pm 0.17$ & $23.84 \pm 0.15$ & $6.31 \pm 0.09$ & $25.07 \pm 0.31$ & $5.48 \pm 0.18$ \\
        BALD mon.-RBF & $27.78 \pm 0.19$ & $13.08 \pm 0.17$ & $16.72 \pm 0.25$ & $9.40 \pm 0.20$ & $22.89 \pm 0.15$ & $5.59 \pm 0.08$ & $25.00 \pm 0.25$ & $5.34 \pm 0.15$ \\
        BALD lin.-add. & $35.78 \pm 0.24$ & $17.36 \pm 0.25$ & $20.12 \pm 0.36$ & $11.85 \pm 0.37$ & $29.23 \pm 0.50$ & $7.78 \pm 0.16$ & $31.71 \pm 0.28$ & $12.35 \pm 0.37$ \\
        BALV RBF  & $27.05 \pm 0.17$ & $10.45 \pm 0.14$ & $17.24 \pm 0.21$ & $8.84 \pm 0.16$ & $20.63 \pm 0.13$ & $5.41 \pm 0.07$ & $27.38 \pm 0.34$ & $7.24 \pm 0.30$ \\
        BALV mon.-RBF & $26.05 \pm 0.17$ & $12.12 \pm 0.15$ & $18.47 \pm 0.22$ & $10.31 \pm 0.21$ & $18.89 \pm 0.10$ & $5.29 \pm 0.05$ & $25.82 \pm 0.29$ & $5.73 \pm 0.16$ \\
        BALV lin.-add. & $40.70 \pm 0.24$ & $19.31 \pm 0.28$ & $18.18 \pm 0.26$ & $9.65 \pm 0.29$ & $29.40 \pm 0.07$ & $8.56 \pm 0.07$ & $32.34 \pm 0.31$ & $9.65 \pm 0.36$ \\
        LSE RBF & $37.68 \pm 0.23$ & $8.35 \pm 0.08$ & $15.82 \pm 0.21$ & $\mathbf{7.16 \pm 0.16}$ & $27.93 \pm 0.20$ & $4.76 \pm 0.04$ & $29.95 \pm 0.47$ & $\mathbf{3.81 \pm 0.20}$ \\
        LSE mon.-RBF & $40.84 \pm 0.22$ & $8.33 \pm 0.09$ & $16.20 \pm 0.22$ & $7.87 \pm 0.16$ & $27.90 \pm 0.21$ & $4.32 \pm 0.04$ & $28.53 \pm 0.61$ & $\mathbf{3.37 \pm 0.11}$ \\
        LSE lin.-add. & $51.06 \pm 0.19$ & $7.45 \pm 0.12$ & $16.49 \pm 0.29$ & $\mathbf{6.88 \pm 0.13}$ & $42.31 \pm 0.13$ & $4.76 \pm 0.04$ & $30.34 \pm 0.34$ & $4.90 \pm 0.23$ \\
        EAVC RBF & $48.12 \pm 0.34$ & $11.23 \pm 0.16$ & $26.23 \pm 0.54$ & $13.81 \pm 0.36$ & $40.79 \pm 0.46$ & $8.33 \pm 0.17$ & $39.68 \pm 1.60$ & $17.65 \pm 2.12$ \\
        GlobalMI RBF & $39.09 \pm 0.31$ & $11.85 \pm 0.15$ & $26.30 \pm 0.59$ & $15.23 \pm 0.38$ & $35.46 \pm 0.27$ & $7.88 \pm 0.16$ & $35.53 \pm 1.61$ & $16.01 \pm 2.04$ \\
        \hline \hline
        \textbf{Method} & \multicolumn{2}{c|}{\textbf{\ac{DN2D} detection}} & \multicolumn{2}{c|}{\textbf{\ac{DN2D} discrimination}} & \multicolumn{2}{c|}{\textbf{\acs{MAX2D} detection}} & \multicolumn{2}{c}{\textbf{\acs{MAX2D} discrimination}} \\
         & \textbf{AUC$_{\mathrm{RMSE}}$} & \textbf{AUC$_{\mathrm{Brier}}$} & \textbf{AUC$_{\mathrm{RMSE}}$} & \textbf{AUC$_{\mathrm{Brier}}$} & \textbf{AUC$_{\mathrm{RMSE}}$} & \textbf{AUC$_{\mathrm{Brier}}$} & \textbf{AUC$_{\mathrm{RMSE}}$} & \textbf{AUC$_{\mathrm{Brier}}$} \\
         \hline
         NEST (ours) & $\mathbf{48.18 \pm 0.40}$ & $\mathbf{25.71 \pm 0.30}$ & $\mathbf{41.37 \pm 0.61}$ & $\mathbf{43.88 \pm 1.09}$ & $16.42 \pm 0.16$ & $\mathbf{4.22 \pm 0.06}$ & $20.81 \pm 0.78$ & $14.56 \pm 0.87$ \\
        QUEST+ & N/A & N/A & N/A & N/A & $\mathbf{7.19 \pm 0.12}$ & N/A & $\mathbf{5.74 \pm 0.11}$ & N/A \\
        BALD RBF & $63.52 \pm 1.40$ & $\mathbf{26.10 \pm 0.61}$ & $52.71 \pm 0.47$ & $49.73 \pm 0.90$ & $29.96 \pm 0.10$ & $8.65 \pm 0.12$ & $32.54 \pm 0.09$ & $10.34 \pm 0.11$ \\
        BALD mon.-RBF & $61.38 \pm 0.62$ & $\mathbf{27.60 \pm 0.39}$ & $42.36 \pm 0.40$ & $62.65 \pm 0.81$ & $32.92 \pm 0.12$ & $10.99 \pm 0.13$ & $32.55 \pm 0.09$ & $8.40 \pm 0.07$ \\
        BALD lin.-add. & $103.45 \pm 0.85$ & $102.52 \pm 1.75$ & $69.99 \pm 0.38$ & $88.79 \pm 0.73$ & $28.18 \pm 0.37$ & $11.03 \pm 0.28$ & $50.09 \pm 0.13$ & $16.60 \pm 0.28$ \\
        BALV RBF  & $70.54 \pm 1.50$ & $31.14 \pm 0.83$ & $57.58 \pm 0.63$ & $53.61 \pm 0.88$ & $25.39 \pm 0.13$ & $8.09 \pm 0.09$ & $29.68 \pm 0.09$ & $9.56 \pm 0.09$ \\
        BALV mon.-RBF & $63.10 \pm 0.91$ & $32.82 \pm 0.65$ & $57.93 \pm 0.58$ & $58.65 \pm 1.01$ & $34.51 \pm 0.08$ & $10.08 \pm 0.09$ & $27.54 \pm 0.10$ & $8.01 \pm 0.07$ \\
        BALV lin.-add. & $150.03 \pm 0.86$ & $101.67 \pm 1.02$ & $73.53 \pm 0.48$ & $103.30 \pm 0.84$ & $52.37 \pm 0.12$ & $24.05 \pm 0.20$ & $52.64 \pm 0.09$ & $21.48 \pm 0.15$ \\
        LSE RBF & $111.15 \pm 1.40$ & $44.08 \pm 0.64$ & $63.36 \pm 0.48$ & $58.34 \pm 0.84$ & $31.18 \pm 0.40$ & $6.21 \pm 0.07$ & $41.63 \pm 0.18$ & $\mathbf{5.61 \pm 0.06}$ \\
        LSE mon.-RBF & $94.03 \pm 2.02$ & $40.93 \pm 0.99$ & $62.86 \pm 0.63$ & $60.44 \pm 1.15$ & $32.44 \pm 0.23$ & $5.99 \pm 0.08$ & $44.41 \pm 0.12$ & $\mathbf{5.17 \pm 0.06}$ \\
        LSE lin.-add. & $130.79 \pm 0.76$ & $102.52 \pm 0.60$ & $76.47 \pm 0.59$ & $99.83 \pm 0.67$ & $65.77 \pm 0.25$ & $8.13 \pm 0.15$ & $75.64 \pm 0.16$ & $8.14 \pm 0.06$ \\
        EAVC RBF & $78.82 \pm 0.62$ & $71.41 \pm 1.38$ & $\mathbf{39.71 \pm 0.50}$ & $64.25 \pm 0.76$ & $30.38 \pm 0.52$ & $7.37 \pm 0.12$ & $51.49 \pm 0.33$ & $9.81 \pm 0.12$ \\
        GlobalMI RBF & $127.67 \pm 0.39$ & $58.00 \pm 0.96$ & $75.53 \pm 0.38$ & $94.32 \pm 1.31$ & $23.64 \pm 0.48$ & $6.45 \pm 0.13$ & $43.09 \pm 0.30$ & $7.85 \pm 0.11$ \\
        \hline \hline 
        \textbf{Method} & \multicolumn{2}{c|}{\textbf{\ac{SIN2D} detection}} & \multicolumn{2}{c|}{\textbf{\ac{SIN2D} discrimination}} & \multicolumn{2}{c|}{\textbf{\ac{HART6} detection}} & \multicolumn{2}{c}{\textbf{\ac{HART6} discrimination}} \\
         & \textbf{AUC$_{\mathrm{RMSE}}$} & \textbf{AUC$_{\mathrm{Brier}}$} & \textbf{AUC$_{\mathrm{RMSE}}$} & \textbf{AUC$_{\mathrm{Brier}}$} & \textbf{AUC$_{\mathrm{RMSE}}$} & \textbf{AUC$_{\mathrm{Brier}}$} & \textbf{AUC$_{\mathrm{RMSE}}$} & \textbf{AUC$_{\mathrm{Brier}}$} \\
         \hline
         NEST (ours) & $66.85 \pm 0.32$ & $\mathbf{25.01 \pm 0.24}$ & $53.40 \pm 0.76$ & $\mathbf{53.35 \pm 1.24}$ & $\mathbf{271.15 \pm 2.00}$ & $\mathbf{158.80 \pm 2.87}$ & $\mathbf{156.26 \pm 0.69}$ & $\mathbf{169.79 \pm 1.42}$ \\
        QUEST+ & $\mathbf{57.73 \pm 1.01}$ & N/A & $\mathbf{28.59 \pm 0.36}$ & N/A & N/A & N/A & N/A & N/A \\
        BALD RBF & $119.34 \pm 0.42$ & $71.95 \pm 0.34$ & $66.30 \pm 0.35$ & $69.31 \pm 0.42$ & $287.09 \pm 1.13$ & $307.12 \pm 5.16$ & $167.50 \pm 1.17$ & $200.30 \pm 2.35$ \\
        BALD mon.-RBF & $116.95 \pm 0.52$ & $68.60 \pm 0.38$ & $63.52 \pm 0.26$ & $68.75 \pm 0.36$ & $288.47 \pm 3.25$ & $307.05 \pm 8.81$ & $170.07 \pm 2.67$ & $195.39 \pm 4.35$ \\
        BALD lin.-add. & $127.21 \pm 0.39$ & $78.15 \pm 0.19$ & $63.67 \pm 0.30$ & $73.44 \pm 0.59$ & $342.06 \pm 0.80$ & $360.31 \pm 4.30$ & $183.45 \pm 1.07$ & $262.51 \pm 4.29$ \\
        BALV RBF  & $121.79 \pm 0.66$ & $75.91 \pm 0.52$ & $66.20 \pm 0.41$ & $70.03 \pm 0.69$ & $287.18 \pm 1.34$ & $280.11 \pm 4.08$ & $177.26 \pm 1.41$ & $219.46 \pm 2.88$ \\
        BALV mon.-RBF & $116.11 \pm 0.52$ & $67.59 \pm 0.41$ & $63.74 \pm 0.36$ & $66.79 \pm 0.57$ & $284.36 \pm 1.74$ & $281.93 \pm 8.21$ & $171.89 \pm 1.84$ & $205.22 \pm 3.91$ \\
        BALV lin.-add. & $120.45 \pm 0.30$ & $78.22 \pm 0.24$ & $64.00 \pm 0.39$ & $69.58 \pm 0.82$ & $329.77 \pm 1.21$ & $347.69 \pm 4.03$ & $185.03 \pm 0.95$ & $267.77 \pm 3.06$ \\
        LSE RBF & $116.62 \pm 0.52$ & $66.99 \pm 0.51$ & $74.29 \pm 0.79$ & $72.85 \pm 0.57$ & $342.44 \pm 1.37$ & $256.03 \pm 4.48$ & $173.81 \pm 1.80$ & $198.32 \pm 3.45$ \\
        LSE mon.-RBF & $117.87 \pm 0.49$ & $67.58 \pm 0.49$ & $69.51 \pm 0.90$ & $62.58 \pm 0.33$ & $335.37 \pm 2.36$ & $266.34 \pm 19.77$ & $178.24 \pm 6.70$ & $195.70 \pm 13.56$ \\
        LSE lin.-add. & $116.31 \pm 0.20$ & $73.10 \pm 0.33$ & $65.91 \pm 0.75$ & $66.29 \pm 0.36$ & $340.71 \pm 1.46$ & $327.82 \pm 6.36$ & $178.88 \pm 1.53$ & $232.13 \pm 3.56$ \\
        EAVC RBF & $125.08 \pm 0.49$ & $83.17 \pm 0.68$ & $67.04 \pm 0.60$ & $72.55 \pm 0.85$ & $333.23 \pm 0.81$ & $216.01 \pm 3.21$ & $191.01 \pm 2.68$ & $195.51 \pm 1.64$ \\
        GlobalMI RBF & $116.67 \pm 0.31$ & $66.53 \pm 0.30$ & $63.95 \pm 0.34$ & $71.26 \pm 0.33$ & $311.79 \pm 0.51$ & $253.46 \pm 4.49$ & $168.42 \pm 1.26$ & $200.23 \pm 2.58$ \\
        \hline \hline 
        \textbf{Method} & \multicolumn{2}{c|}{\textbf{PS8D detection}} & \multicolumn{2}{c|}{\textbf{PS8D discrimination}} \\
         & \textbf{AUC$_{\mathrm{RMSE}}$} & \textbf{AUC$_{\mathrm{Brier}}$} & \textbf{AUC$_{\mathrm{RMSE}}$} & \textbf{AUC$_{\mathrm{Brier}}$} \\
         \hline
         NEST (ours) & $370.68 \pm 0.87$ & $204.16 \pm 1.01$ & $215.94 \pm 0.51$ & $272.73 \pm 1.23$ \\
        QUEST+ & N/A & N/A & N/A & N/A \\
        BALD RBF & $375.09 \pm 2.16$ & $290.86 \pm 2.34$ & $215.40 \pm 1.28$ & $278.04 \pm 4.13$ \\
        BALD mon.-RBF & $401.96 \pm 1.95$ & $313.53 \pm 1.86$ & $220.64 \pm 1.70$ & $292.94 \pm 4.99$ \\
        BALD lin.-add. & $394.19 \pm 1.01$ & $282.32 \pm 1.32$ & $215.85 \pm 0.90$ & $275.89 \pm 2.43$ \\
        BALV RBF & $367.19 \pm 1.97$ & $282.28 \pm 2.00$ & $222.59 \pm 1.43$ & $291.17 \pm 3.94$ \\
        BALV mon.-RBF & $395.95 \pm 1.56$ & $307.94 \pm 2.14$ & $227.02 \pm 2.01$ & $304.46 \pm 4.70$ \\
        BALV lin.-add. & $395.71 \pm 1.32$ & $285.70 \pm 1.44$ & $224.10 \pm 1.46$ & $295.52 \pm 3.65$ \\
        LSE RBF & $405.42 \pm 2.54$ & $244.03 \pm 3.38$ & $229.19 \pm 1.62$ & $296.55 \pm 3.89$ \\
        LSE mon.-RBF & $407.75 \pm 3.50$ & $242.43 \pm 4.99$ & $233.93 \pm 2.08$ & $304.42 \pm 4.59$ \\
        LSE lin.-add. & $379.24 \pm 1.23$ & $202.10 \pm 1.81$ & $208.69 \pm 1.48$ & $233.22 \pm 3.04$ \\
        EAVC RBF & $\mathbf{328.22 \pm 0.49}$ & $\mathbf{183.26 \pm 0.71}$ & $\mathbf{187.66 \pm 0.57}$ & $\mathbf{197.17 \pm 1.33}$ \\
        GlobalMI RBF & $332.21 \pm 0.56$ & $204.72 \pm 0.93$ & $192.67 \pm 0.60$ & $211.61 \pm 1.49$ \\

    \end{tabular}
    \caption{The mean \ac{AUC} $\pm$ standard error of the \ac{RMSE} and Brier metrics for the \ac{NV2D}, \ac{ON2D}, \ac{DN2D}, \ac{MAX2D}, \ac{SIN2D}, \ac{HART6}, and \ac{PS8D} detection and discrimination functions. The best method and methods that are statistically similar ($p \geq 0.05$, Games-Howell test) to it are shown in bold typeface for each metric. }
    \label{tab:full_2D_AUC_results}
\end{table*}

\begin{table*}[hb]
    \centering
    \setlength{\tabcolsep}{0.25em}
    \footnotesize
    \begin{tabular}{l|c|c|c|c|c|c|c|c}
     \textbf{Method} & \multicolumn{2}{c|}{\textbf{WEI1D detection}} & \multicolumn{2}{c|}{\textbf{WEI1D discrimination}} & \multicolumn{2}{c|}{\textbf{WEI2D detection}} & \multicolumn{2}{c}{\textbf{WEI2D discrimination}} \\
         & \textbf{AUC$_{\mathrm{RMSE}}$} & \textbf{AUC$_{\mathrm{Brier}}$} & \textbf{AUC$_{\mathrm{RMSE}}$} & \textbf{AUC$_{\mathrm{Brier}}$} & \textbf{AUC$_{\mathrm{RMSE}}$} & \textbf{AUC$_{\mathrm{Brier}}$} & \textbf{AUC$_{\mathrm{RMSE}}$} & \textbf{AUC$_{\mathrm{Brier}}$} \\
        \hline
        NEST (ours) & $10.17 \pm 0.18$ & $\mathbf{2.68 \pm 0.07}$ & $10.19 \pm 0.46$ & $\mathbf{5.71 \pm 0.32}$ & $\mathbf{15.76 \pm 0.20}$ & $\mathbf{3.10 \pm 0.08}$ & $23.21 \pm 0.83$ & $\mathbf{12.37 \pm 0.67}$ \\
        QUEST+ & $\mathbf{4.99 \pm 0.17}$ & N/A & $\mathbf{4.58 \pm 0.16}$ & N/A & $\mathbf{16.35 \pm 0.51}$ & N/A & $\mathbf{10.97} \pm 0.32$ & N/A \\
        BALD RBF & $17.45 \pm 0.13$ & $5.36 \pm 0.11$ & $31.68 \pm 0.44$ & $10.75 \pm 0.50$ & $31.10 \pm 0.26$ & $12.18 \pm 0.16$ & $45.13 \pm 0.48$ & $18.73 \pm 0.54$ \\
        BALD mon.-RBF & $29.61 \pm 0.13$ & $8.84 \pm 0.10$ & $39.89 \pm 0.62$ & $20.38 \pm 0.68$ & $26.01 \pm 0.12$ & $6.81 \pm 0.08$ & $40.35 \pm 0.54$ & $16.68 \pm 0.53$ \\
        BALD lin.-add. & $173.89 \pm 0.48$ & $135.80 \pm 0.38$ & $79.48 \pm 0.92$ & $118.74 \pm 4.35$ & $68.04 \pm 0.16$ & $31.05 \pm 0.34$ & $50.78 \pm 0.64$ & $28.50 \pm 0.91$ \\
        BALV RBF & $20.33 \pm 0.44$ & $7.15 \pm 0.22$ & $22.12 \pm 0.38$ & $8.94 \pm 0.40$ & $28.71 \pm 0.10$ & $8.76 \pm 0.07$ & $48.34 \pm 0.61$ & $26.98 \pm 0.85$ \\
        BALV mon.-RBF & $23.56 \pm 0.23$ & $7.35 \pm 0.13$ & $48.01 \pm 0.57$ & $25.07 \pm 0.90$ & $24.87 \pm 0.12$ & $6.86 \pm 0.09$ & $50.19 \pm 0.57$ & $20.94 \pm 0.84$ \\
        BALV lin.-add. & $177.48 \pm 0.36$ & $117.09 \pm 1.12$ & $64.05 \pm 0.33$ & $104.15 \pm 2.68$ & $57.36 \pm 0.24$ & $25.29 \pm 0.21$ & $50.78 \pm 0.64$ & $38.78 \pm 1.22$ \\
        LSE RBF & $28.06 \pm 0.26$ & $\mathbf{2.74 \pm 0.04}$ & $35.93 \pm 1.43$ & $7.98 \pm 0.43$ & $35.75 \pm 0.26$ & $5.70 \pm 0.06$ & $56.72 \pm 1.24$ & $\mathbf{9.91 \pm 0.43}$ \\
        LSE mon.-RBF & $38.19 \pm 0.20$ & $5.61 \pm 0.08$ & $57.19 \pm 1.59$ & $11.50 \pm 0.53$ & $31.16 \pm 0.20$ & $3.48 \pm 0.04$ & $49.73 \pm 1.01$ & $\mathbf{11.28 \pm 0.45}$ \\
        LSE lin.-add. & $155.03 \pm 0.02$ & $144.65 \pm 3.05$ & $70.46 \pm 0.21$ & $122.14 \pm 2.01$ & $74.33 \pm 0.22$ & $16.94 \pm 0.14$ & $59.87 \pm 1.51$ & $15.93 \pm 0.72$ \\
        EAVC RBF & $28.17 \pm 0.21$ & $2.90 \pm 0.04$ & $30.15 \pm 1.14$ & $7.49 \pm 0.46$ & $31.93 \pm 0.17$ & $5.74 \pm 0.05$ & $46.34 \pm 0.92$ & $16.21 \pm 0.66$ \\
        GlobalMI RBF & $13.93 \pm 0.33$ & $3.29 \pm 0.07$ & $23.63 \pm 0.82$ & $14.20 \pm 0.65$ & $26.42 \pm 0.17$ & $5.70 \pm 0.06$ & $43.69 \pm 0.49$ & $12.58 \pm 0.41$ \\
        \hline \hline
        \textbf{Method} & \multicolumn{2}{c|}{\textbf{WEI3D detection}} & \multicolumn{2}{c|}{\textbf{WEI3D discrimination}} & \multicolumn{2}{c|}{\textbf{WEI4D detection}} & \multicolumn{2}{c}{\textbf{WEI4D discrimination}} \\
         & \textbf{AUC$_{\mathrm{RMSE}}$} & \textbf{AUC$_{\mathrm{Brier}}$} & \textbf{AUC$_{\mathrm{RMSE}}$} & \textbf{AUC$_{\mathrm{Brier}}$} & \textbf{AUC$_{\mathrm{RMSE}}$} & \textbf{AUC$_{\mathrm{Brier}}$} & \textbf{AUC$_{\mathrm{RMSE}}$} & \textbf{AUC$_{\mathrm{Brier}}$} \\
         \hline
         NEST (ours) & $\mathbf{31.52 \pm 0.61}$ & $\mathbf{4.62 \pm 0.17}$ & $50.67 \pm 0.78$ & $\mathbf{14.38 \pm 0.53}$ & $\mathbf{43.85 \pm 0.80}$ & $\mathbf{8.14 \pm 0.36}$ & $\mathbf{75.69 \pm 1.13}$ & $\mathbf{20.20 \pm 0.99}$ \\
        QUEST+ & $55.27 \pm 1.82$ & N/A & $\mathbf{34.35 \pm 1.26}$ & N/A & $110.74 \pm 3.82$ & N/A & $\mathbf{82.76 \pm 3.02}$ & N/A \\
        BALD RBF & $104.44 \pm 0.65$ & $62.23 \pm 0.75$ & $77.34 \pm 0.67$ & $126.80 \pm 1.01$ & $77.91 \pm 0.25$ & $18.89 \pm 0.17$ & $102.88 \pm 0.80$ & $35.79 \pm 0.62$ \\
        BALD mon.-RBF & $42.39 \pm 0.19$ & $8.59 \pm 0.11$ & $77.02 \pm 0.91$ & $25.79 \pm 0.83$ & $76.88 \pm 0.32$ & $23.73 \pm 0.22$ & $99.57 \pm 1.59$ & $33.55 \pm 1.25$ \\
        BALD lin.-add. & $108.59 \pm 0.51$ & $44.66 \pm 0.39$ & $72.29 \pm 0.51$ & $34.00 \pm 0.34$ & $106.47 \pm 0.60$ & $23.77 \pm 0.06$ & $86.25 \pm 0.86$ & $28.84 \pm 0.47$ \\
        BALV RBF  & $122.76 \pm 0.37$ & $58.02 \pm 0.15$ & $74.63 \pm 1.10$ & $116.07 \pm 1.92$ & $100.33 \pm 0.30$ & $24.24 \pm 0.27$ & $105.09 \pm 1.12$ & $29.84 \pm 0.81$ \\
        BALV mon.-RBF & $42.38 \pm 0.24$ & $8.43 \pm 0.14$ & $81.23 \pm 1.37$ & $32.23 \pm 1.27$ & $57.46 \pm 0.63$ & $10.46 \pm 0.25$ & $100.57 \pm 1.79$ & $23.71 \pm 0.73$ \\
        BALV lin.-add. & $112.89 \pm 0.31$ & $59.06 \pm 0.26$ & $80.92 \pm 0.73$ & $45.40 \pm 0.83$ & $129.73 \pm 0.71$ & $39.18 \pm 0.22$ & $109.41 \pm 1.73$ & $23.26 \pm 0.68$ \\
        LSE RBF & $87.72 \pm 1.24$ & $43.90 \pm 0.40$ & $64.50 \pm 1.92$ & $93.42 \pm 2.70$ & $73.82 \pm 0.32$ & $\mathbf{7.31 \pm 0.06}$ & $121.59 \pm 1.74$ & $\mathbf{21.92 \pm 0.34}$ \\
        LSE mon.-RBF & $53.68 \pm 0.36$ & $7.48 \pm 0.07$ & $82.46 \pm 2.21$ & $14.88 \pm 0.40$ & $119.14 \pm 3.57$ & $27.15 \pm 0.96$ & $119.14 \pm 3.57$ & $\mathbf{19.85 \pm 0.55}$ \\
        LSE lin.-add. & $97.14 \pm 0.35$ & $18.44 \pm 0.21$ & $78.35 \pm 1.30$ & $28.27 \pm 0.44$ & $145.64 \pm 2.58$ & $42.47 \pm 1.26$ & $112.87 \pm 1.43$ & $34.57 \pm 0.54$ \\
        EAVC RBF & $73.47 \pm 0.47$ & $14.76 \pm 0.24$ & $79.85 \pm 1.19$ & $\mathbf{13.89 \pm 0.45}$ & $78.81 \pm 0.52$ & $10.61 \pm 0.16$ & $109.41 \pm 1.73$ & $23.64 \pm 0.41$ \\
        GlobalMI RBF & $53.14 \pm 0.30$ & $11.66 \pm 0.24$ & $73.19 \pm 1.25$ & $\mathbf{12.86 \pm 0.31}$ & $102.60 \pm 0.41$ & $22.31 \pm 0.35$ & $103.80 \pm 1.83$ & $26.15 \pm 0.48$

    \end{tabular}
    \caption{The mean \ac{AUC} $\pm$ standard error of the \ac{RMSE} and Brier metrics for the WEI1D-4D functions. The best method and methods that are statistically similar ($p \geq 0.05$, Games-Howell test) to it are shown in bold typeface for each metric. }
    \label{tab:weibull_AUC_results}
\end{table*}

\begin{table}[b!]
    \ssmall
    \caption{Simulation results for different neural network architectures for the test function \ac{NV2D}. The AUC scores are normalized based on the number of trials.}
    \centering
    \begin{tabular}{c|c|c|c|c}
        \textbf{Architecture} & \textbf{$\text{AUC}_{\mathrm{RMSE}}$} & \textbf{$\text{p}^{*}_{\mathrm{RMSE}}$} & \textbf{$\text{AUC}_{\mathrm{Brier}}$} & \textbf{$\text{p}^{*}_{\mathrm{Brier}}$} \\ \hline
        $W_{0} = 128, d = 2$ & 0.0884 & 0.05 & 0.0238 & 0.1 \\
        $W_{0} = 256, d = 2$ & \textbf{0.0780} & 0.05 & 0.0225 & 0.05 \\
        $W_{0} = 512, d = 2$ & 0.0812 & 0.2 & 0.0221 & 0.05 \\
        $W_{0} = 128, d = 3$ & 0.0827 & 0.05 & 0.0226 & 0.05 \\
        $W_{0} = 256, d = 3$ & 0.0833 & 0.1 & 0.0227 & 0.1 \\
        $\mathbf{W}_{\mathrm{experiments}}$ & 0.0833 & 0.15 & 0.0235 & 0.1 \\
        $W_{0} = 512, d = 3$ & 0.0852 & 0.2 & 0.0229 & 0.2 \\
        $W_{0} = 128, d = 4$ & 0.0862 & 0.05 & 0.0241 & 0.05 \\
        $W_{0} = 256, d = 4$ & 0.0871 & 0.1 & \textbf{0.0217} & 0.05 \\
        $W_{0} = 512, d = 4$ & 0.0921 & 0.2 & 0.0225 & 0.1 \\
        $W_{0} = 128, d = 5$ & 0.0910 & 0.05 & 0.0248 & 0.05 \\
        $W_{0} = 256, d = 5$ & 0.0874 & 0.15 & 0.0233 & 0.1 \\
        $W_{0} = 512, d = 5$ & 0.0970 & 0.15 & 0.0233 & 0.1 \\
        Best GP result & 0.0868 & N/A & 0.0248 & N/A \\
    \end{tabular}
    \label{tab:architecture_novel_test_results}
\end{table}

\begin{table}[b!]
    \ssmall
    \caption{Simulation results for different neural network architectures for the test function \ac{SIN2D}. The AUC scores are normalized based on the number of trials.}
    \centering
    \begin{tabular}{c|c|c|c|c}
        \textbf{Architecture} & \textbf{$\text{AUC}_{\mathrm{RMSE}}$} & \textbf{$\text{p}^{*}_{\mathrm{RMSE}}$} & \textbf{$\text{AUC}_{\mathrm{Brier}}$} & \textbf{$\text{p}^{*}_{\mathrm{Brier}}$} \\ \hline
        $W_{0} = 128, d = 2$ & 0.3081 & 0.05 & 0.1510 & 0.05 \\
        $W_{0} = 256, d = 2$ & 0.2358 & 0.05 & 0.0901 & 0.05 \\
        $W_{0} = 512, d = 2$ & 0.1827 & 0.05 & 0.0656 & 0.05 \\
        $W_{0} = 128, d = 3$ & 0.2418 & 0.05 & 0.0954 & 0.05 \\
        $W_{0} = 256, d = 3$ & 0.1835 & 0.05 & 0.0666 & 0.05 \\
        $\mathbf{W}_{\mathrm{experiments}}$ & 0.1946 & 0.05 & 0.0752 & 0.05 \\
        $W_{0} = 512, d = 3$ & \textbf{0.1663} & 0.05 & 0.0612 & 0.1 \\
        $W_{0} = 128, d = 4$ & 0.2375 & 0.05 & 0.0961 & 0.05 \\
        $W_{0} = 256, d = 4$ & 0.1736 & 0.05 & 0.0656 & 0.05 \\
        $W_{0} = 512, d = 4$ & 0.1672 & 0.1 & \textbf{0.0602} & 0.1 \\
        $W_{0} = 128, d = 5$ & 0.2950 & 0.05 & 0.1447 & 0.05 \\
        $W_{0} = 256, d = 5$ & 0.1792 & 0.05 & 0.0698 & 0.05 \\
        $W_{0} = 512, d = 5$ & 0.1725 & 0.1 & 0.0616 & 0.1 \\
        Best GP result & 0.3870 & N/A & 0.2233 & N/A \\
    \end{tabular}
    \label{tab:architecture_sinusoid_results}
\end{table}

\begin{table}[b!]
    \ssmall
    \caption{Simulation results for different neural network architectures for the test function \ac{DN2D}. The AUC scores are normalized based on the number of trials.}
    \centering
    \begin{tabular}{c|c|c|c|c}
        \textbf{Architecture} & \textbf{$\text{AUC}_{\mathrm{RMSE}}$} & \textbf{$\text{p}^{*}_{\mathrm{RMSE}}$} & \textbf{$\text{AUC}_{\mathrm{Brier}}$} & \textbf{$\text{p}^{*}_{\mathrm{Brier}}$} \\ \hline
        $W_{0} = 128, d = 2$ & 0.2805 & 0.05 & 0.1372 & 0.05 \\
        $W_{0} = 256, d = 2$ & 0.1904 & 0.05 & 0.0808 & 0.05 \\
        $W_{0} = 512, d = 2$ & 0.1488 & 0.05 & 0.0696 & 0.05 \\
        $W_{0} = 128, d = 3$ & 0.2015 & 0.05 & 0.0858 & 0.05 \\
        $W_{0} = 256, d = 3$ & 0.1491 & 0.05 & 0.0685 & 0.1 \\
        $\mathbf{W}_{\mathrm{experiments}}$ & 0.1626 & 0.05 & 0.758 & 0.05 \\
        $W_{0} = 512, d = 3$ & 0.1423 & 0.15 & \textbf{0.0633} & 0.25 \\
        $W_{0} = 128, d = 4$ & 0.2197 & 0.05 & 0.1032 & 0.05 \\
        $W_{0} = 256, d = 4$ & 0.1458 & 0.05 & 0.0677 & 0.1 \\
        $W_{0} = 512, d = 4$ & 0.1489 & 0.15 & 0.0642 & 0.1 \\
        $W_{0} = 128, d = 5$ & 0.2824 & 0.05 & 0.1620 & 0.05 \\
        $W_{0} = 256, d = 5$ & 0.1533 & 0.05 & 0.0704 & 0.05 \\
        $W_{0} = 512, d = 5$ & 0.1519 & 0.15 & 0.0655 & 0.1 \\
        Best GP result & 0.2034 & N/A & 0.0849 & N/A \\
    \end{tabular}
    \label{tab:architecture_donut_results}
\end{table}

\begin{table}[b!]
    \ssmall
    \caption{Simulation results for different neural network architectures for the test function \ac{WEI4D}. The AUC scores are normalized based on the number of trials.}
    \centering
    \begin{tabular}{c|c|c|c|c}
        \textbf{Architecture} & \textbf{$\text{AUC}_{\mathrm{RMSE}}$} & \textbf{$\text{p}^{*}_{\mathrm{RMSE}}$} & \textbf{$\text{AUC}_{\mathrm{Brier}}$} & \textbf{$\text{p}^{*}_{\mathrm{Brier}}$} \\ \hline
        $W_{0} = 128, d = 2$ & 0.0555 & 0.05 & 0.00650 & 0.15 \\
        $W_{0} = 256, d = 2$ & 0.0558 & 0.15 & 0.00895 & 0.2 \\
        $W_{0} = 512, d = 2$ & 0.0551 & 0.3 & 0.00889 & 0.3 \\
        $W_{0} = 128, d = 3$ & 0.0525 & 0.05 & 0.00607 & 0.2 \\
        $W_{0} = 256, d = 3$ & \textbf{0.0456} & 0.3 & \textbf{0.00587} & 0.3 \\
        $\mathbf{W}_{\mathrm{experiments}}$ & 0.0484 & 0.2 & 0.00693 & 0.2 \\
        $W_{0} = 512, d = 3$ & 0.0622 & 0.3 & 0.01036 & 0.3 \\
        $W_{0} = 128, d = 4$ & 0.0576 & 0.05 & 0.00884 & 0.15 \\
        $W_{0} = 256, d = 4$ & 0.0504 & 0.2 & 0.00664 & 0.05 \\
        $W_{0} = 512, d = 4$ & 0.0605 & 0.25 & 0.00753 & 0.25 \\
        $W_{0} = 128, d = 5$ & 0.0671 & 0.05 & 0.00895 & 0.05 \\
        $W_{0} = 256, d = 5$ & 0.0593 & 0.2 & 0.00938 & 0.05 \\
        $W_{0} = 512, d = 5$ & 0.0622 & 0.3 & 0.00795 & 0.05 \\
        Best GP result & 0.0766 & N/A & 0.00975 & N/A \\
    \end{tabular}
    \label{tab:architecture_weibull_4D_results}
\end{table}

\begin{table}[b!]
    \ssmall
    \caption{Simulation results for different neural network architectures for the test function \ac{HART6}. The AUC scores are normalized based on the number of trials.}
    \centering
    \begin{tabular}{c|c|c|c|c}
        \textbf{Architecture} & \textbf{$\text{AUC}_{\mathrm{RMSE}}$} & \textbf{$\text{p}^{*}_{\mathrm{RMSE}}$} & \textbf{$\text{AUC}_{\mathrm{Brier}}$} & \textbf{$\text{p}^{*}_{\mathrm{Brier}}$} \\ \hline
        $W_{0} = 128, d = 2$ & 0.2375 & 0.15 & 0.1067 & 0.15 \\
        $W_{0} = 256, d = 2$ & 0.2265 & 0.15 & 0.1056 & 0.15 \\
        $W_{0} = 512, d = 2$ & 0.2515 & 0.2 & 0.1182 & 0.2 \\
        $W_{0} = 128, d = 3$ & 0.2333 & 0.25 & 0.1075 & 0.25 \\
        $W_{0} = 256, d = 3$ & 0.2133 & 0.3 & \textbf{0.0968} & 0.3 \\
        $\mathbf{W}_{\mathrm{experiments}}$ & 0.1019 & 0.25 & 0.0587 & 0.25 \\
        $W_{0} = 512, d = 3$ & 0.2358 & 0.3 & 0.1066 & 0.3 \\
        $W_{0} = 128, d = 4$ & 0.2103 & 0.15 & 0.0969 & 0.15 \\
        $W_{0} = 256, d = 4$ & 0.2192 & 0.3 & 0.1008 & 0.3 \\
        $W_{0} = 512, d = 4$ & 0.2448 & 0.3 & 0.1117 & 0.3 \\
        $W_{0} = 128, d = 5$ & 0.2279 & 0.15 & 0.1028 & 0.15 \\
        $W_{0} = 256, d = 5$ & 0.2275 & 0.25 & 0.1053 & 0.25 \\
        $W_{0} = 512, d = 5$ & 0.2482 & 0.3 & 0.1100 & 0.25 \\
        Best GP result & 0.2844 & N/A & 0.2159 & N/A \\
    \end{tabular}
    \label{tab:architecture_hartmann6_results}
\end{table}

\begin{figure*}
    \centering
    \includegraphics[width=\textwidth]{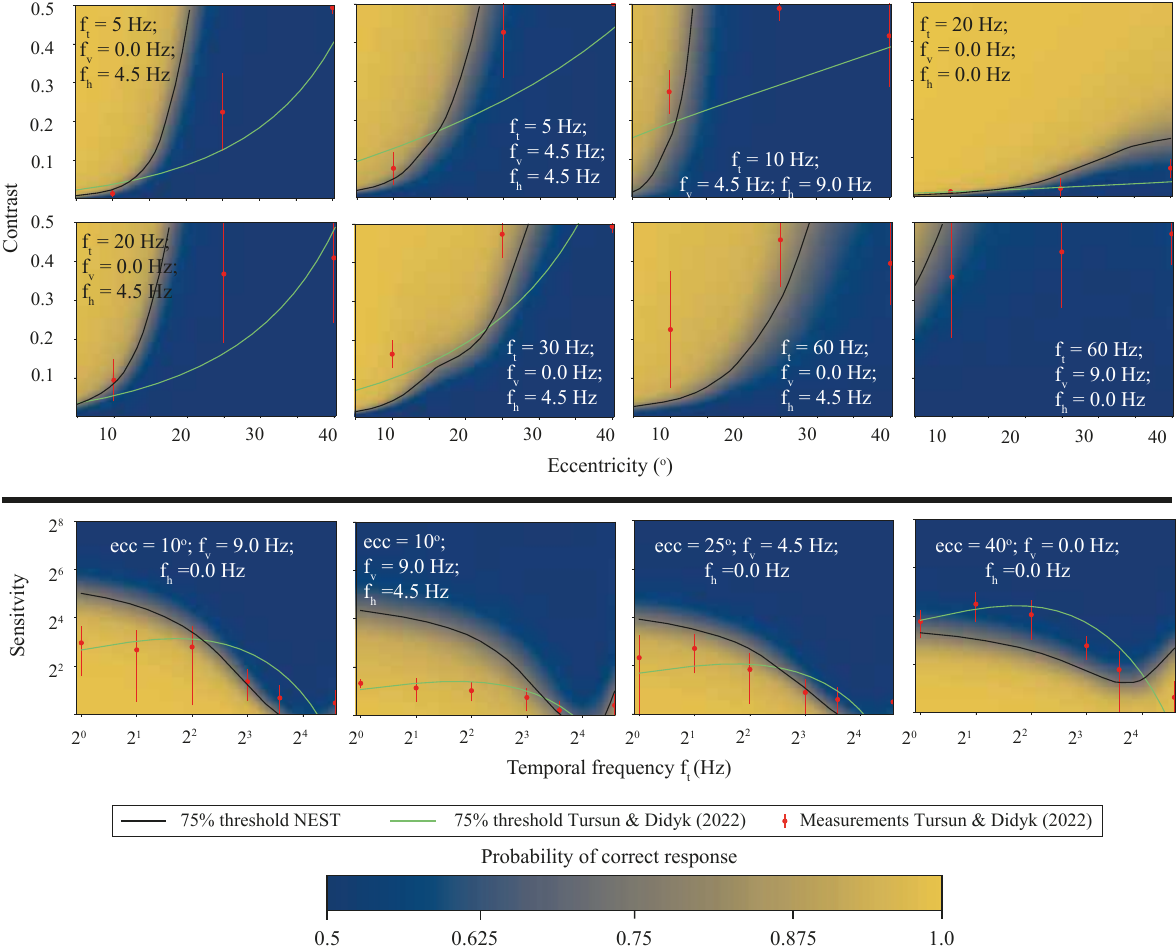}
    \caption{Results of the spatio-temporal experiment for participant P2.}
    \label{fig:spatio_temporal_P2}
\end{figure*}

\begin{figure*}
    \centering
    \includegraphics[width=\textwidth]{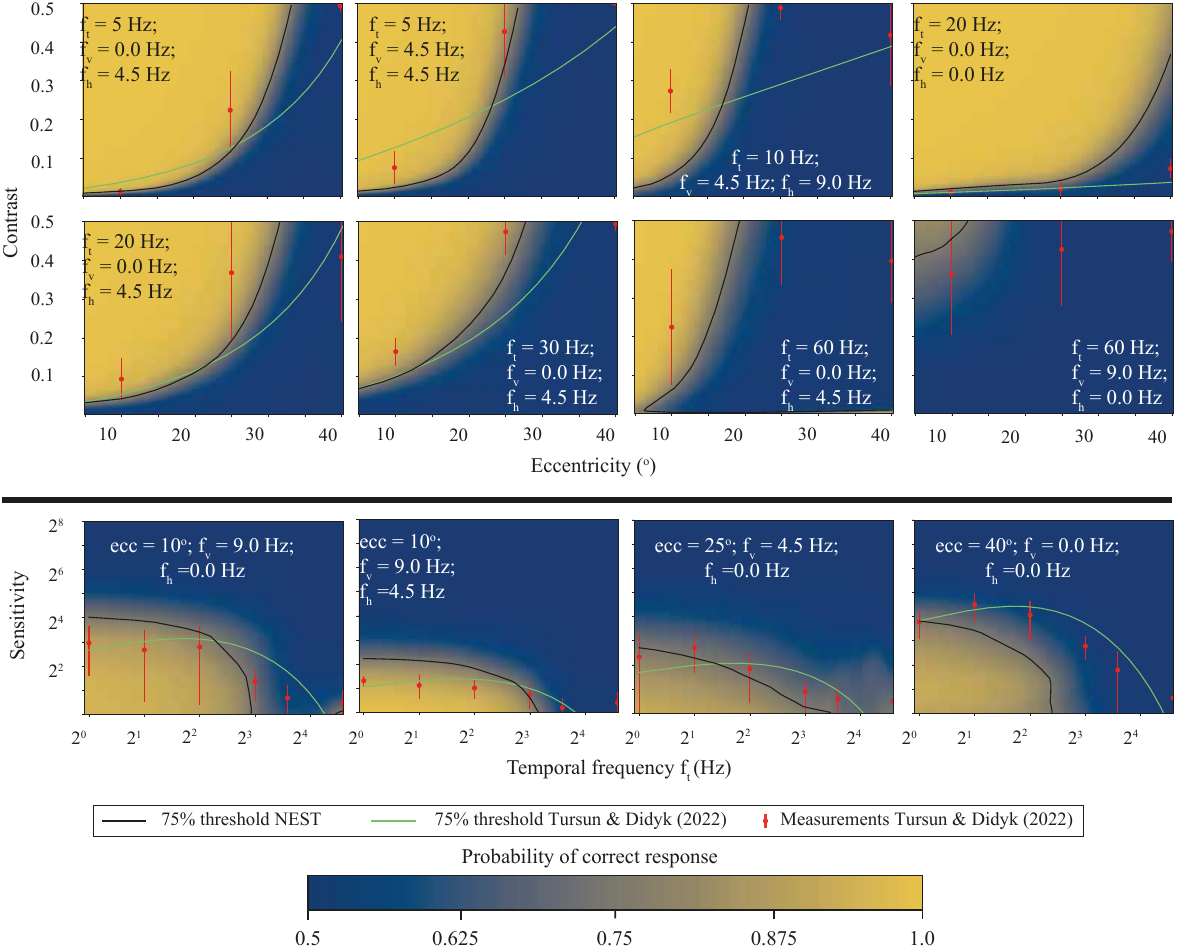}
    \caption{Results of the spatio-temporal experiment for participant P3.}
    \label{fig:spatio_temporal_P3}
\end{figure*}

\begin{figure*}
    \centering
    \includegraphics[width=\textwidth]{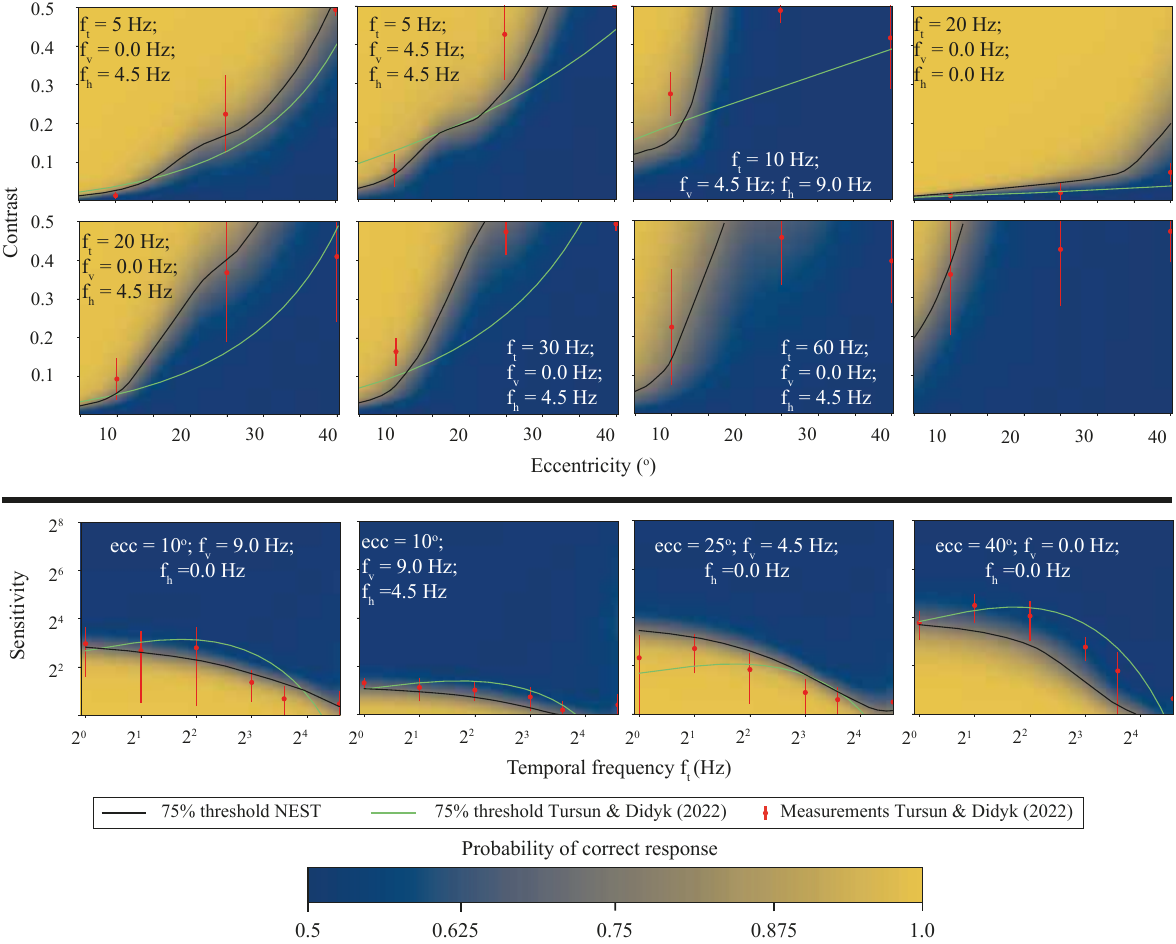}
    \caption{Results of the spatio-temporal experiment for participant P4.}
    \label{fig:spatio_temporal_P4}
\end{figure*}

\begin{figure*}
    \centering
    \includegraphics[width=\textwidth]{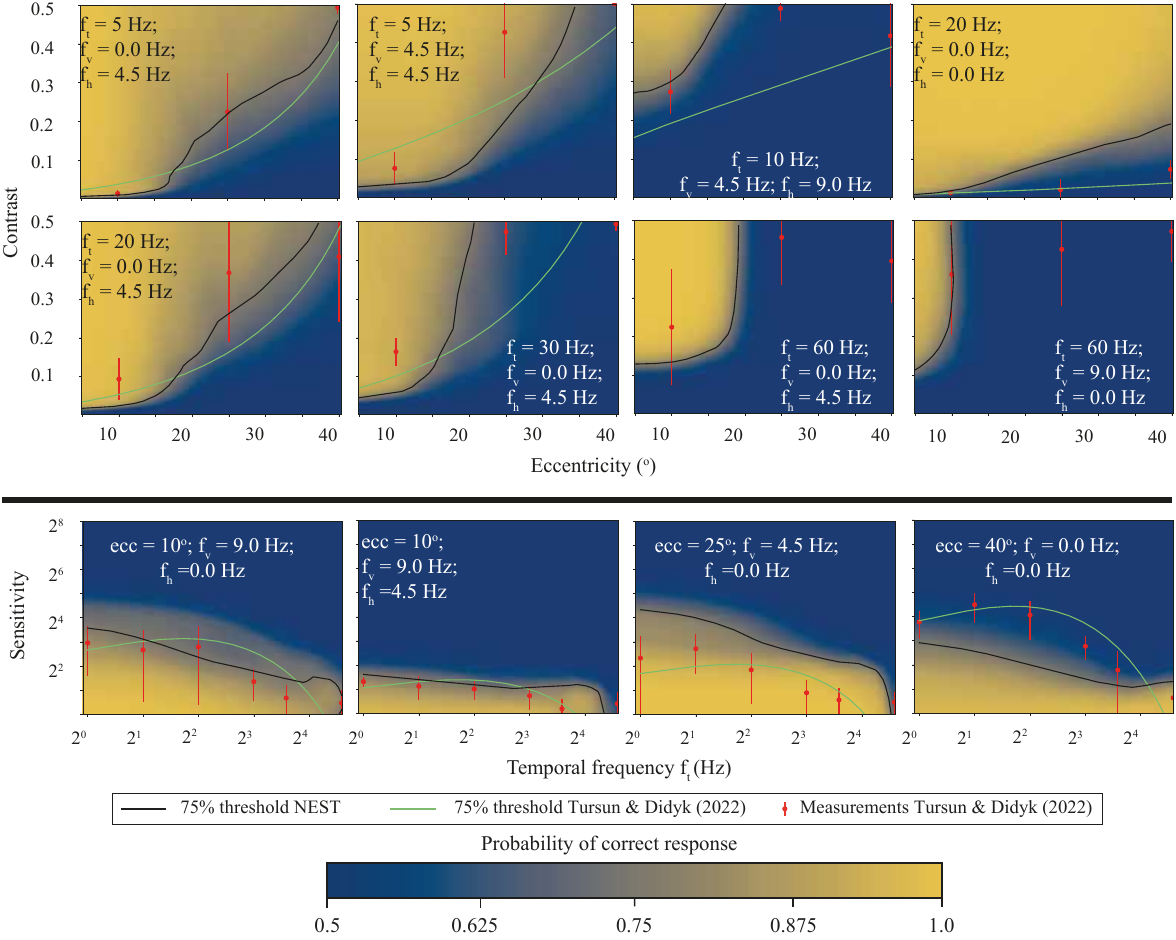}
    \caption{Results of the spatio-temporal experiment for participant P5.}
    \label{fig:spatio_temporal_P5}
\end{figure*}

\begin{figure*}
    \centering
    \includegraphics[width=\textwidth]{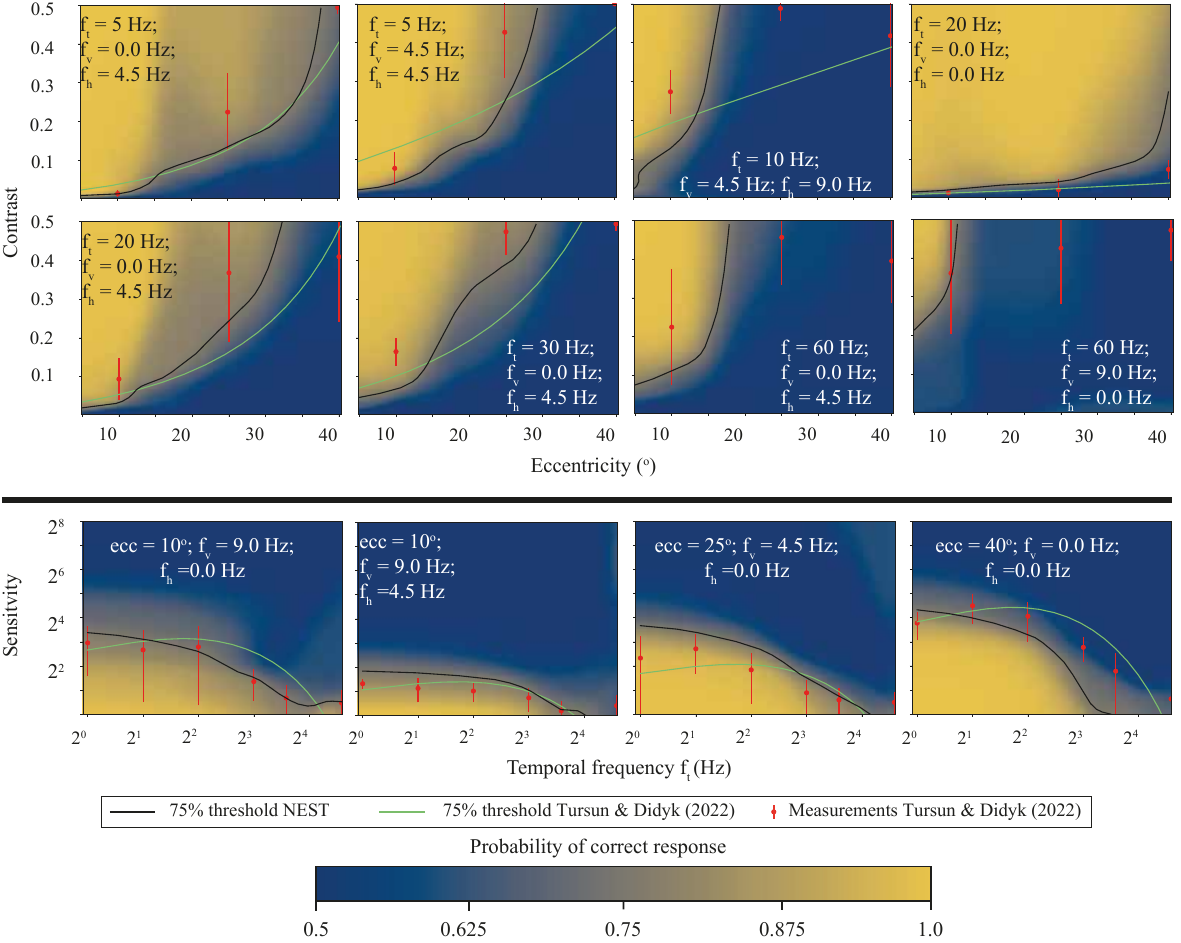}
    \caption{Combined results of the spatio-temporal experiment for all participants.}
    \label{fig:spatio_temporal_combined_full}
\end{figure*}
